\documentstyle[fleqn]{article}
\input epsf

\begin{document}
\baselineskip=12pt
\setcounter{page}{1}

\noindent
{\LARGE\bf
Monte Carlo Simulations of Interfaces 

\vspace{0.2cm}

\noindent
in Polymer Blends
}

\vspace{0.5cm}

\noindent
Marcus M\"uller and Friederike Schmid

\vspace{0.5cm}
\noindent
{\em Institut f\"ur Physik, Universit\"at Mainz, D-55099 Mainz, FRG}

\vspace{1.5cm}

\noindent
{\bf Abstract.}
We review recent simulation studies of interfaces between immiscible
homopolymer phases. Special emphasis is given to the presentation of 
efficient simulation techniques and powerful methods of data analysis, 
such as the analysis of capillary wave spectra. Possible reasons for polymer
incompatibility and ways to relate model dependent interaction parameters
to an effective Flory Huggins parameter $\chi$ are discussed. 
Various interfaces are then considered and characterised with respect
to their microscopic structure and thermodynamic properties. In
particular, interfaces between homopolymers of equal or disparate
stiffness are studied, interfaces containing diblock copolymers, and 
interfaces confined in thin films. The results are related to
the phase behaviour of ternary homopolymer/copolymer systems, and
to wetting transitions in thin films.

\vspace{1cm}

\section{ Introduction}

Blending chemically different polymers is a cheap and relatively 
straightforward way of creating new materials, and polymeric alloys are 
therefore industrially and technologically omnipresent. Prominent examples
are, {\em e.g.}, rubber toughened plastics. Their widespread use 
notwithstanding, polymer mixtures are seldom homogeneous at temperatures of
practical interest. Any slight incompatibility of the monomers, as is usually 
present between different organic molecules, is amplified by the number of 
monomers in the macromolecule, and cannot be balanced by the entropy of 
mixing for typical molecular weights \cite{degennes}. On a mesoscopic scale, 
such materials consist of numerous microdroplets of one phase, which are 
finely dispersed in the other phase. The material properties thus depend 
sensitively on the structure and the properties of the interfaces between 
different phases. Correlations between interfacial and bulk properties
are present on various length scales. 
On the one hand, the local interfacial structure -- the conformation of 
polymers, the interfacial width, which is closely related to the number of 
entanglements between polymers of different type, enrichment of chain
ends or solvent at the interface etc -- has a fundamental influence on the
mechanical stability of the alloy. 
On the other hand, the morphology of the blend at given conditions of 
preparation ({\em e.g.}, given stirring rate during mechanical mixing) is 
basically determined by the interfacial tension: An old theoretical argument 
due to Taylor\cite{taylor1} balances the viscous stress (caused by stirring) 
and the interfacial tension, and predicts that the droplet size should be 
directly proportional to the latter. This law is indeed found 
experimentally\cite{MORPHOLOGY2}. 

A huge number of interesting questions are connected with the general
subject of polymer interfaces. For example, interfacial properties
can be tailored by adding a small amount of a third substance to the 
blend\cite{BATEST}.
In particular, copolymeric surfactants containing both types of monomers
are often used as effective compatibiliser. Their effect on the morphology 
of the blend is twofold. First, they reduce directly the interfacial tension: 
Being compatible with both components, they aggregate at interfaces, thereby 
reducing the number of direct contacts between homopolymers of different 
type\cite{COPINTER}. Second, recent experimental\cite{beck} and 
theoretical\cite{MILNER} studies indicate that their presence at the droplet 
surface prevents the coalescence of the droplets brought into collision in 
the course of mixing. The effect can be related to the two-dimensional 
compression modulus of the copolymer film\cite{MILNER}. Since droplets
can break up, but do not merge any more, one obtains a particularly fine
dispersion. In addition, copolymers improve the mechanical 
properties of interfaces. The interfacial width increases, and likewise
the number of entanglements, they adhesive attraction and the
fracture toughness\cite{BROWN}. At high enough copolymer concentrations,
additional copolymer rich phases emerge which display a variety of
structures ordered on a mesoscopic scale\cite{phdexp,BATES,phdth,PHILIP}. 
These mesoscopically structured materials promise to possess unique and useful
materials properties\cite{BATES,SCIENCE1,SCIENCE2}.

Another complex of important problems refers to confined interfaces, 
{\em i.e.}, interfaces interacting with one or two surfaces. The
presence of a surface influences the interface on all length scales --
the local structure is affected as well as long wavelength fluctuations 
of the interface position. Depending on the interactions of the surface
with the different components of the blend, the interface may bind or unbind, 
giving rise to a whole diversity of wetting phenomena\cite{WET1,WET2,WET3}. 

From the point of view of basic science, inhomogeneous polymer systems
are interesting because of the different length scales involved
(polymer gyration radius vs. monomer size), and because of the additional 
conformational degrees of freedom of polymers as opposed to smaller molecules. 
Interestingly, these apparent complications have partly the effect of 
simplifying the physics: Since polymers interact with so many other polymers, 
microscopic details of the chemical structure of monomers wash out to a large 
extent, and can be absorbed quite successfully into a few number of effective 
parameters. Furthermore, the effective interaction range -- roughly the 
polymer gyration radius -- is very large, and the region in which
critical concentration fluctuations become important is extremely small 
as a consequence (Ginzburg criterion, see Ref. \cite{GINZBURG}).
Polymer blends are thus unusually well described by mean field type theories. 

On the other hand, the treatment on the mean field level in itself is
already very involved, especially if one attempts to account for local
correlations, and is interested in local structures. A number of mean field 
type approaches have been established in the past years, which differ 
by level of coarse-graining and by the type of questions they address. 
Among these, we quote Landau-Ginzburg \cite{holyst,sharon} and scaling 
approaches\cite{degennes}, which coarse-grain over the microscopic and
to some extent even over the chain conformational structure; 
Self-consistent field theories\cite{SCF1,SCF2,SCF3,scheutjens} and
density functional theories \cite{FREED1}, which treat chains as random
walks in a mean field environment, mostly ignoring the monomer structure and 
local chain correlations; lattice-based theories such as the famous Flory-Huggins
theory\cite{FH,flory} and subsequent more refined extensions 
(e.g, \cite{FREED2,FREED3,KIKUCHI}); and finally Schweizer's and Curro's P-RISM 
theory\cite{SCHWEIZERR}, which incorporates the local liquid structure into
a theory of polymer melts, using concepts from integral equation theories 
for simple liquids. The highly coarse-grained theories have the advantage
of relative simplicity, thus allowing in many cases for an analytical
treatment. However, the information they can provide on local structure
properties is, {\em a priori}, very limited. Furthermore, they require 
a number of ``effective'' parameters as input, which cannot be determined
from microscopic parameters within the theory. On the other hand, the
more microscopic details are incorporated into a mean field theory, the
more involved the treatment gets, and the more additional approximations
have to be made in order to make it tractable at all. At the lowest level of 
coarse-graining, for example, the P-RISM equations totally neglect chain
end effects, but their truly self-consistent solution in a one component 
melt nonetheless requires a series of single chain Monte Carlo 
simulations\cite{SCHWEIZERR,SCHWEIZER}.

The universal aspect of the physics of polymeric alloys, and the close 
relationship between the local structure of interfaces and the
global material properties, make them particularly suited for computer 
simulations\cite{binder1,tanny}. These provide simultaneously a detailed 
microscopic picture
of the interfacial structure, and information on the thermodynamics of the 
interfaces. When compared to experiments, they serve as a test of the 
microscopic model which has been used. When compared to theories, they
serve as a test of the theory, within a well-defined microscopic model.
In addition, they may provide structural information which may not yet
be accessible experimentally or theoretically. They can thus contribute
substantially towards a deeper understanding of the connections between the
microscopic parameters, the microscopic structure and the macroscopic 
properties of a material. 

Obviously, there are limitations. With the present computational resources,
full calculations of polymer melts in atomistic detail are currently far 
beyond reach. It is thus necessary to take again advantage of the
universality idea\cite{degennes,flory,freedr}, and represent real polymeric
system by idealised polymer models. In this spirit,
a number of chemical monomer units are mapped onto one
effective monomer with a much simpler structure. Even within such a simplified 
model, only melts of polymers of rather modest chain lengths 
can currently be equilibrated and subsequently studied. 
Depending on the questions that one wishes to study, models of different 
levels of idealisation have to be chosen. The careful choice of a suitable 
model is thus crucial for the success of an investigation. We will discuss this 
important point in more detail in Sec.3. The reader interested in a general
overview over the use of computer simulations in polymer science is referred 
to, {\em e.g.}, the set of excellent reviews in Ref.\cite{binder1}.

In this contribution, we will review some recent simulation studies of polymer
interfaces in polymer blends. We shall restrict ourselves to interfaces at 
thermodynamical equilibrium, and to studies of static properties. 
Our survey will focus on the insight that simulation studies can
provide into the local structure of interfaces, and the implications for 
the global thermodynamics of the systems. We will start with the 
closely related issue of the general relationship between local correlations 
and thermodynamic miscibility in binary blends. Then, we will discuss
interphase boundaries in binary homopolymer blends. The effect of adding
diblock copolymers on the interface and on the phase diagram is
considered.  Finally the behaviour in a thin film is examined, with
special emphasis on the effect of the interfacial fluctuations on
the measured profiles.

\section{Polymer incompatibility and Flory Huggins parameter}

We begin with a discussion of polymer incompatibility: Immiscibility
in polymer blends can be caused by several factors.
First, one has usually a direct relative repulsion between monomers
of different type. In nonpolar molecules, for example, the van der Waals
attraction between monomers $i$ and $j$ is proportional to the product
of their polarisabilities $\alpha_i \alpha_j$. Thus the interaction
between unlike monomers is smaller than the arithmetic mean of the interactions
between like monomers -- which gives rise to a relative repulsion
$\propto (\delta_A - \delta_B)^2$, where $\delta_i \propto \alpha_i$ is
the Hildebrand solubility parameter\cite{hilde}. This enthalpic incompatibility
is inversely proportional to the temperature $T$. 
It may be supplemented by entropic effects: 
If the monomers have different shapes, like monomers tend to pack more 
efficiently than unlike monomers. This effect has been studied in detail
within a lattice model by Freed and coworkers\cite{FREED2}. Similarly, 
stiffness disparities tend to favor demixing, as has been shown within P-RISM 
theory\cite{singh}: When mixed with stiff polymers, the more 
flexible polymers loose conformational entropy, and even though the stiff 
polymers win entropy in return, the net effect turns out to be negative. 
Fredrickson, Liu, and Bates pointed out that phase separation in blends with 
components of different flexibility or architecture is also promoted by long 
range composition correlations\cite{liu}. 
The entropic contribution to the incompatibility of polymers has no direct 
temperature dependence. An indirect temperature dependence may enter through
the chain stiffness. The sum of entropic
and enthalpic contributions will thus generally lead to a complicated
temperature behaviour\cite{FREED2}.

The incompatibility of polymers in a binary A/B-blend is often described in 
terms of a single Flory-Huggins parameter $\chi$. It has originally been 
derived from a simple lattice model on the base of three 
assumptions\cite{degennes}:
\begin{itemize}
\item[(i)] The distribution of polymer conformations does not depend on the
composition of the blend.
\item[(ii)] Composition correlations are neglected.
\item[(iii)] All monomers have equal size (one lattice site per monomer),
  and the melt is incompressible ({\em i.e.}, the lattice is fully 
  occupied by monomers).
\end{itemize}
The approximation (ii) implies, first, that nonlocal correlations induced by 
the chain connectivity (the ``correlation hole'') are ignored, and second,
that short range composition correlations related to local demixing tendencies
are disregarded (``random mixing''). 
One obtains the free energy of mixing per site
\begin{equation}
\label{flory}
F_{FH}/(k_B T)  = \frac{\phi_A}{N_A} \ln(\phi_A) + \frac{\phi_B}{N_B} \ln(\phi_B)
+ \chi \phi_A \phi_B,
\end{equation}
where $\phi_i$ is the volume fraction and $N_i$ the chain length of component 
$i$, and the last term describes the loss of enthalpy upon mixing. Assuming 
that neighbour monomers $i$ and $j$ interact with the interaction energy 
$\epsilon_{ij}$, the Flory-Huggins parameter $\chi$ is given by 
\begin{equation}
\label{chig}
\chi = \frac{z-2}{k_B T} \; \left(\epsilon_{AB}- \frac{\epsilon_{AA}+\epsilon_{BB}}{2} \right),
\end{equation}
with the coordination number $z$ of the lattice. Equation (\ref{chig}) takes 
into account that the interaction of a monomer with its two neighbours along 
the same chain should not contribute to the energy of mixing. 
The free energy (\ref{flory}) describes usual demixing 
behaviour, with a miscibility gap at $\chi$ parameter values larger than
\begin{equation}
 \chi_c = \frac{1}{2}\left(\sqrt{\frac{1}{N_A}}+\sqrt{\frac{1}{N_B}}\right)^2.
 \label{eqn:unmix}
\end{equation}
In the one phase region, the $\chi$ parameter can be determined, {\em e.g.},
from the small angle structure factor\cite{higgins}.
As long as $\chi$ is used as an adjustable, heuristic parameter, this simple 
model is found to perform extremely well in comparison with experiments. 
The value of $\chi$ itself is theoretically less accessible. For example,
neutron scattering data reveal in apparent contradiction with Eqn.(\ref{chig}) 
that $\chi$ may depend significantly on the 
composition of the blend\cite{bates1,krishna,taylor2}.

However, this apparent failure is not surprising, since the crudest 
version of the Flory theory disregards a number of effects which influence
real polymeric fluids. In particular, the relative repulsion between 
polymers of different type depends on the number and the character of contacts 
between polymers. One can distinguish between different factors.

First, the total volume (at given constant pressure) or the pressure 
(at given constant volume) of the mixture depends on the composition of the 
blend. This effect is commonly referred to as the ``equation-of-state''
effect. Experiments are usually conducted under constant pressure conditions, 
whereas simulations and analytic calculations often use a simulation box of fixed 
volume.  Since the thermodynamics of mixtures is conceptually simpler at constant
pressure $p$, we shall mostly discuss the $NpT$ ensemble
in this section. We will relate our conclusions to constant-volume simulations 
later. At a given pressure, the different components will generally have 
different monomer densities $\varrho_A^{*}$ and $\varrho_B^{*}$ in the pure 
states, and the volume per monomer in the mixture is to lowest approximation 
the appropriately weighted average of $1/\varrho_A^{*}$ and $1/\varrho_B^{*}$. 
In addition, the incompatibility of unlike chains often leads to an 
``excess volume on mixing'' $v_{\rm exc}$. As we shall see, the composition 
dependence of $v_{\rm exc}$ is roughly parabolic, 
$v_{\rm exc} \approx \tilde{v} \rho (1-\rho)$.  All taken together, 
the density $\varrho$ of the mixture is given by 
\begin{equation}
\label{rho}
1/\varrho = 
\rho/\varrho_A^{*} + (1-\rho)/\varrho_B^{*} + \tilde{v} \rho 
(1-\rho),
\end{equation}
where $\rho$ is the {\em number} fraction of monomers $A$.
Since the number of contacts per polymer is proportional to the density,
the composition dependence of $\varrho$ translates directly into a composition 
dependence of the $\chi$ parameter in blends of polymers with dissimilar
monomer structure, $\varrho_A^{*} \ne \varrho_B^{*}$. Thus $\chi$ should depend
linearly on the composition of the blend. The linear contribution vanishes in 
blends of monomers with very similar monomer structure, {\em e.g.}, isotopic
blends or saturated hydrocarbon mixtures, and the remaining composition 
dependence is parabolic\cite{taylor2}. This is indeed observed 
experimentally\cite{bates1,krishna}. Equation-of-state effects are thus
apparently responsible for the linear part of the composition dependence 
of $\chi$. However, the magnitude of experimentally observed parabolic 
contributions cannot be explained by Eqn.(\ref{rho}) 
alone\cite{kumar1}.

A second important factor is the local structure of the fluid, {\em i.e.},
the form of the correlation functions. Let us first note that the demixing
is mainly driven by the {\em inter}molecular contacts between monomers
of {\em different} chains. The {\em intra}molecular contacts contribute
to the conformational free energy of single chains, which does not change
very much upon mixing, since conformational changes are generally not
very high\cite{VILGIS2}. In the simplest approximation, the pair distribution of 
monomers $i$ and $j$ from different chains is assumed to have the form
\begin{equation}
\label{rpm}
 \varrho^{(2),inter}_{ij}(r) = \varrho_i \varrho_j \; g_{ij}^{inter}(r)
\qquad \mbox{with} \qquad
g_{ij}^{inter} (r) \equiv g^{inter}(r),
\end{equation}
where the pair correlation function $g^{inter}(r)$ depends neither on the type
$i$ and $j$ of the monomers nor on the composition. It is normalised such that
$g(r) \to 1$ for $r \to \infty$. In real fluids,
this approximation will fail in two respects:
First, the local {\em packing} of chains depends on the chain species, either
directly due to monomer structure differences (monomer size etc.), or as a 
more subtle result of chain structure differences (chain architecture,
chain stiffness). For example, the position of the peak in the
correlation function which corresponds to the first coordination shell 
depends on the size of the central monomer\cite{WEINHOLD}. The reasons for non-random 
packing are generally entropic: Excluded volume effects (related to the 
effective monomer sizes), effects of chain conformational entropy\cite{STIFF1} etc. 
Energetic interactions usually do not affect the packing very much. 
A particularly strong effect on the demixing behaviour can be expected if 
like monomers pack closer than unlike monomers. Monte Carlo studies have 
shown that such ``non-additive packing'' alone is sufficient to bring about 
phase separation\cite{STIFF1,GARY1}.
Second, even if the local fluid structure is preserved, {\em i.e.}, the sum
\mbox{$\sum_j \varrho^{(2),inter}_{ij}(r)/\varrho_i$} is independent of $i$, one 
still expects local composition fluctuations. Such ``non-random-mixing'' 
also affects the demixing behaviour\cite{kumar1}, especially very close to the 
critical point\cite{Deutsch}. According to the Ginzburg 
criterion\cite{GINZBURG}, however, the random-mixing approximation becomes 
better upon increasing the chain length\cite{M0}.

To summarize, demixing in polymer blends occurs for energetic and entropic
reasons. The energetic factors include: Energetic incompatibility
of monomers, and shifts of the ratio between inter- and intramolecular
monomer contacts, caused by conformational changes. 
The entropic factors include: Entropic incompatibility of monomers 
({\em e.g.}, due to non-additive packing), packing inhomogeneities
due to the different chain structure, and conformational changes of
the chain. 
A huge amount of
theoretical work has been devoted to elucidate the effect and the importance
of the different contributions\cite{FREED2,SCHWEIZERR, Schweizer93}. In most cases, the energetic or 
entropic incompatibility of {\em monomers} dominates the demixing behaviour. 
On the other hand, details of the monomer interactions are irrelevant
on the scale of whole chains. It is thus reasonable to retain the spirit
of the Flory theory and absorb the microscopic details into a few
effective parameters, {\em e.g.}, the $\chi$ parameter and the compressibility.
These can be used as input parameters into theories of more complex
systems such as polymer interfaces and surfaces. In addition, they allow 
to relate simulations of coarse-grained polymer models to experimental 
systems and idealised theories. 

The problem remains to calculate $\chi$ for a given simulation model. 
Obviously, Eqn.(\ref{chig}) cannot be used for any model 
different from the Flory lattice model. However, the extension of the
Flory theory to continuous space models or more sophisticated lattice
models is relatively straightforward. This shall be demonstrated in the 
following. We emphasise that we do not aim to present a complete theory of the 
$\chi$ parameter, nor to review the state of the art of generalised Flory
theories. Rather, we wish to present a simple ``recipe'' for the calculation 
of $\chi$, one which takes into account the dominant contributions, and
thus gives good results for most practical purposes. In fact, our type
of approach has usually provided a good quantitative understanding
of simulation data in the past\cite{STIFF1,M0,MBO,SM,STIFF2}.

We consider a mixture of $n_A$, $n_B$ polymers of length $N_A$, $N_B$. 
Polymers $k$ are characterised by their center of mass position
$\vec{R}_k$ and the relative coordinates of the monomers 
$\vec{u}_{j,k} = \vec{r}_{j,k}-\vec{R}_k$. The general partition function 
of such a system in the given volume $V$ can be written in the form
\begin{equation}
{\cal Z} = \frac{(V/V_0)^{n_A+n_B}}{n_A! n_B!} \exp(-m \: f(\rho,\varrho)),
\end{equation}
where $m = n_A N_A + n_B N_B$ is the total number of monomers in the system,
$V_0$ is an arbitrary reference volume, and $f(\rho,\varrho)$ is defined by
\begin{equation}
e^{-m \: f(\rho,\varrho))} = 
\prod_{k=1}^{n_A+n_B} \int_{\Omega} d^3 [\frac{\vec{R}_k}{V^{1/3}}]
\Big\{
\prod_{j=1}^{N_k} \int d^3 \vec{u}_{j,k} 
\Big\}
e^{-\beta U_{}[\{\vec{r}_{j,k}\}]} 
\end{equation}
with $\beta = 1/k_B T$ and the total energy $U$. Note that the center of 
mass positions $\vec{R}_k$ have been rescaled such that the integration
volume $\Omega$ does not depend on the volume $V$ any more.
The Helmholtz free energy thus reads
\begin{equation}
\beta F = n_A \ln (\frac{n_A}{V/V_0}) + n_B \ln (\frac{n_B}{V/V_0}) 
+ m\: f(\rho,\varrho).
\end{equation}
The first two terms describe the combinatorial entropy of mixing and the
translational entropy of the center of mass of polymers. The last term 
subsumes the remaining contributions to the free energy, {\em i.e.}, the  internal energy and the
conformational entropy. Since both are proportional to 
the number of monomers $m$, it is conveniently expressed in terms of
a ``monomer free energy'' $f(\rho,\varrho)$, which depends on the
total density $\varrho$ and the number fraction of $A$ monomers $\rho$. The 
pressure $p$ at volume $V$ is given by
\begin{equation}
\label{press}
\beta p = - \frac{\partial}{\partial V} \beta F
= \varrho (\frac{\rho}{N_A} + \frac{1-\rho}{N_B} 
+ \varrho \frac{\partial f}{\partial \varrho}).
\end{equation}
and the compressibility $\kappa$ by
\begin{equation}
\label{comp}
\beta \: \kappa{}^{-1} = \varrho \frac{\partial \beta p}{\partial \varrho} 
= 2 \beta p + \varrho^3 \frac{\partial^2 f}{\partial \varrho^2},
\end{equation}
where terms of order $1/N$ (ideal gas contributions) have been neglected.
Equation (\ref{press}) can be used as an implicit expression for the density
$\varrho(\rho,p)$, as a function of composition $\rho$ and pressure $p$.
At constant pressure, the Gibbs free energy per monomer $\mu$ then reads
\begin{equation}
\beta \mu(\rho,p) = 
\frac{\rho}{N_A} \ln (\frac{\rho \varrho V_0}{N_A}) + 
\frac{1-\rho}{N_B} \ln (\frac{(1-\rho) \varrho V_0}{N_B}) + f(\rho,\varrho)
+ \frac{\beta p}{\varrho}.
\end{equation}
One obtains the excess free energy of mixing
\begin{eqnarray}
\label{mexc0}
\beta \mu^{\rm exc} &\equiv& \beta \mu(\rho,p) 
- \rho \beta \mu(1,p) - (1-\rho) \beta \mu(0,p) \\
&=& \frac{\rho}{N_A} \ln (\rho \frac{\varrho}{\varrho_A^*}) 
+ \frac{1-\rho}{N_B} \ln ((1-\rho) \frac{\varrho}{\varrho_B^*})
+ \beta p v_{\rm exc}\nonumber\\
&&+ [f(\rho,\varrho) - \rho \: f(1,\varrho_A^*) - (1-\rho) \: f(0,\varrho_B^*) ]
\end{eqnarray}
with the densities of the pure component melt $\varrho_A^*$ and $\varrho_B^*$, 
and the excess volume on mixing per monomer $v_{\rm exc}$. 
$\mu^{\rm exc}$ 
can be used to define an effective Flory Huggins parameter 
$\chi_{\mbox{\small eff}}$ 
\begin{equation}
\label{mexc1}
\beta \mu^{\rm exc}  = 
\frac{\rho}{N_A} \ln \phi + \frac{1-\rho}{N_B} \ln(1-\phi)
+ \chi_{\mbox{\small eff}} \: \rho (1-\rho),
\end{equation}
where $\phi$ is the {\em volume} fraction of monomers $A$. 
\begin{equation}
\phi = \frac{\rho}{\varrho_A^*}
\Big/(\frac{\rho}{\varrho_A^*} + \frac{1-\rho}{\varrho_B^*}).
\end{equation}
The temperature dependence and the phase behaviour results from the interplay 
between the enthalpic and entropic contributions. Since the 
entropy of mixing is proportional to the inverse chain length, 
entropic contributions of different origin to the effective
Flory-Huggins parameter $\chi_{\rm eff}$ are much more important in 
macromolecular blends than in mixtures of small molecules.

In order to proceed, it will be useful to expand $f(1,\varrho_A^*)$ and 
$f(0,\varrho_B^*)$ around $\varrho$. Using Eqns.(\ref{press}) and (\ref{comp}),
and dropping terms of order $1/N$, this yields
\begin{equation}
f(\rho,\varrho_i^*) \approx f(\rho,\varrho) + 
\frac{\beta p}{\varrho_i^*} (1-\frac{\varrho}{\varrho_i^*}) +
\frac{1}{2 \varrho_i^*} (2 \beta p - \beta \kappa_i^{-1}) 
(1-\frac{\varrho}{\varrho_i^*})^2.
\end{equation}
with the compressibilities of the pure component systems $\kappa_i$.
Furthermore, we define the difference between volumes per monomer in the pure 
$A$ and $B$ melt, $\delta = 1/\varrho_A^* - 1/\varrho_B^*$, and assume that 
both $\delta$ and the volume change upon mixing 
$ v_{\rm exc} = 1/\varrho-\rho\varrho_A^* - (1-\rho)/\varrho_B^* $
are small. 
To second order in 
$\varrho \delta$ and $\varrho v_{\rm exc}$, one obtains
\begin{equation}
\label{mexc2}
\beta \mu^{\rm exc}  = 
\frac{\rho}{N_A} \ln \phi + \frac{1-\rho}{N_B} \ln(1-\phi)
+ \Delta f(\rho,\varrho) + \beta \mu^{\rm exc}_{comp}
\end{equation}
\begin{displaymath}
\hspace*{-0.8cm} \mbox{with} \qquad
\Delta f(\rho,\varrho) = 
 [f(\rho,\varrho)-\rho \: f(1,\varrho) - (1-\rho) \: f(0,\varrho)] 
\end{displaymath}
and the equation-of-state contribution
\begin{eqnarray}
\label{com}
 \mu^{\rm exc}_{comp} & \approx &
 \frac{1}{2} \varrho\delta^2 \: \rho (1-\rho) \: 
[(1-\rho) \kappa_A^{-1} + \rho \kappa_B^{-1}]  \nonumber \\
&&  + \frac{1}{2} \varrho v_{\rm exc}^2 \:
[\rho \kappa_A^{-1} + (1-\rho)\kappa_B^{-1}] \nonumber \\
&& 
+ \delta v_{\rm exc}\: \rho (1-\rho)\: [\kappa_B^{-1} - \kappa_A^{-1}]. 
\end{eqnarray}

So far, these are general thermodynamic considerations. The next step is to
evaluate $\Delta f(\rho,\varrho)$ and $v_{\rm exc}$. It is at this point that 
we introduce the generalised Flory Huggins approximation. First, we adopt
the Flory assumption (i) and neglect conformational changes upon mixing; 
hence intrachain interactions do not contribute to the free energy.
Second, we assume that species dependent interactions between monomers can be 
treated as perturbations of a reference system "0" of polymers made of 
identical monomers\cite{FREDDI}. 

We suppose that we can choose a reference system with a short range
monomer potential $u_0(r)$, {\em e.g.}, a hard core potential,
which does not distinguish between monomers of different type. 
We further suppose that the correlation functions $g_{ij}^{inter,0}(r)$ 
for interchain contacts and $g_{ii}^{intra,0}(r)$ for intrachain
contacts, and the ``bare'' incompatibility $\Delta f^0(\rho,\varrho)$ in the 
reference system, are known. The correlation functions are related to the 
monomer pair distribution $\varrho^{(2)}(r)$ {\em via}
\begin{equation}
\varrho_{ij}^{(2)}(r)
= \varrho^2 \rho_i \rho_j \: g_{ij}^{inter,0}(r)
 + \varrho \: \rho_i \delta_{ij} \: g_{ii}^{intra,0}(r),
\end{equation}
where $\delta_{ij}$ is the Kronecker symbol.
They are independent of the composition $\rho$ in symmetrical systems, but they 
may have a (weak) composition dependence if the polymers $A$ and $B$ have 
different architecture or stiffness. In practice, $\Delta f^0(\rho,\varrho)$ is 
very small and can usually be neglected, and $g_{ij}^0(r)$ can be approximated by 
the correlation functions $g_{ij}(r)$ measured in simulations of the ``full'' 
system \cite{M0,STIFF2}. 

Next, monomer specific ``nonbonded interactions'' are turned on. 
One has to distinguish between two possible factors: The monomers may have 
different size, {\em i.e.}, the excluded volume interactions may depend 
on the type of the monomers, and additional energetic interactions may
be present which also distinguish between monomers. The excluded volume
part of the interaction will be strong, but short ranged, whereas the
energetic part may be extended, but weak. It is thus advantageous to 
separate the total monomer potential $u_{ij}(r)$ into three parts 
\begin{equation}
u_{ij}(r) = u_0(r) + v_{ij}(r) + w_{ij}(r),
\end{equation}
where $u_0(r)$ is the potential of the reference system between 
identical monomers of radius $d$, $v_{ij}(r)$ is short ranged and nonzero 
only in a small region $\xi d$ around $r=d$, and $w_{ij}(r)$ is weak. 
We will borrow concepts from perturbation expansions in the theory of simple 
liquids\cite{hansen} and proceed in two steps: First, we include the
potential $v_{ij}(r)$ and expand to lowest order in $\xi$. Then,
we use this system as the new reference system and expand to lowest order
in $w_{ij}$.

In order to do so, we introduce the indirect correlation function
for interchain correlations\cite{hansen} 
\begin{equation}
y_{ij}^{inter,0}(r) = g_{ij}^{inter,0}(r) \exp(\beta u_0(r)),
\end{equation}
and, analogously, the indirect correlation function for intrachain correlations
$y_{ii}^{intra,0}(r)$. In simple liquids, $y_{ij}(r)$ has the advantage 
of being continuous even in hard core systems. We shall assume that this holds 
also for polymer melts. In systems of polymers with equal stiffness, in 
particular, the interchain correlation functions in the polymer melt can be 
related to the corresponding monomer correlation function $y_{\rm mono}(r)$ 
{\em via} 
\cite{degennes,SCHWEIZERR,M0}
\begin{equation}
y_{ij}^{inter,0}(r) = y_{\rm mono}(r) 
\left(1-\frac{1}{r} \tilde{g}_{ij}\left(\frac{r}{\sqrt{N}}\right)\right),
\end{equation}
where $\tilde{g}_{ij}$ are smooth functions (see also Sec.4). If the
monomers have hard core interactions, $y_{ij}(r)$ outside the core is simply 
the pair correlation function, and good estimates for the value of 
$y_{\rm mono}$ inside can be looked up in the literature \cite{hansen}.
We now consider a ``primed'' intermediate system,
in which $v_{ij}(r)$ is turned on, and $w_{ij}(r)$ is still set to zero.
The perturbation expansion to lowest order in $\xi$ yields the monomer 
free energy
\begin{eqnarray}
f'(\rho,\varrho) &=& f^0(\rho,\varrho)  + 
\frac{1}{2} \int dr \: \{e^{-\beta u_0(r)}-e^{-\beta (u_0(r)+v_{ij}(r))}\} 
\nonumber \\
\label{fp1}
&& \times \Big[ \varrho \sum_{i,j}  \rho_i \rho_j \: y_{ij}^{inter,0}(r) 
 + \sum_i \rho_i y_{ii}^{intra,0}(r)
\Big]
\end{eqnarray}
and the indirect correlation function 
$y_{ij}'(r) \approx y_{ij}^0(r) + {\cal O}(\xi^2)$ (cf \cite{hansen}).
Here the sum runs over $i,j=A,B$ and $\rho_i$ is the number fraction of 
component $i$, {\em i.e.}, $\rho_A = \rho$ and $\rho_B = (1-\rho)$.
Corrections due to the chain connectivity\cite{YETH}  are of order 
$1/\sqrt{N}$ and will be ignored here. The pair correlation function in the 
primed system thus takes the form 
$g_{ij}'(r) = y_{ij}^0(r) \exp[-\beta (u_0(r)+v_{ij}(r))]$.

We use the primed system as the new reference system and 
expand $f(\rho,\varrho)$ in lowest order of $w_{ij}(r)$ 
\begin{equation}
\label{fp2}
f(\rho,\varrho) = f'(\rho,\varrho) + \frac{\beta}{2} 
\! \int \!\! dr \: w_{ij}(r) \: 
\{ \varrho \sum_{i,j} \rho_i \rho_j  g_{ij}'{}^{inter}(r) \: 
+ \sum_i \rho_i g_{ii}'{}^{intra}(r) \}.
\end{equation}
After putting everything together, we get to lowest order
\begin{equation}
\label{fp}
f(\rho,\varrho) = f^0(\rho,\varrho) + 
\frac{\varrho}{2} \sum_{i,j} \rho_i \rho_j \: \gamma_{ij}^{inter}(\rho)
+ \frac{1}{2} \sum_i \rho_i \: \gamma_{ii}^{intra}(\rho)
\end{equation}
with effective interaction parameters $\gamma_{ij}(\rho)$
\begin{equation}
\label{gamma}
\gamma_{ij}(\rho)= \int dr \: y_{ij}^0(r) 
\{\exp(-\beta u_0(r) - \exp(-\beta u_{ij}(r)) \}.
\end{equation}
The latter may depend weakly on the composition $\rho$ {\em via} the pair 
correlation function $g_{ij}^0(r)$. It will prove useful to introduce the
quantity
\begin{eqnarray}
\label{X}
{\cal X} &=& 2 
\: \gamma_{AB}^{inter}(\rho) - \gamma_{AA}^{inter}(\rho) - 
\gamma_{BB}^{inter}(\rho) \\
&& + [ (\gamma_{AA}^{total}(\rho)-\gamma_{AA}^{total}(1))/(1-\rho)
   + (\gamma_{BB}^{total}(\rho)-\gamma_{BB}^{total}(0))/\rho ], \nonumber
\end{eqnarray}
where $\gamma_{ii}^{total}(\rho)$ is defined as the sum
$\gamma_{ii}^{total}(\rho)=
\gamma_{ii}^{inter}(\rho)+\gamma_{ii}^{intra}(\rho)/\varrho$.
Note that the bracketed term vanishes in symmetrical mixtures. In general,
it is small and approaches a constant as $\rho \to 0$ or $\rho \to 1$. 

We are now in the position to calculate the volume change upon mixing 
$v_{\rm exc}$ and the volume difference $\delta$ between the two pure component 
systems. Eqns.(\ref{fp}) and (\ref{comp}) inserted into Eqn.(\ref{press}) yield
the composition dependence of the density
\begin{equation}
\varrho = \varrho_0 \:
\Big( 1-\frac{\kappa}{2 \beta} \Big[ 
\varrho_0^2 \sum_{i,j} \rho_i \rho_j \: \gamma_{ij}^{inter}(\rho)
+ \varrho_0 \sum_i \rho_i \: \gamma_{ii}^{intra}(\rho) \Big] \: \Big)
,
\end{equation}
where $\varrho_0$ is the density in the reference system. Using this 
expression, we obtain
\begin{eqnarray}
\label{delta}
\delta = \frac{1}{\varrho_A^*} - \frac{1}{\varrho_B^*} \qquad \qquad 
&\approx& \frac{\varrho \kappa}{2 \beta} \: 
(\gamma_{BB}^{total}(0) -\gamma_{AA}^{total}(1))\\
\label{vexc}
v_{\rm exc} = \frac{1}{\varrho}-\frac{\rho}{\varrho_A^*} 
- \frac{1-\rho}{\rho_B^*}
&\approx& 
\frac{\varrho \kappa}{2 \beta} \: {\cal X} \: \rho (1-\rho) .
\end{eqnarray}
In particular, we recover the roughly parabolic composition dependence of 
$v_{\rm exc}$ announced above. Note that these equations relate 
equation-of-state effects at constant pressure to quantities which are 
accessible in constant-volume simulations.  
Dudowicz and Freed \cite{FREED2} have studied
the volume change of mixing in detail for a lattice model within lattice 
cluster theory. They find that $v_{\rm exc}$ is slightly asymmetric, but the 
approximation by a parabolic law is still reasonable. Unfortunately, they
do not calculate the compressibility and the correlation functions,
thus we cannot relate Eqn.(\ref{vexc}) quantitatively to their work.

These results can be inserted into Eqn.(\ref{mexc1}). The effective
Flory Huggins parameter then takes the form
\begin{equation}
\label{chip}
\chi_{\mbox{\small eff}} = \frac{\varrho}{2} \: {\cal X} 
+ \frac{\Delta f^0(\rho,\varrho)}{\rho(1-\rho)} 
+ \frac{\varrho^3 \kappa}{8 \beta} \: [
(\gamma_{AA}^{total}(0) - \gamma_{BB}^{total}(1))^2 + {\cal X}^2 \: \rho (1-\rho)
].
\end{equation}
The first term comprises the incompatibility of monomers and usually
dominates by far. It contains equation-of-state effects in part 
{\em via} the prefactor $\varrho$. The second term results from packing 
inhomogeneities due to the different chain structure, and the last term 
subsumes additional compressibility effects. Effects of conformational
changes on mixing are disregarded, and so are non-random-mixing effects,
which would be of higher order in the perturbation expansion.
Since the monomer contribution is generally much stronger than the other,
the simplified version of Eqn.(\ref{chip}) 
\begin{equation}
\label{chi}
\chi_{\mbox{\small eff}} \approx \frac{\varrho}{2} \: {\cal X}
\end{equation}
yields a good estimate of the Flory Huggins parameter in most cases.

Most simulations as well as analytical self-consistent field 
\cite{SCF1,SCF2,SCF3} and P-RISM\cite{SCHWEIZERR} calculations are performed at 
constant volume. Here, the situation is slightly different:
The free energy of mixing is the difference between the Helmholtz free energy 
of a mixed system at density $\varrho$, and between the sum of free energies of 
the corresponding pure component systems at densities $\varrho_A^*$ and 
$\varrho_B^*$, where the average density is $\varrho$, but the two pure
systems are at {\em pressure} equilibrium with each other. In other words, the
pressure of the pure systems $p^*$ is different from that of the mixed system, 
$p$, and chosen such that there is no volume change on mixing.  
Thus Eqn.(\ref{mexc2}) has to be replaced by
\begin{eqnarray}
\frac{F_{\rm exc}}{m} & = &
\frac{\rho}{N_A} \ln \phi + \frac{1-\rho}{N_B} \ln(1-\phi) \\
 && + \; \; [\; f(\rho,\varrho) 
- \rho f(1,\varrho_A^*) - (1-\rho) f(0,\varrho_B^*) \; ],
\nonumber
\end{eqnarray}
where $\varrho_A^*=\varrho(\rho=1,p^*)$ and 
$\varrho_B^*=\rho(\rho=0,p^*)$ are evaluated from Eqn.(\ref{press}), 
and the pressure $p^*$ is determined implicitly through the constraint
\begin{equation}
\label{bal}
\frac{\rho}{\varrho_A^*} + \frac{1-\rho}{\varrho_B^*} = \frac{1}{\varrho}.
\end{equation}
The effective $\chi$ parameter can then be calculated in the same way as 
before (in Sec.2.1). 
One obtains an expression similar to (\ref{chip}), with different 
compressibility contributions. The situation is particularly 
simple for mixtures with $\varrho_A^* = \varrho_B^*$. Eqn (\ref{bal}) then 
implies $\varrho = \varrho_A^* = \varrho_B^*$, and the effective 
$\chi$ parameter is given by
\begin{equation}
\label{chiv}
\chi_{\mbox{\small eff}} = \frac{\varrho}{2} \: {\cal X} 
+ \frac{\Delta f^0(\rho,\varrho)}{\rho(1-\rho)} .
\end{equation}
We will present some applications of these considerations to simulation
models in section 4.1.

\section{Models and technical aspects}

Polymeric materials are characterised by widely spread length and time 
scales\cite{B1,B2,B3}. They range from the size of the chemical subunits 
(about $\AA$) to the coil size of the order $100 \AA$. In mixtures, the
correlation length of composition fluctuations becomes even larger very 
close to the critical demixing point. Finally, the length scale which 
characterises the morphology of an immiscible blend is macroscopic,
and crucially determines the properties of the material. The corresponding 
time scales range from the vibrations of chemical bonds ($10^{-14}s$) to 
the time needed for macroscopic phase separation, {\em i.e.}, hours, days
or even longer.

A unified theoretical treatment spanning the whole range of length scales 
is clearly not feasible. Fortunately, polymeric materials share a variety of 
different dynamic and static properties, which can be traced back to a few 
relevant polymeric properties. This justifies the description of polymer
mixtures on a coarse-grained length scale, using appropriate coarse-grained
parameters. One of them is the $\chi$ parameter, which has been discussed in 
the previous section, others will be introduced shortly.
On that level, the study of coarse-grained polymer models can contribute to a 
better understanding of the blend properties in two ways:

First, one needs to establish the relation between the chemical structure of 
polymers and the coarse-grained parameters. Unfortunately, a quantitative prediction 
of, {\em e.g.}, the Flory-Huggins parameter $\chi$, from the atomistic structure is 
extremely difficult. This is because the typical energy scale for intermolecular 
interactions on the monomer scale is of the order $k_BT$, whereas the Flory-Huggins 
parameter is typically 3-5 orders of magnitude smaller in dense polymer blends. 
Here, coarse-grained models which incorporate some degree of fluid-like structure 
or conformational asymmetry  provide useful qualitative information about the 
relation between the microscopic structure and the coarse-grained parameters.

Second, the models can be used to study the interplay between the coarse-grained 
parameters and the bulk thermodynamics and interfacial structure. Such
investigations yield valuable insight into the universal material properties in 
polymeric systems. They can be compared to predictions of mean-field theories or 
experimental results. This aspect is the focus of most simulations presented 
here.

A minimal set of relevant polymeric properties comprises: The connectivity of the 
macromolecules along the backbone, the excluded volume of the segments, and
short range thermal interactions. Coarse-grained models, which retain these 
macromolecular properties, have proven extremely efficient in investigating the 
universal thermodynamic properties of polymeric multicomponent systems. 
Only a small number of parameters is required to compare experiments, simulations 
and theory quantitatively on this coarse-grained level.

These are, for example, the $\chi$ parameter, the compressibility, and the chain 
length. In inhomogeneous polymer melts an addition parameter is required
which sets the length scale in the system. A natural choice is the extension of 
single chains, {\em i.e.}, the end to end distance $R_e$ of a bulk 
chain, or equivalently, the gyration radius $R_g$. In a dense melt, sufficiently
long and flexible chains are well described by random walks\cite{degennes}. 
Thus $R_e$ and $R_g$ depend on the degree of polymerisation $N$ like 
$R_e = b \sqrt{N}$ and $R_g = b \sqrt{N/6}$, where the proportionality constant 
$b$ is called the statistical segment length. The two parameters $\chi N$ and 
$R_e$ are generally sufficient to describe inhomogeneous systems of long, flexible 
polymers with weak monomer interactions (Gaussian chain model). Here, ``weak'' 
means that the length scale for composition changes is several times the 
statistical segment length, {\em i.e.}, $\chi\ll 1$. If the chains are very 
stiff\cite{MORSE}, or the incompatibility is quite large\cite{STIFF2}, a second 
length scale may come into play, the persistence length $\xi$, which 
characterises the length of the chain over which the monomers are still 
strongly correlated\cite{WORM}. 

A minimal polymer model, which retains the above properties, is the lattice 
model of Flory and Huggins\cite{FH}. A small number of chemical subunits is 
represented by a single lattice site on a simple cubic lattice. The excluded 
volume property is mimicked by the constraint of single site occupancy, and 
bonded segments occupy nearest neighbour sites on the lattice. Phase separation 
is caused by unfavourable interactions between neighbour segments of different 
type. In spite of its extreme simplicity, this model and its mean-field 
analysis by Flory and Huggins is usually the basis for discussions of polymer 
miscibility in melts, and it already yields many important informations on the 
bulk thermodynamics (see section 2.). In particular, the Flory Huggins theory 
predicts a linear increase of the critical temperature of a binary polymer 
blend with the degree of polymerisation, in agreement with 
experiments\cite{Gehlsen} and simulations\cite{Deutsch}.

Lattice models which resemble the Flory-Huggins-model have been used in 
simulations of polymer melts and solutions. One prominent example is
the Larson model, which has found wide application for the study
of amphiphilic systems (for a recent review see\cite{tanny}).

However, many important properties of polymer melts cannot be represented in 
such a simple model\cite{Deutsch2}: The small number of bond angles does not 
allow for a realistic modelling of the bending rigidity or other structural 
asymmetries. Monomers and vacancies have the same size, hence the underlying 
monomer fluid displays no packing effects\cite{PACK}. These are however crucial 
ingredients to the Flory-Huggins parameter, as discussed above. With the 
constraint that the bond length is always one lattice unit, the implementation 
of a local diffusive monomer dynamics is not possible. The latter is needed 
when studying the collective dynamics of phase separation ({\em i.e.}, spinodal 
decomposition), or the relation between the thermodynamic state and the single 
chain dynamics. 

The bond fluctuation model (BFM) of Carmesin and Kremer\cite{BFM} overcomes the 
above mentioned limitations while retaining the computational advantages of a 
lattice model: Each monomer blocks eight corners of a whole unit cell in a 
simple cubic lattice from further occupancy. Monomers along a polymer are 
connected by one of 108 bond vectors of length $2,\sqrt{5},\sqrt{6},3$, or $10$. 
Due to the size difference between vacancies and monomers, one gets fluid-like 
packing effects. The extended monomer size and the variety of possible bonds
helps washing out the underlying lattice structure. A local monomer hopping 
dynamics can be implemented, which leads to a Rouse-like dynamics in dilute 
solutions. 

At the volume fraction $0.5$ or monomer number density $\varrho = 0.5/8$ (the 
factor 8 reflects the monomer volume), the model mimics many properties of a 
concentrated polymer solution or melt. The excluded volume interaction is 
screened down to about 6 lattice spacings, {\em i.e.}, the chains have Gaussian 
statistics on larger length scales. The diffusion of very long chains exhibits 
the signature of reptation-like motion: the entanglement length is roughly 32 
monomers\cite{Paul}. Due to the relatively high number of remaining vacancies, 
the polymer conformations still equilibrate reasonably fast. Real polymer 
systems can be mapped onto the BFM\cite{MAP1,MAP2}. One monomer then corresponds
to to 3-5 chemical repeat units, and one lattice units to roughly 2 $\AA$. 
In the following, all lengths shall be given in units of the lattice constant.

The BFM has been widely used to study the statics and dynamics of polymer 
solutions\cite{SOLVENT,Paul}, melts\cite{Paul,RING}, and glasses\cite{JOERG}. 
Hence, a variety of properties of the model are known, and this established 
model is well suited to test theories of polymer miscibility and polymer 
interfaces. 

Due to the high number of possible bond vectors, one has 87 different bond angles.
This allows for a rather realistic modelling of a local stiffness along the
chain.  Unless noted otherwise, all of the polymers are complete flexible. 
In some of our studies\cite{STIFF1,STIFF2}, we impose on the $B$ component
of the blend an intramolecular potential, which favours straight bond angles:
$E(\Theta) = f \cos(\Theta)$, where $\Theta$ denotes the complementary angle 
between successive bonds. On increasing the strength of this potential
from $f=0$ (flexible) to $f=2$, the chain extension increases by a factor 1.5 
(for chain length $N=32$).

We shall consider binary polymer mixtures containing $n_A$ $A$-polymers of length 
$N_A$ and $n_B$ $B$-polymers of length $N_B$, and ternary systems with 
additionally $n_C$ symmetric $A$B diblock copolymers of length $N_C$. 
In some applications, the stiffness of the $B$ component is tuned through the 
bond angle potential described above. The two kinds of monomer species, $A$ and 
$B$, interact {\em via} a square well potential. The range of the potential 
corresponds roughly to the first coordination shell of the monomer fluid, and 
includes the 54 nearest lattice sites. Unless noted otherwise, monomers of the 
same type attract each other, whereas contacts between unlike species increase 
the energy
\begin{equation}
\label{eps}
-\epsilon_{AA}=-\epsilon_{BB}= \epsilon_{AB} \equiv {\epsilon}.
\label{eqn:energy}
\end{equation}
In the following, we measure all energies in units of $k_BT$. Hence, $\epsilon$ 
is proportional to the inverse temperature. 

One might argue that purely attractive monomer interactions of different strength
would mimic the experimental situation in a more realistic way. The choice
(\ref{eps}) is motivated by the following consideration: Even the pure 
one component system ($\rho=0$ or $\rho=1$) is in fact a mixture of homopolymers
and vacancies. For strong attractive interactions between monomers, one hence 
encounters a liquid-vapour phase separation into a homopolymer-rich and 
vacancy-rich phase. The temperature of this ``$\Theta$ transition'' is independent 
of the chain length ($\epsilon_\Theta \approx 0.495(5)$\cite{SOLVENT}), whereas
the temperature of the AB unmixing transition increases like $N$. For long chains,
these two temperature scales are thus well separated. For short chains, however,
the choice of purely attractive monomer interactions would lead to a rather
complicated interplay between both transitions. It is thus expedient to choose
the interactions such that the demixing occurs already at rather weak absolute
values of $\epsilon_{ij}$. Simulations of the Bond fluctuation model
have also been performed using a slightly
smaller interaction range. The results for strictly symmetric interactions
are qualitatively similar\cite{Deutsch,M0}.

Even though complex lattice models capture many of the relevant properties of 
polymer mixtures, there are limitations: Constant pressure simulations of
lattice models\cite{CIFRA1} are difficult, whereas their implementation is
relatively straightforward for off-lattice models. Keeping the pressure constant
is desirable, for example, when the blend exhibits a pronounced volume change upon
mixing (cf. section 2.). Even at constant volume, off-lattice models have the 
advantage that the pressure and the interfacial tension can be measured 
{\em via} the virial equation\cite{GARY1,VIRIAL}. Furthermore, they lend 
themselves
to  molecular dynamics simulations, which are able to capture hydrodynamic flow.
The latter is crucial, for example, to describe the final stage of spinodal 
decomposition\cite{SPINODAL}. On the other hand, off-lattice simulations are more 
demanding with respect to computational resources. In many respects, the BFM 
simulations are remarkably consistent with off-lattice predictions, {\em e.g.}, 
with respect to the packing contributions to the $\chi$ parameter (see Sec.4.1).

In studies of the thermodynamics and structure of polymer blends, two statistical 
ensembles are mainly used: In the canonical ensemble, the temperature, the volume 
and the global composition of the mixture are fixed, and the Monte Carlo moves 
update the polymer conformations. In simulations of the BFM, local monomer 
hops\cite{BFM} and slithering snake-like movements\cite{ASYM} have been applied.
The former allow an interpretation of the 
Monte Carlo simulation in terms of an overdamped, purely diffusive dynamics, 
the latter relax the polymer conformations faster by a factor $N$. 
In off-lattice models, MD integration\cite{GARY1} is often employed to propagate 
the system. The canonical ensemble is useful for the study of dynamic properties,
the local interfacial structure and the phase behaviour of systems with strong 
structural asymmetries\cite{STIFF2}.

In systems with weak or moderate structural asymmetry, however, the phase 
behaviour is investigated more efficiently in the semi-grandcanonical 
ensemble\cite{SARIBAN}: The temperature, the volume, and the total number of 
monomers are held constant, while the composition of the mixture is allowed to 
fluctuate. In binary mixtures, it is controlled by just one parameter, the
exchange monomer potential $\Delta \mu$ (see Eqn. (\ref{eos}). In ternary
mixtures, a second variable $\delta \mu$ has to be introduced, which couples to
the copolymer content in the mixture. The Monte Carlo scheme includes moves which 
turn polymers of type $A$ into polymers of type $B$ and vice versa, and 
additional moves in ternary systems which switch between homopolymers and
copolymers\cite{COP}. This algorithm has first been introduced by Sariban and
Binder\cite{SARIBAN}, and applied to the simple Flory-Huggins lattice model
of strictly symmetric binary blends. In that case, the semi-grandcanonical 
identity switches are merely a change of the polymer label. It literally carries
over to off-lattice models of symmetric mixtures\cite{GARY1}.
The scheme can also be applied to asymmetric mixtures with different dispersion 
forces\cite{HPDASYM} , with chain length asymmetry\cite{M0,ASYM}, with stiffness 
disparity\cite{STIFF1}, and non-additive monomer shapes\cite{STIFF1}. For even 
stronger asymmetries, configurational bias\cite{SOLVENT} or gradual inserting 
schemes can be envisaged.  For very strong conformational asymmetries, however, the 
semi-grandcanonical moves become increasingly inefficient\cite{STIFF2}.

The advantages of the semi-grandcanonical ensemble are threefold\cite{STIFF1,COP}:

First, since the global composition is not conserved, large length scale 
composition fluctuations relax much faster than in the canonical ensemble. 
The attempt to observe an unmixing transition driven just by the diffusion of 
polymers poses a substantial challenge in terms of computational 
resources\cite{Deutsch2}. Furthermore, a thorough analysis of finite-size effects 
in the canonical ensemble ({\em e.g.}, via a subbox analysis\cite{ROVERE}) requires
formidably large system sizes. The semi-grandcanonical ensemble concentrates the 
computational effort on the composition fluctuations, the slowest mode in the 
blend. Sophisticated finite-size scaling methods have been applied to study
the crossover from mean-field to Ising critical behaviour in strictly symmetric 
polymer mixtures\cite{Deutsch}, and mixed-field finite-size techniques
have been used to accurately determine the critical point in asymmetric 
mixtures\cite{SOLVENT,MMFSS,NIGEL}. These techniques can be combined with a
multi-histogram extrapolation of the joint energy-composition 
distribution\cite{Deutsch2,HISTO} and hence allow to determine the phase 
behaviour accurately from simulations of systems of rather modest size 
({\em i.e.}, comprising only a few hundred polymers).

Second, the semi-grandcanonical equation of state, which relates the composition 
$\rho$ to the chemical potential difference $\Delta \mu$, can be measured in
semi-grandcanonical simulations. This gives information about the thermodynamics 
of mixing, and provides the most direct connection to mean-field theories. 
In particular, it can be used to relate simulations to the considerations
of Sec. 2 (see also Sec. 4.1).

Third, the semi-grandcanonical ensemble can be combined with multi-canonical 
reweighting schemes\cite{REWEIGHT}. Rather than generating configurations
following the Boltzmann distribution, the simulation then samples the phase space 
according to the probability $\exp(-(E-n_A \Delta \mu)/k_BT)/W(\rho)$, where $E$ 
denotes the total energy of the binary system. By choosing the preweighting factors $W$ 
proportional to the equilibrium  probability distribution in the 
semi-grandcanonical ensemble $P(\rho)$, one achieves uniform sampling of all 
compositions $\rho$. An excellent estimate for the equilibrium  probability distribution
can be obtained from histogram extrapolations of previous runs.  Below the critical temperature, this Monte Carlo 
scheme\cite{MBO,COP} encourages the system to explore configurations in which the 
two phases coexist in the simulation box.

\begin{figure}[htbp]
    \begin{minipage}[t]{110mm}%
           \mbox{
	              \setlength{\epsfxsize}{9cm}
	              \epsffile{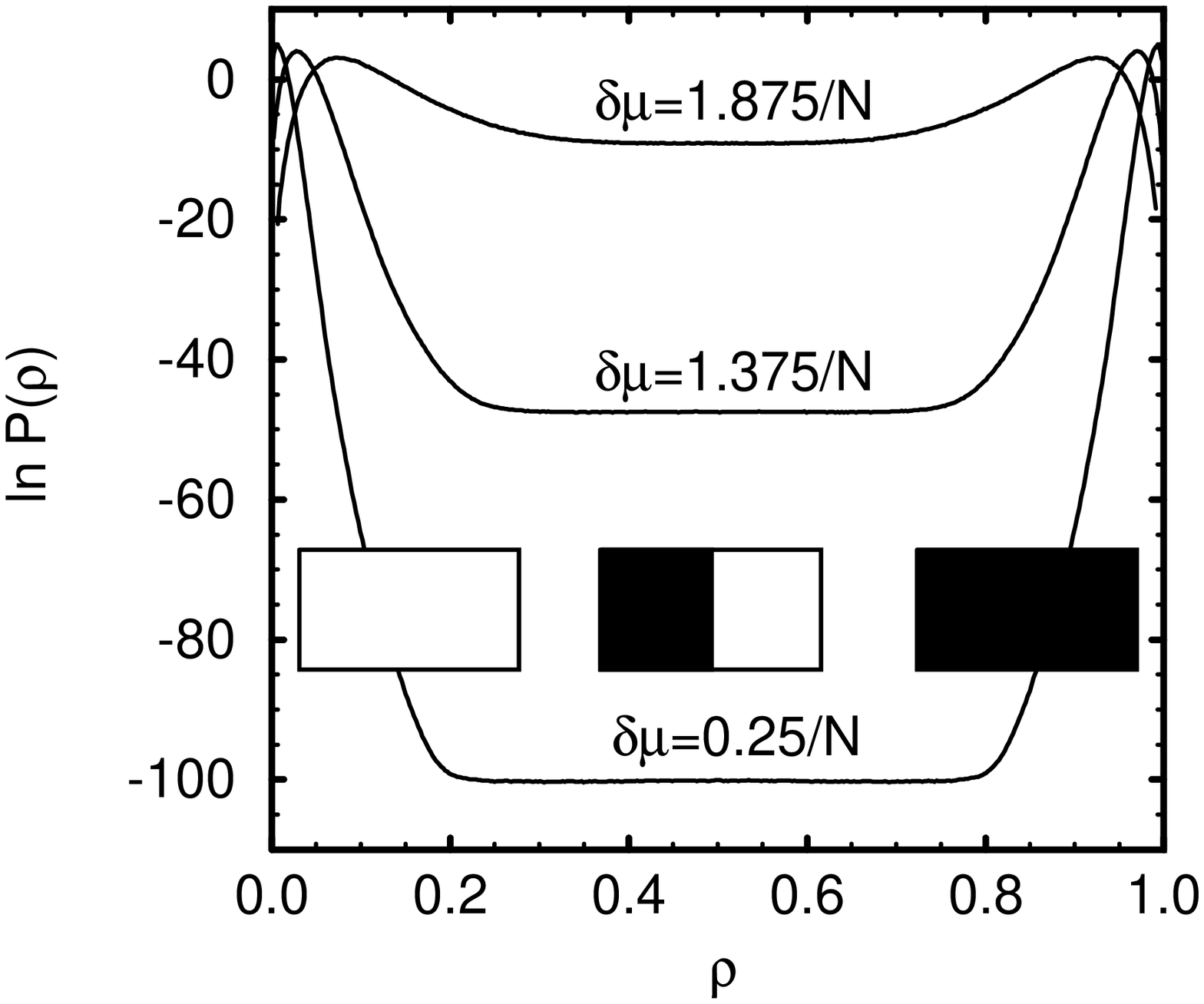}
	        }\\
           \mbox{
	              \setlength{\epsfxsize}{9cm}
	              \epsffile{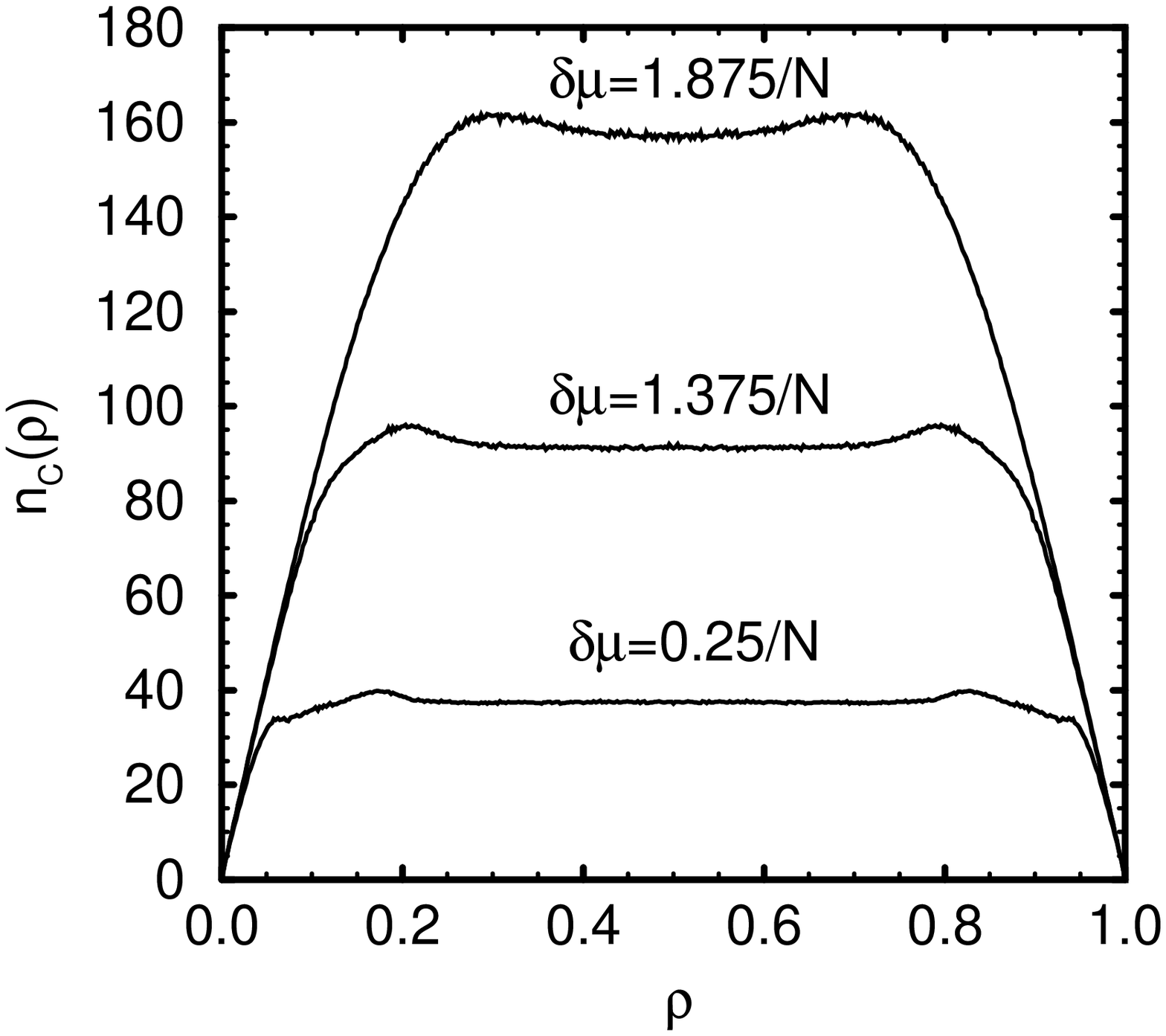}
	        }							         
     \end{minipage}%
     \hfill%
     \begin{minipage}[b]{110mm}%
\caption{({\bf a}): Probability distribution of the composition in a ternary 
	homopolymer-copolymer mixture at $\epsilon=0.054$ and system size 
	$48 \times 48 \times 96$. Upon increasing the chemical potential 
	$\delta \mu$ of the copolymer, the ``valley'' in the probability 
	distribution becomes shallower, indicating that the copolymers decrease 
	the interfacial tension. 
	({\bf b}): Average number of copolymers $n_C(\rho)$ as a function of the composition 
	for the same system as in ({\bf a}).
	}
	From Ref. \cite{COP}.
	   \label{fig:reweight}
     \end{minipage}%
\end{figure}

We shall illustrate the last point with the simulation data for a symmetric 
ternary homopolymer/copolymer blend. The probability distribution of the 
composition as a function of the chemical potential $\delta \mu$ of the copolymers
is presented in Fig.\ref{fig:reweight} on a logarithmic scale. We have chosen an
elongated simulation cell\cite{COP} of $xy$-cross section $L^2$ and extension in 
$z$-direction $2L=96$. The distributions exhibit two pronounced peaks 
corresponding to the composition of the two coexisting phases. In the two-domain 
state between these coexistence compositions, the probability is strongly reduced. 
For large enough system sizes, the probability distribution there is dominated 
by configurations containing two interfaces parallel to the $xy$ plane, as 
sketched schematically in the figure (periodic boundary conditions are applied).
The two interfaces of area $L^2$ separate the two bulk coexisting phases. 
The probability of these configurations with respect to the bulk is reduced by 
the Boltzmann weight $\exp(-2L^2\sigma/k_BT)$ of the interfacial free energy 
cost\cite{SIGMA}, where $\sigma$ is the interfacial tension between the 
coexisting phases. Hence, the interfacial tension can be calculated from
\begin{equation}
\frac{\sigma}{k_BT} = \frac{1}{2L^2} \ln \left( 
\frac{P(\rho_{\rm coex})}{P(1/2)}\right).
\end{equation}
The plateau in the probability distribution around $\rho=1/2$ shows, that the two 
interfaces can vary their mutual distance, thereby changing the composition, 
without free energy costs. This important consistency check indicates that the
the interfaces can be described as flat and non-interacting. The free energy of 
the interface is lowered upon increasing the copolymer chemical potential.
Within the same simulation run, we can determine how the number of copolymers 
at fixed chemical potential $\delta \mu$ depends on the composition.
(cf. Fig. \ref{fig:reweight} (b)). Not surprisingly, configurations with two 
interfaces contain more copolymers. From this we can extract the excess of 
copolymers at the interface. It grows upon increasing the chemical potential $\delta \mu$ of the copolymers.
Similarly, we can monitor the energy as a function 
of the composition, and extract the excess energy of the interface. In 
addition, a careful analysis of the probability distribution yields information 
about the interaction potential (joining pressure) between two interfaces.  
Hence, bulk and excess interfacial properties are simultaneously accessible in 
the semi-grandcanonical ensemble\cite{COP}. 

The same combination of techniques has been applied to thin film, where both flat 
walls attract the same component\cite{MBWET}. In this case the coexistence 
chemical potential $\Delta \mu_{coex}$ is shifted from its bulk value
$\Delta \mu_{coex}^b=0$, but nevertheless reweighting techniques permit to locate
the phase coexistence accurately even for low temperatures.

Even with a careful choice of the statistical ensemble and the use of recently
developed simulation and analysis techniques, the accurate measurement of
structural and thermodynamical properties in the computer simulation poses
huge computational demands. One the one hand, very large system sizes are required
to investigate, {\em e.g.}, the effect of capillary-wave broadening. In our simulations\cite{MBO,
andreas,COP2} we vary the lateral extension of the simulation box from  64 to 512 lattice units. 
The largest simulation cell contains more than a million monomers. Only massively parallel
computers like the CRAY T3D/T3E make the simulations of such large systems feasible.
Using a two-dimensional geometric parallelisation scheme, we have efficiently 
implemented the local monomer movements. Even for small simulation cells,
 the program scales practically linear with the number of processors up 
to 256\cite{MCOMP}.
On the other hand, many systems are characterised by protracted long time scales:
Increasing the chain length $N$ increases dramatically the relaxation times of 
the chain conformations. Deutsch et al.\ have employed chains of up to 512 
segments to investigate the crossover between mean-field and Ising critical 
behaviour. Their investigation 
required several thousand hours of CRAY Y-MP time\cite{Deutsch}. The equilibration of long wave length
interfacial fluctuations is also difficult because of its slow dynamics. Another example is
the relaxation time in the framework of the multicanonical reweighting scheme\cite{REWEIGHT}. 
In the most ideal case, the system performs a random walk on all compositions. Therefore
the correlation time scales with the system size $L$ like $L^6$. The total amount of CPU time 
investigated into the results presented in the following exceeds the equivalent of $10^5$
hours on a single CRAY T3D processor.

In blends with strong structural asymmetries ({\em e.g.}, large stiffness disparity, 
polymer solvent systems), the semi-grandcanonical identity switches between 
different polymer types are rather inefficient\cite{STIFF1,STIFF2}. Hence, 
methods for measuring the interfacial tension in the canonical ensemble have been 
explored. In off-lattice models, the interfacial tension can be measured via the 
anisotropy of the pressure tensor. This method has been successfully applied to
determine the free energy costs of a hard wall\cite{VIRIAL} in a concentrated 
polymer solution and the interfacial tension\cite{GARY1} in a binary polymer 
blend. However, the generalisation to lattice models is difficult\cite{CIFRA1}.

\begin{figure}[htbp]
    \begin{minipage}[t]{110mm}%
           \mbox{
	              \setlength{\epsfxsize}{9cm}
	              \epsffile{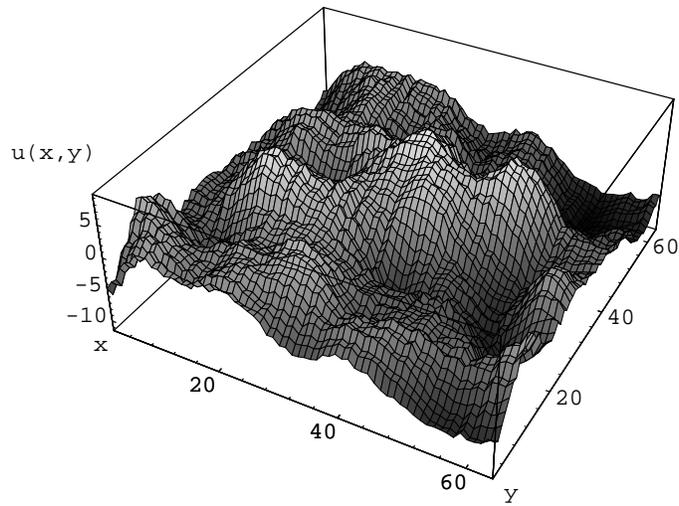}
	        }							         
     \end{minipage}%
     \hfill%
     \begin{minipage}[b]{110mm}%
\caption{
Typical snapshot of the local interfacial position $u(x,y)$ for a system 
of lateral extension $L=64$, $\epsilon=0.03$ and $N=32$. The coarse-graining 
size $B=8$ corresponds roughly to the chain's radius of gyration.
Long wavelength fluctuations are clearly visible.
	}
	From Ref. \cite{COP2}.
	   \label{fig:6}
     \end{minipage}%
\end{figure}

An alternative way of measuring the interfacial tension is the analysis of the
capillary fluctuation spectrum\cite{STIFF2,COP,PORE}: In general, interfaces are 
not flat, but exhibit long-wavelength capillary-wave fluctuations. A typical 
snapshot of the local interfacial position in a binary polymer blend is presented 
in Fig.\ref{fig:6}. The local interfacial position has been laterally averaged 
over a size comparable to the polymer's radius of gyration, and large length 
scale fluctuations are clearly visible\cite{andreas}. Let us consider a
flat interface, neglect bubbles and overhangs, and describe the local deviations 
of the interfacial position from its lateral average by the single valued 
function $u(x,y)$. Due to the local deviations, the interfacial area is
increased compared to the corresponding planar interface. The effect on the
thermodynamics of the interface is commonly described by the capillary-wave 
Hamiltonian for long-wavelength interfacial fluctuations\cite{BUFF,HELFRICH}
\begin{equation}
{{\cal H}} = \int dx\;dy\; \left\{ \frac{\sigma}{2} (\nabla u)^2 \right\},
\label{eqn:capwave}
\end{equation}
where $\sigma$ denotes the effective interfacial tension. It can be
diagonalised very naturally with a simple Fourier transform 
\begin{equation}
u(x) \sim \frac{a_0}{2} + \sum_{l=0}^{L/2-1} 
\left\{ a(q_l)\cos(q_lx) + b(q_l)\sin(q_lx) \right\},
\end{equation}
with $q_l = 2 \pi l/L$. The equipartition theorem shows that the Fourier 
components have a Gaussian distribution with Gaussian width
\begin{equation}
\frac{2}{L^2\langle a(q)^2\rangle} = 
\frac{2}{L^2\langle b(q)^2\rangle}  = \frac{\sigma}{k_BT} q^2.
\label{eqn:spectrum}
\end{equation}
\begin{figure}[htbp]
    \begin{minipage}[t]{110mm}%
           \mbox{
	              \setlength{\epsfxsize}{9cm}
	              \epsffile{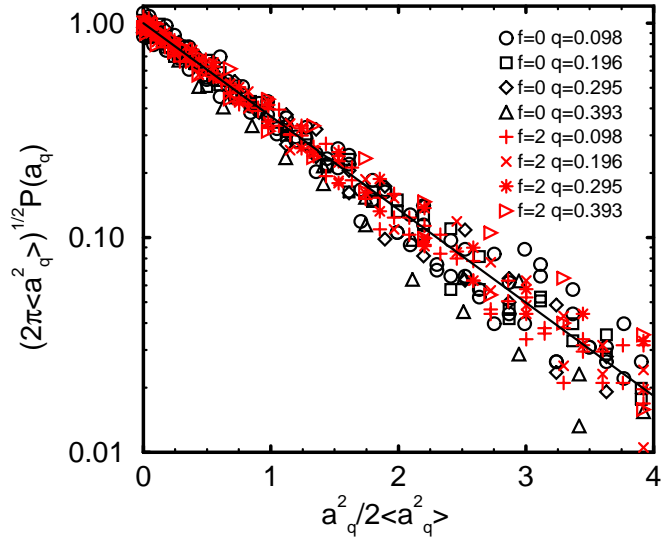}
	        }							         
     \end{minipage}%
     \hfill%
     \begin{minipage}[b]{110mm}%
\caption{Probability distribution of Fourier components $a(q)$ and $b(q)$ of the 
	local interfacial position for the four smallest wave-vectors $q$ in a
	system with system of lateral size $L=64$, bending energies $f=0,2$, and chain length 
        $N=32$. For the data collapse  the $1/q^2$ dependence of the variance has 
	been exploited. The solid line shows the expected Gaussian distribution.
	}
	From Ref. \cite{STIFF2}.
	   \label{fig:6b}
     \end{minipage}%
\end{figure}
The probability distribution for the Fourier components of the local interfacial 
position are presented in Fig.\ref{fig:6b} for a symmetric blend and a blend with 
strong stiffness disparity\cite{STIFF2}. In both cases, the long-wavelength 
interfacial fluctuations are well described by the quadratic interfacial 
Hamiltonian. The interfacial tension can be determined from the inverse width of 
the distribution. For the symmetric blend, this value can be compared with the
independent measurement in the semi-grandcanonical ensemble. The agreement is
very good. The interfacial fluctuation spectrum thus provides an efficient 
alternative way of measuring the interfacial tension in structurally asymmetric 
systems. For example, the method has been successfully applied to measure the 
interfacial tension of amphiphilic bilayers\cite{PORE}.

We note that the relative increase of the interfacial area is proportional to 
$\ln L/L^2$. Hence interfacial excess quantities\cite{COP} ({\em e.g.}, the
interfacial tension, the excess energy of the interface, the enrichment of 
copolymers or vacancies) are rather insensitive to capillary fluctuations.
Interfacial profiles are affected more dramatically\cite{andreas}.
``Apparent'' profiles obtained {\em via} lateral averaging depend sensitively 
on the extension of the system and are significantly broader than the 
``intrinsic'' profiles.  This capillary-wave broadening has to be accounted
for when comparing simulation results analytical predictions, and is also 
relevant for the interpretation of experiments. The fluctuations of the local 
interfacial position $\langle u^2(x,y) \rangle$ are obtained by integrating over 
the lateral Fourier components. The power law spectrum in 
Eqn.(\ref{eqn:spectrum}) leads to logarithmic divergencies for $q \to 0$ and 
$q \to \infty$. We can remove these divergencies by introducing heuristic cut-offs 
$q_{\rm max}$ and $q_{\rm min}$, and obtain
\begin{equation}
\langle u^2(x,y) \rangle = \frac{k_BT}{2\pi \sigma} 
\ln \left( \frac{q_{\rm max}}{q_{\rm min}}\right).
\label{eqn:capwave2}
\end{equation}
The lower cut-off $q_{\rm min}$ is simply determined by the system size, 
{\em i.e.}, $q_{\rm min} =  \frac{2\pi}{L}$. The value of the upper cutoff 
$q_{\rm max}$, however, is less obvious. In addition to long-wavelength capillary 
fluctuations, one has composition fluctuations on short length scales at the 
interface ({\em i.e.}, intrinsic fluctuations), which determine the shape of the 
intrinsic interfacial profile. The upper cutoff $q_{\rm max}$ is hence related to 
the lateral length scale which separates the intrinsic fluctuations from the 
fluctuations of the center of the intrinsic profile. For small molecules and
far from the critical region, one usually estimates $q_{\rm max} = 2 \pi/ a$, 
where $a$ denotes the molecular extension. In polymeric systems\cite{andreas}, 
however, three possible microscopic length scales have to be considered: the 
statistical segment length $b$, the width of the intrinsic interfacial profile 
$w_0$, which is controlled by the Flory-Huggins parameter, and the radius of 
gyration, which scales like $\sqrt{N}$. Semenov\cite{SEM_CAP} proposed to use 
$q_{\rm max}=2/w_0$ as the upper cutoff.  In principle, Monte Carlo simulations 
could address this problem by studying the dependence of the broadening on the 
chain length $N$ and the incompatibility $\chi$. 

A finite bending rigidity $\kappa$ of the interface also generates a smooth 
cutoff of the fluctuation spectrum on small length scales. 
This effect can be described by the Helfrich Hamiltonian\cite{HELFRICH}
\begin{equation}
{{\cal H}} = \int dx\;dy\; 
\left\{ \frac{\sigma}{2} (\nabla u)^2 + \frac{\kappa}{2} 
(\triangle u)^2\right\}.
\label{eqn:helf1}
\end{equation}
Including the bending rigidity $\kappa$ removes the divergency for 
$q \to \infty$.  A similar analysis as above yields
\begin{equation}
\hspace*{-0.5cm}
\frac{2}{L^2\langle a(q)^2\rangle}  = 
\frac{\sigma}{k_BT} q^2 + \frac{\kappa}{k_BT} q^4 \quad \mbox{and} \quad
\langle u^2(x,y) \rangle = \frac{k_BT}{2\pi \sigma} 
\ln \left( \frac{\sqrt{q^2_{\rm min}+\sigma/\kappa}}{q_{\rm min}}\right).
\label{eqn:helf2}
\end{equation}
Thus the crossover length $\sqrt{\sigma/\kappa}$ acts as upper cutoff 
$q_{\rm max}$. This effect is important in interfaces with absorbed amphiphilic 
molecules\cite{COP}. At high concentrations, they oppose themselves to
being squeezed together on one side of a bent interface, thus generating a 
positive contribution to the bending rigidity (cf. Sec. 4.3).

The laterally averaged ``apparent'' profile can be approximated by the 
convolution of the intrinsic profile $\rho_0(z)$ with the Gaussian distribution 
$P(u)$ of the local interfacial position
\begin{equation}
\rho(z) = \int du\; \rho_0(z-u) P(u).
\end{equation}
For the slope at the center of the profile, this convolution approximation
yields in the case of weak broadening\cite{JASNOV}
\begin{equation}
\frac{d\rho}{dz} \Big|_{z=0} \approx 
\frac{d\rho_0}{dz} \Big|_{z=0} + \frac{k_BT}{4 \pi \sigma} 
\frac{d^3\rho_0}{dz^3} \Big|_{z=0} \ln \left( 
\frac{q_{\rm max}}{q_{\rm min}}\right).
\end{equation}
The slope at the center ($z=0$) of the profile can be used to define the inverse
interfacial width $\frac{d\rho}{dz} \Big|_{z=0} = \frac{1}{2w}$. Hence we obtain
for the broadening of the apparent profile\cite{andreas}
\begin{equation}
w^2 \approx w_0^2 + \frac{k_BT}{\pi \sigma} \left| 
\frac{d^3\rho_0}{dz^3} \Big|_{z=0}\right|  w_0^3 \ln 
\left( \frac{q_{\rm max}}{q_{\rm min}}\right) 
= w^2_0 + \frac{k_BT}{4\sigma} \ln \left(\frac{q_{\rm max}}{q_{\rm min}}\right).
\label{eqn:apparent}
\end{equation}
where we have approximated the shape of the intrinsic profile by an error function
in the last step in order to calculate the numerical prefactor.

Since analytic mean-field theories generally yield intrinsic profiles, one
has to take due account of the capillary-wave broadening when comparing them
with simulation data. In order to reduce the broadening effect, we can define 
``reduced'' profiles by laterally dividing the system into subsystems of size 
$B \times B$\cite{STIFF2,andreas}. In each subsystem, we localise the interfacial 
position and average profiles with respect to the local midpoint of the profile. 
In principle, the lateral length scale $B$ can be chosen as to match the 
width of the reduced profile with the theoretically predicted value. However, once 
$B$ is chosen in this way, other profiles ({\em e.g.}, segmental orientations, 
chain ends, etc.) can be compared to theoretical predictions without any 
adjustable parameter. 

\section{Simulation results}

Many of the simulation and analysis techniques presented in the previous sections
can be applied to lattice models of polymers as well as to off-lattice models, and
also to non-polymeric models. In this section, we will show how they have
been applied specifically to simulations of the bond fluctuation model, and 
compare the results to experiments and to analytical predictions.

\subsection{Local packing and miscibility}

We will begin with a discussion of the bulk thermodynamics of polymer mixtures.
The knowledge of the phase behaviour is required for a discussion of 
interfacial properties\cite{MBO}, and of considerable interest in 
itself\cite{Deutsch}. 

In the past, special emphasis has been given to simulations of symmetric, 
binary polymer blends\cite{Deutsch}. Experimentally, nearly symmetric blends can 
be realised by mixing partially deuterated and protonated polymers\cite{Gehlsen}, 
or as blends of statistical copolymers\cite{KLEIN}. Simulations by Deutsch and 
Binder\cite{Deutsch} have confirmed the linear scaling\cite{FH} of the critical 
temperature by investigating polymers with up to 512 monomers. Using a 
sophisticated crossover finite-size scaling technique, they accurately located 
the critical temperature and investigated the crossover between mean-field 
behaviour to 3D Ising critical behaviour, which prevails in the vicinity of 
the critical point. Upon increasing the chain length, the Ising critical region 
shrinks, in accord with the Ginzburg criterion\cite{GINZBURG} and with 
experimental observations\cite{Gehlsen,EXP_GINZ}.

\begin{figure}[htbp]
    \begin{minipage}[t]{110mm}%
           \mbox{
	              \setlength{\epsfxsize}{9cm}
	              \epsffile{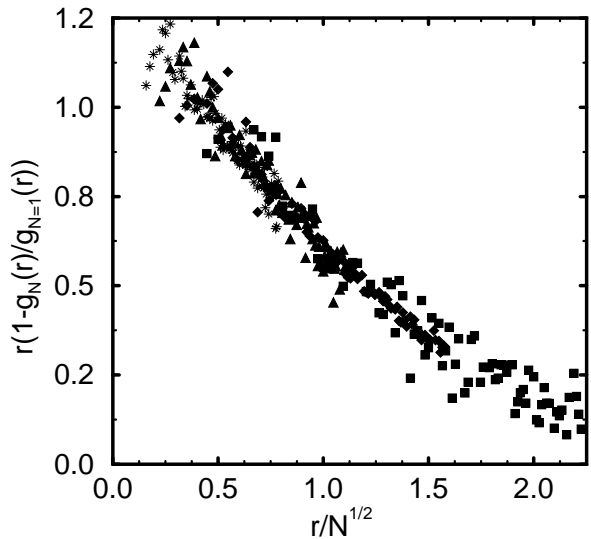}
	        }							         
     \end{minipage}%
     \hfill%
     \begin{minipage}[b]{110mm}%
\caption{Scaling of the reduced correlation function $g_N(r)/g_{N=1}(r)$ 
	with chain length. $N=20$ (squares), $N=40$ (diamonds), $N=80$ 
	(triangles), and $N=160$ (stars) for athermal ($\epsilon=0$) melts.
        }
	From Ref. \cite{M0}.
	   \label{fig:2}
     \end{minipage}%
\end{figure}

The phase behaviour of the bulk is determined by the local structure of the 
polymeric fluid, as reflected by the fact that the Flory-Huggins parameter 
$\chi$ can be related to the intermolecular pair correlation function\cite{M0} 
(see Sec. 2.). In the simplest case of an  athermal melt ({\em i.e.}, 
$\epsilon=0$) 
the intermolecular packing is driven by two effects: On the one hand, the size 
disparity between vacancies and monomers gives rise to fluid-like packing effects
which result in a highly structured correlation function at short distances. 
On the other hand, the extended structure of the macromolecules causes a reduction
of contacts with other chains on intermediate length scales. In order to separate 
the monomeric, local packing effects from the universal behaviour of the 
polymeric correlation hole, we consider the ratio between the intermolecular 
pair correlation function and its monomeric equivalent. As shown in 
Fig. (\ref{fig:2}),
it is found to be largely independent from the packing of the monomers.
Since the correlation hole comprises $N-1$ monomers and the only length scale of 
a flexible Gaussian polymer is its radius of gyration $R_g \sim \sqrt{N}$, the 
reduced correlation function exhibits the following scaling behaviour
\begin{equation}
\frac{g_{N}(r)}{g_{N=1}(r)} = 
1 - \frac{1}{r}\tilde{g}\left(\frac{r}{\sqrt{N}}\right).
\end{equation}
Fig.\ref{fig:2} demonstrates for various chain lengths\cite{M0}, that such
a scaling works very well for flexible chains. We note that the situation
is somewhat more complicated in systems of semiflexible chains, since the 
correlation hole is characterised by two length scales, the persistence length
and the radius of gyration\cite{STIFF2}. 

The deepening of the correlation hole with increasing chain length induces
a chain length dependence of the effective number of intermolecular contacts
\begin{equation}
z = z^{\infty} + \frac{{\rm const}}{\sqrt{N}}
\end{equation}
and hence presents a correction of relative order $1/\sqrt{N}$ to the scaling 
of the critical temperature for flexible chains (see below).

In the following we will present examples for different contributions to the 
effective $\chi$-parameter. In particular, we will consider the 
``semi-grandcanonical equation of state'', which relates the composition of the 
blend $\rho$ to the exchange chemical potential $\Delta \mu$,
\begin{equation}
\label{eos}
\beta \Delta \mu = \frac{\partial F_{\rm exc}/m}{\partial \rho}
= \frac{1}{N_A}\ln\phi - \frac{1}{N_B} \ln(1-\phi) - 
\chi_{\mbox{\small eff}} (2 \phi -1) + C,
\end{equation}
where $C$ is a constant. Since this thermodynamic relation is directly accessible 
in the simulations of the semi-grandcanonical ensemble, we can extract an 
effective $\chi$ parameter. This value can be then compared to Eqn.(\ref{chiv}), 
which relates the local fluid structure to the incompatibility $\chi$.

\begin{figure}[htbp]
    \begin{minipage}[t]{110mm}%
           \mbox{
	              \setlength{\epsfxsize}{9cm}
	              \epsffile{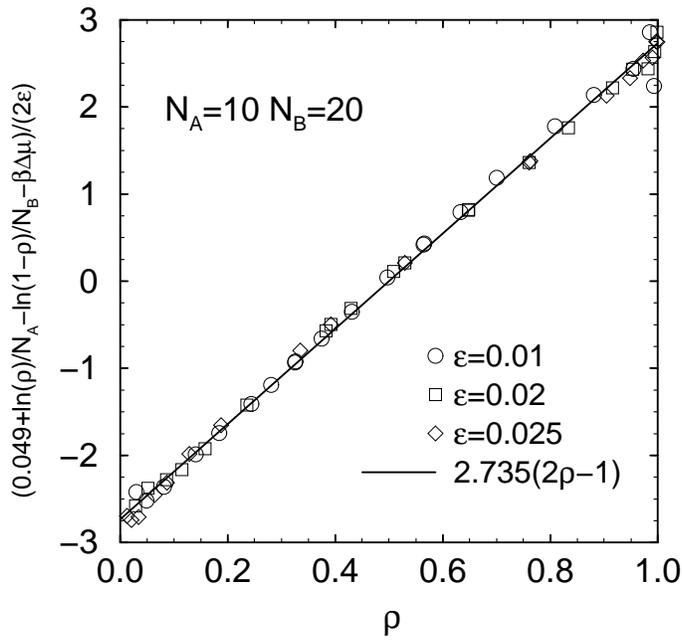}
	        }							         
     \end{minipage}%
     \hfill%
     \begin{minipage}[b]{110mm}%
\caption{Equation of state for an asymmetric polymer mixture $N_A=10$ and 
	$N_B=20$ for various temperatures above the critical point. 
	$\epsilon=0.01$ (circles), $\epsilon=0.02$ (squares), $\epsilon=0.025$ 
	(diamonds). $\epsilon_c= 0.0320(1)$. The straight line corresponds to the prediction with 
	$z_c=2.735$.
        }
	From Ref. \cite{M0}
	   \label{fig:eqs1}
     \end{minipage}%
\end{figure}

Our first example is a blend of polymers of different chain length 
$(N_A \ne N_B)$, which demix for energetic reasons, {\em i.e.}, unlike
monomers repell each other according to Eqn.(\ref{eqn:energy}).
Here the athermal ($\epsilon=0$) system of non-interacting monomers suggests itself
as the reference system. The bare contribution $\Delta f^0$ to the $\chi$ 
parameter is negligible, and the pair correlation functions in the reference 
system do not depend on the identity of the monomers. The combination of
Eqns.(\ref{gamma}), (\ref{X}), and (\ref{chiv}) then yields the
effective Flory Huggins parameter
\begin{equation}
\chi_{\mbox{\small eff}} = 2 z_c \epsilon, \qquad \mbox{where} \qquad
z_c = \varrho \int_0^{\sqrt{6}} dr \: g(r)
\end{equation}
defines an ``effective coordination number'' $z_c$ in the spirit of the 
original Flory-Huggins theory. We recall the $g(r)$ is the intermolecular
pair correlation function.  Fig.\ref{fig:eqs1} demonstrates that the 
equation of state obtained with this value of $\chi_{\mbox{\small eff}}$ is in 
excellent agreement with the simulation data \cite{M0}, for temperatures well 
above the unmixing transition.

The second example illustrates demixing driven by entropic repulsion
between monomers. One particularly simple realisation is non-additive packing.
In real polymer mixtures, it might arise from disparities in the monomer shape or
steric hindrances\cite{FREED2,Honeycutt}. It is also used by Grest and coworkers 
to induce phase separation in an off-lattice model\cite{GARY1}. 
In our case, it is modelled as follows: Monomers of different type are subject 
to the additional constraint that they must not come closer than $\sqrt{5}$ 
lattice constants, whereas monomers of the same type can approach each other up 
to $2$ lattice constants, as before. The natural reference system is again the 
athermal system of non-overlapping monomers. 
Eqns (\ref{gamma}), (\ref{X}) and (\ref{chiv}) 
lead to the entropic $\chi$ parameter
\begin{equation}
\chi_{\mbox{\small eff}} = z_6 \qquad \mbox{with} 
\qquad z_6 = \varrho \int_0^{\sqrt{5}} dr \: g(r).
\end{equation}
This non-additive packing thus gives rise to a rather large, positive 
contribution to the effective Flory-Huggins parameter, and the system 
demixes. In addition, energetic interactions can be introduced as before.
The total effective $\chi$ parameter is then given by
\begin{equation}
\chi_{\mbox{\small eff}} = z_6 + 2 \epsilon z_c.
\label{eqn:za}
\end{equation}
The case of {\em attractive} interaction between unlike monomers 
({\em i.e.}, $\epsilon<0$) is particularly interesting. Here one encounters a 
Lower Critical Solution Point (LCSP) upon heating the system, {\em i.e.}, 
reducing the strength of the thermal interaction $\epsilon$. 

\begin{figure}[htbp]
    \begin{minipage}[t]{110mm}%
           \mbox{
	              \setlength{\epsfxsize}{9cm}
	              \epsffile{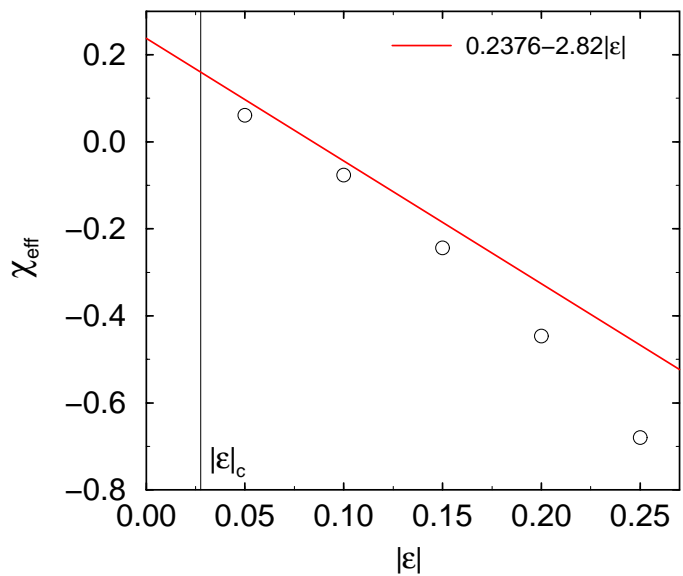}
	        }							         
     \end{minipage}%
     \hfill%
     \begin{minipage}[b]{110mm}%
\caption{
Effective Flory-Huggins parameter extracted from the semi-grandcanonical 
	equation of state for a non-additive polymer mixture of chain length 
	$N=20$ at a total monomer density of $ 8 \varrho = 0.35$. There is an 
	attraction between unlike monomers ($\epsilon<0$). Upon heating 
	{\em i.e.}, reducing the thermal interactions, the blend phase separates. 
	The location of the Lower Critical Solution point is indicated as a 
	vertical line.
        }
	From Ref. \cite{STIFF1}.
	   \label{fig:eqs}
     \end{minipage}%
\end{figure}

Using the intermolecular pair correlation function of a reference system 
($|\epsilon|=0.05$, additive), we obtain $z_6=0.238(2)$ and $z_c=1.41(2)$ at 
the monomer volume fraction $8 \varrho = 0.35$, which corresponds to a 
concentrated solution. Fig.\ref{fig:eqs} compares the effective $\chi$ parameter 
extracted from the semi-grandcanonical equation of state in the one phase region 
({\em i.e.}, at low temperatures) with the above estimate.
Eqn.(\ref{eqn:za}) describes the simulation data quite successfully \cite{STIFF1}. 
We note that Eqn. (\ref{eqn:za}) predicts that the critical temperature of the 
LCSP does not increase upon increasing the chain length, but converges to 
$1/T_c = |\epsilon_c| = z_6/2z_c$. This is indeed found in the simulations, and 
in qualitative agreement with experiments of Schwahn and co-workers\cite{Schwahn} 
on the phase behaviour of PVME/PS blends.  

\begin{figure}[htbp]
    \begin{minipage}[t]{110mm}%
           \mbox{
	              \setlength{\epsfxsize}{9cm}
	              \epsffile{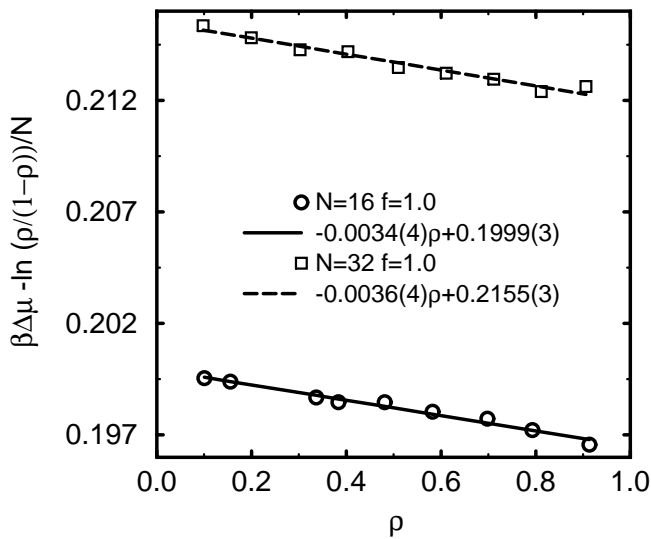}
	        }							         
     \end{minipage}%
     \hfill%
     \begin{minipage}[b]{110mm}%
\caption{Semi-grandcanonical equation of state for a polymer mixture of flexible 
	($f=0$) and semi-flexible ($f=1$) polymers without thermal interactions 
	($\epsilon=0$). The translational entropy is already accounted for and 
	the composition dependence indicates the stiffness-induced contribution 
	to the Flory-Huggins parameter.
	}
	From Ref. \cite{STIFF1}.
	   \label{fig:3}
     \end{minipage}%
\end{figure}

The last example presents a situation where polymer incompatibility is 
caused on the level of whole chains as a result of different chain
stiffness. Monomers are taken to be identical otherwise. Hence we
are left with the ``bare'' contribution to the $\chi$ parameter,
$\chi = \Delta f^0(\rho,\varrho)/[\rho(1-\rho)]$, which results
from packing inhomogeneities due to the different chain structure.
The explicit evaluation of this term requires more elaborate 
techniques\cite{singh,liu} than introduced in Sec.2. 
Fig.\ref{fig:3} shows the semi-grandcanonical equation of state for an athermal 
({\em i.e.}, $\epsilon=0$) mixture of flexible ($f=0$) and semiflexible ($f=1$) polymers.
A composition dependence in the figure indicates a non-zero Flory-Huggins parameter.
The simulation data clearly show, that the purely entropic packing differences 
give rise to a small, chain-length independent, positive Flory-Huggins parameter 
$\chi$. Though the effect is quite small, it might lead to phase separation for 
very long chain lengths $N > {\cal O}(1000)$. Of course, one could suspect, 
that these subtle packing effects are strongly influence by the structure of 
the underlying lattice. However, field theoretical calculations by Liu and 
Fredrickson\cite{liu} as well as recent P-RISM calculations by Singh and 
Schweizer\cite{singh} predict effects of similar magnitude within 
off-lattice models. Off-lattice simulations by Weinhold et al.\ \cite{WEINHOLD} 
also confirm the order of magnitude of the packing induced incompatibility.

Since purely athermal systems of flexible and stiff chains with chain lengths 
accessible to simulations usually don't demix, additional, repulsive monomer 
interactions have to be turned on in order to induce demixing. 
The ``athermal'' system of flexible and stiff chains then has to be used as 
the reference system\cite{STIFF2} (cf. Eqn.(\ref{gamma})). 
We note that the reference system is asymmetric,
and the effective coordination numbers depend on the identity of the monomers 
and the composition of the blend. This is because stiffer chains are more
extended, and folding back is less probable. Hence the number 
of intermolecular contacts increases and,  as a result, the interaction
parameter $\gamma_{ii}$ in (\ref{gamma}) is larger for the stiffer species, 
even if the interaction potentials are chosen symmetric. The effect on the
$\chi$ parameter is comparable in magnitude to the stiffness induced (athermal) 
contribution discussed above\cite{STIFF1} for temperatures of the order $T_c$.
For strongly incompatible blends, the enthalpic contribution to $\chi$ dominates.

\begin{figure}[htbp]
    \begin{minipage}[t]{110mm}%
           \mbox{
	              \setlength{\epsfxsize}{9cm}
	              \epsffile{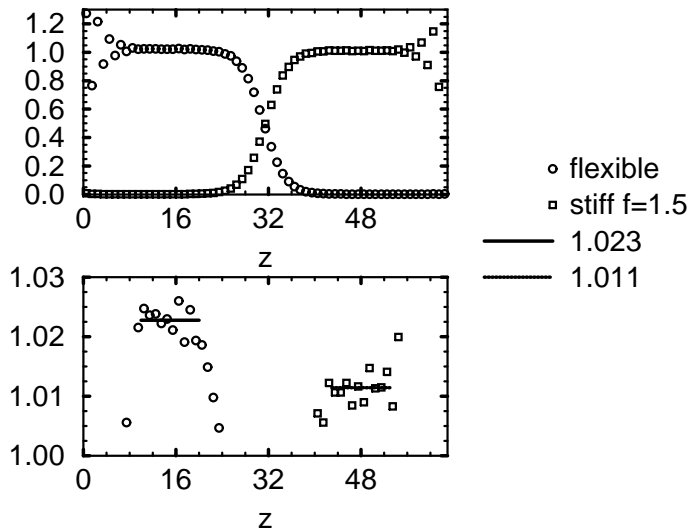}
	        }							         
     \end{minipage}%
     \hfill%
     \begin{minipage}[b]{110mm}%
\caption{Polymer blend of flexible ($f=0$) and semi-flexible ($f=1.5$) polymers ($N=16$)
	between two hard and neutral walls well below the critical temperature
	$T=0.346 T_c$. The normalised monomer density profile shows pronounced
	packing effects at the wall. The blow-up shows that vacancies are 
	enriched in the stiffer phase (right hand side).
	           }
	From Ref. \cite{STIFF1}.
	   \label{fig:stiffeq}
     \end{minipage}%
\end{figure}

Fig.\ref{fig:stiffeq} shows an interface between
two demixed phases in a blend of flexible $A$ and stiff $B$ chains, confined
between hard walls. The two phases are at pressure equilibrium, and
vacancies are enriched in the stiffer phase. 
A similar effect has been observed in an athermal two-dimensional 
system\cite{ROD}. It is due to the fact, that the stiffer chains pack less 
efficiently and hence their osmotic pressure is slightly higher. The
order of magnitude of the effect is however only 1 \%.
The oscillations of the density directly at the walls are packing effects 
and reflect directly the monomer correlation functions. 

Simulation studies of polymer compatibility in continuous space and at
constant pressure have been performed recently by Kumar\cite{KUMAR_PD}. 
He studies miscible and immiscible blends of Lennard-Jones chains, and
finds volume changes of mixing with the expected parabolic composition 
dependence. He also measures incremental chemical potentials of both components 
and relates them to an expression similar in spirit to Eqn.(\ref{chi}). 
Inserting correlation functions which include both interchain and intrachain 
contacts gives worse results than using correlation functions for interchain 
contacts only, in agreement with the considerations of Sec. 2. In particular,
he studies the special case of a blend with 
interactions chosen such that ${\cal X}$ happens to cancel to zero\cite{kumar2}. 
As expected, the volume change of mixing practically disappears. Nevertheless, 
the blend demixes, if the chains are sufficiently long. The effect can be 
attributed in part to compressibility effects ({\em i.e.}, the remaining third 
term in Eqn. (\ref{chip})), but probably also involves higher-order effects, 
such as non-random-mixing and chain-stretching effects. A systematic 
investigation of the scaling of the critical temperature with chain length 
would help to separate the different contributions, but has not yet been 
attempted.

In the above examples we have related the structure of the polymeric fluid to 
its thermodynamical properties far from the unmixing transition, and obtained a 
reasonable quantitative agreement. Within mean-field theory, the unmixing 
temperature is given by Eqn.\ref{eqn:unmix}. 

\begin{figure}[htbp]
    \begin{minipage}[t]{110mm}%
           \mbox{
	              \setlength{\epsfxsize}{9cm}
	              \epsffile{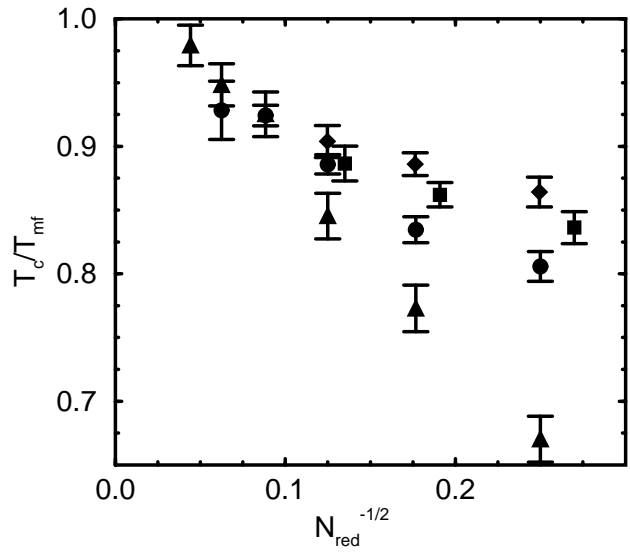}
	        }							         
     \end{minipage}%
     \hfill%
     \begin{minipage}[b]{110mm}%
\caption{Chain length dependence of the ratio between the critical temperature of 
	a binary polymer blend and the mean-field estimate, which takes the local 
	fluid structure into account. Symmetric mixtures ($N_A=N_B$) are 
	presented by circles and triangles (for a reduced range of the thermal 
	interactions). Asymmetric mixtures with $N_B=k N_A$ are shown as squares 
	($k=2$) and diamonds ($k=3$). 
	$N_{\rm red}=4 N_A N_B/(N_A^{1/2}+N_B^{1/2})^2$.  
	}
	From Ref. \cite{M0}.
	   \label{fig:4}
     \end{minipage}%
\end{figure}

However, even in the simplest possible systems -- strictly symmetric, binary 
polymer blends --  mean-field theory overestimates the critical temperature 
at short chain lengths\cite{M0}. This discrepancy  between the mean-field theory 
and the Monte Carlo results can be attributed to  composition fluctuations. 
The fact that we observe 3D Ising critical behaviour at the unmixing transition
already indicates that the deviations from the random-mixing approximation 
involved in the mean-field theory are quite strong. The effect is well known in 
mixtures of small molecules, {\em i.e.}, the small chain length limit of polymer 
blends. P-RISM calculations\cite{Schweizer93} and calculations of fluctuation 
corrections to  mean-field by Holyst and Vilgis\cite{VILGIS} suggest that these 
non-random-mixing effects die out with $1/\sqrt{N}$. Thus, the ratio between the 
actual critical temperature and the mean-field value should converge to 1 with a 
$1/\sqrt{N}$ correction. This is illustrated in Fig.\ref{fig:4} for strictly 
symmetric models with different interaction range, and for blends with chain 
length asymmetry\cite{M0}. The relative deviation between the critical 
temperature and its mean-field estimate is the larger, the smaller the 
interaction range. When the mean-field theory is used to fit Monte
Carlo data on the critical isotherm, non-random mixing effects thus
lead to a strong parabolic composition dependence of the effective $\chi$ 
parameter close to the critical point\cite{Deutsch,M0}.

\subsection{Homopolymer interfaces}

We shall now discuss some general concepts on the structure of interfaces 
between demixed homopolymer phases. For the sake of simplicity, we will 
consider a symmetrical system of polymers $A$ and $B$ with
equal chain length $N$ and statistical segment length $b$. Generalised
expressions for asymmetric systems will be given later if necessary.
Within the Gaussian chain model, the behaviour of the interface is basically 
driven by the relative polymer incompatibility $\chi N$. One distinguishes
the two limiting cases $\chi N \to \infty$ (strong segregation limit)
and $\chi N \to (\chi N)_c$ (weak segregation limit). The two quantities
in which we will be interested are the interfacial width and the
interfacial tension. 

The intrinsic interfacial width $w$ results from an interplay between entropic 
factors and ``energetic'' factors related to the monomer incompatibility, 
{\em i.e.}, the $\chi$ parameter.  The interfacial tension, $\sigma$,
is closely related. It turns out to be roughly proportional to $m^2/w$, 
where $m=\rho_A^{coex}-\rho_B^{coex}$ is the width of the miscibility gap.
In the limit of very long chains, the conformations of chains are strongly
perturbed in the vicinity of the interface. Polymers can win conformational 
entropy by allowing chain portions to loop into the unfavourable side. 
The entropy gain due to the formation of such loops has to be balanced with 
the energy loss due to the unfavourable contacts of the loops. Since the total 
chain length does not enter this argument, the interfacial width is independent
of chain length, and so is the interfacial tension. The strong segregation 
theory ($2/N \ll \chi \ll 1$) yields \cite{SCF1}
\begin{equation}
w_{\rm SSL} = b/\sqrt{6 \chi}, \qquad \beta \sigma_{\rm SSL} = \sqrt{\chi/6} 
\: \varrho \; b.
\end{equation}
In the limit of short chains or weak incompatibility, the conformations
of chains are only weakly perturbed. Here, polymers as a whole may move 
over to the unfavourable side, thereby gaining translational entropy. The
latter has again to be balanced with the energy loss, and the resulting
interfacial width and interfacial tension depend on the distance 
of $\chi N$ from the critical point $(\chi N)_c$.
In the weak segregation limit, one thus gets
\begin{equation}
w_{\rm WSL} \propto (1-\chi_c/\chi )^{-\nu}, \qquad 
\sigma_{\rm WSL} \propto (1-\chi_c/\chi)^{\mu},
\end{equation}
with the critical exponents $\nu=1/2$, $\mu=3/2$ and the critical point 
$\chi_c=2/N$ in the mean-field regime.
Experiments are usually well described by mean-field exponents, except in
the ultimate vicinity of the critical point. In 
simulations, the chains in the blend are often too short to prevent
critical fluctuations, and one finds Ising exponents.
In the intermediate region between the weak segregation limit and
the strong segregation limit, the contribution of loops competes with
an increasing influence of chain ends. Interpolation schemes between
the two limits have been suggested by Tang and Freed\cite{tang} and
more recently by Ermoshkin and Semenov\cite{ER}. Numerically,
the problem can be treated, {\em e.g.}, within the self-consistent
field theory\cite{SCF1,SCF2,SCF3}.



From this discussion, it is clear that the properties of interfaces
are strongly influenced by their microscopic structure. This interplay
has attracted abiding theoretical\cite{SCF1,SCF2,SCF3,scheutjens,FREED1,tang}
and experimental\cite{SANCHEZ,RUSSELL,COP_EXP,NOSE,EX1,EX2,EX3,EX4} interest.
Apart from the factors already mentioned, packing and compressibility effects 
are again important, as well as chain orientations and chain-end distributions. 
These are all accessible to computer simulations. 
Unfortunately, simulation studies of polymer
interfaces are computationally extremely demanding. Only recently have 
such simulations been carried out by different groups. The first is due to 
Reiter and coworkers and treats polydisperse chains on a lattice model using
a bond breaking algorithm\cite{REITER}. More recently, Ypma et al
have performed simulations of a polymer interface confined between
a hard wall and a free surface\cite{CIFRA1}. Grest and coworkers 
study a polymer interface in a continuous model, in which demixing
is driven by non-additive packing\cite{GARY2}. Similar to our 
simulations\cite{andreas,COP}, they also put particular emphasis on the 
analysis of the capillary-wave broadening.
In general, the results from these simulations agree with our own
results obtained with the bond-fluctuation model , which shall be 
discussed in some detail in the following sections.
Simulations of copolymers embedded in a homopolymer interface are even
more scarce. In many studies, the homopolymer interface has been
approximated by an external potential\cite{cop,JENSUWE}. Simulations of a 
copolymer/homopolymer mixture have been performed by Pan et al\cite{pan}, 
however, the homopolymer concentration in their study is less than 10 \%.

\begin{figure}[htbp]
    \begin{minipage}[t]{110mm}%
           \mbox{
	              \setlength{\epsfxsize}{9cm}
	              \epsffile{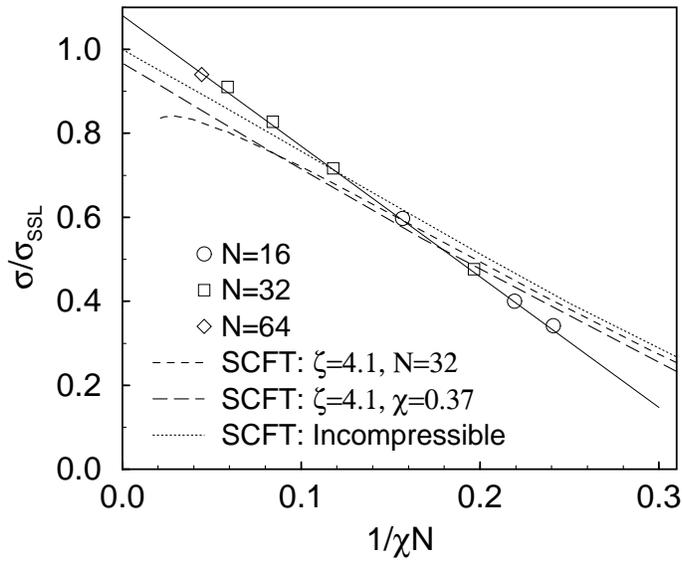}
	        }							         
     \end{minipage}%
     \hfill%
     \begin{minipage}[b]{110mm}%
\caption{Temperature dependence of the interfacial tension for chain lengths 
	$N=16,32,64$. The ratio of the interfacial tension (measured using the 
	reweighting scheme) and the analytical prediction is presented to 
	investigate chain end corrections. The lines correspond to self-consistent
	field calculations for compressible and incompressible systems.
	}
	From Ref. \cite{SM}.
	   \label{fig:8}
     \end{minipage}%
\end{figure}

We begin with the discussion of the interfacial properties of a homopolymer 
interface in a strictly symmetric binary blend. For a structurally symmetric 
polymer blend, the interfacial tension is accessible {\em via} the reweighting 
technique. At the critical point, it vanishes with a critical power law of the 
form $\sigma = \hat{\sigma} (\epsilon - \epsilon_c)^\mu$, where the critical 
exponent $\mu = 2 \nu =1.258$ is given by the 3D Ising universality class, and 
the critical amplitude takes the value $\hat{\sigma}=2.9$ for $N=32$. 
In the strong segregation limit, the self-consistent field theory for 
Gaussian chains predicts in the long chain limit
\begin{equation}
\sigma_{\rm SSL} = \varrho \sqrt{\chi/6} 
\left( \frac{2}{3} \frac{b_A^2+b_Ab_B+b_B^2}{b_A+b_B}\right),
 \label{eqn:tension}
\end{equation}
where $b_A$ and $b_B$ denote the statistical segment length of the components. 
For finite chain length $N$, however, the interfacial tension is reduced 
by a factor $1 - \alpha_\sigma \frac{2}{\chi N}$.
There exist various analytical predictions for the constant $\alpha_\sigma$:
$\alpha_\sigma = \ln 2$\cite{1}, $\alpha_\sigma = \pi^2/12$\cite{2}, 
$\alpha_\sigma = 1.35$\cite{tang}, and  $\alpha_\sigma = 2 \ln 2$\cite{ER}. 
In Fig.\ref{fig:8} we present the ratio between the measured interfacial tension 
and the strong segregation estimate, and compare it to the self-consistent field 
theory\cite{SM}. The deviations from the strong segregation result are indeed
well described by a correction of the proposed form, and the simulation data as 
well as the SCF results agree best with a constant $2\ln 2$.

\begin{figure}[htbp]
    \begin{minipage}[t]{110mm}%
           \mbox{
	              \setlength{\epsfxsize}{9cm}
	              \epsffile{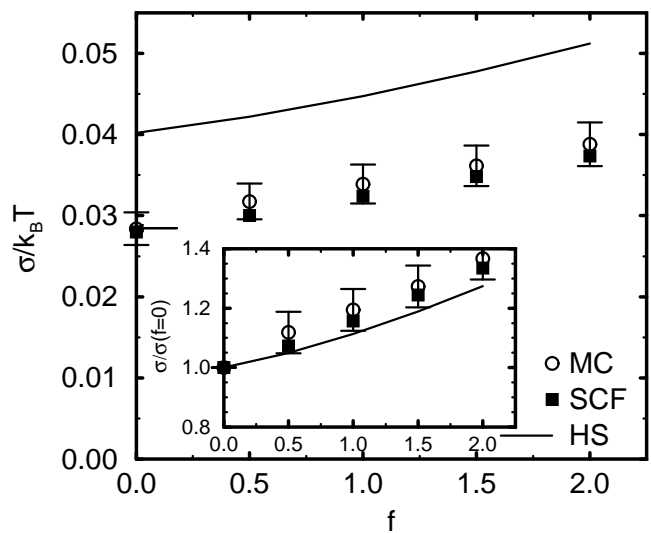}
	        }							         
     \end{minipage}%
     \hfill%
     \begin{minipage}[b]{110mm}%
\caption{Stiffness dependence of the interfacial tension for chain length 
	$N=32$ for rather high incompatibility $\epsilon=0.05$. 
	Circles denote estimates via an analysis of the spectrum of interfacial 
	fluctuations. The arrow marks the independently measured value ({em via}
	the reweighting scheme) for the symmetric blend. The data are compared to 
	self-consistent field calculations which take due account of the chain 
	architecture. Also shown is the strong segregation prediction of the 
	Gaussian chain model for blends with different statistical segment length 
	by Helfand and Sapse (HS). The inset presents the relative effect of 
	increasing the stiffness $f$ of the $B$ component.
	}
	From Ref. \cite{STIFF2}.
	   \label{fig:stiffs}
     \end{minipage}%
\end{figure}

The influence of architectural asymmetry is illustrated in Fig.\ref{fig:stiffs}. 
Upon increasing the persistence length of the $B$ component, the interfacial 
tension increases\cite{STIFF2}. The arrow marks the result for symmetrical 
mixtures obtained {\em via} the reweighting technique. Note that packing effects, 
which are induced by the stiffness disparity, increase the incompatibility by a 
factor of $\Delta \chi/\chi \approx 7 \; 10^{-3}$ for $f=1$\cite{STIFF1}. Hence, 
they are not strong enough to explain the increase of the interfacial tension 
for this combination of chain length, enthalpic repulsion and stiffness disparity.
The simulation data in the figure are compared to numerical self-consistent field 
calculations (SCF), which take due account of the chain conformations on all 
lengths scales, and to the analytical prediction for infinitely long Gaussian 
chains of different statistical segment lengths (cf.\ Eqn.\ \ref{eqn:tension}). 
The self-consistent field theory describes the dependence on the chain 
architecture almost quantitatively, whereas the analytical calculation captures
the qualitative trend, but fails to predict the correct absolute values.

\begin{figure}[htbp]
    \begin{minipage}[t]{110mm}%
           \mbox{
	              \setlength{\epsfxsize}{9cm}
	              \epsffile{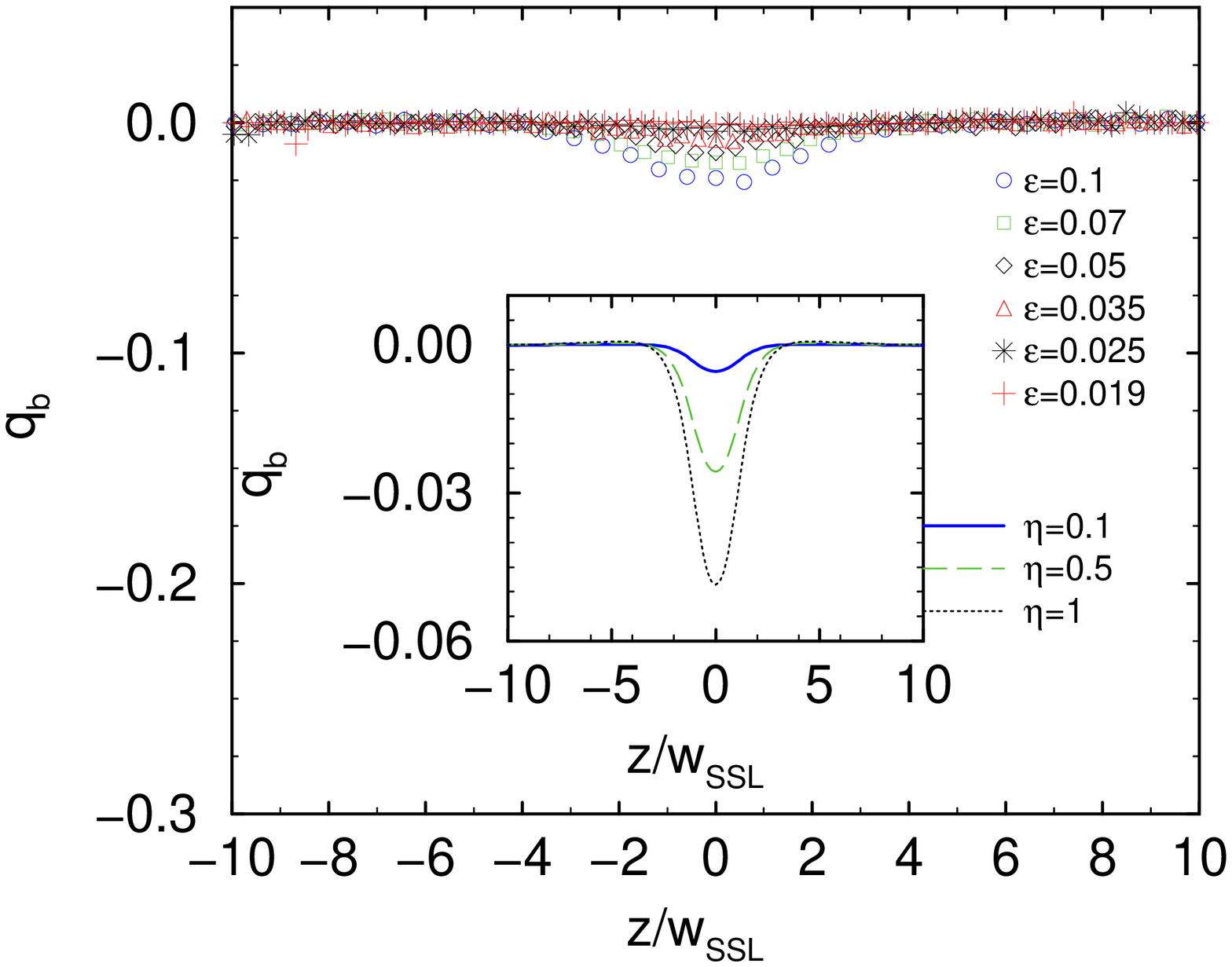}
	        }         \\
           \mbox{
	              \setlength{\epsfxsize}{9cm}
	              \epsffile{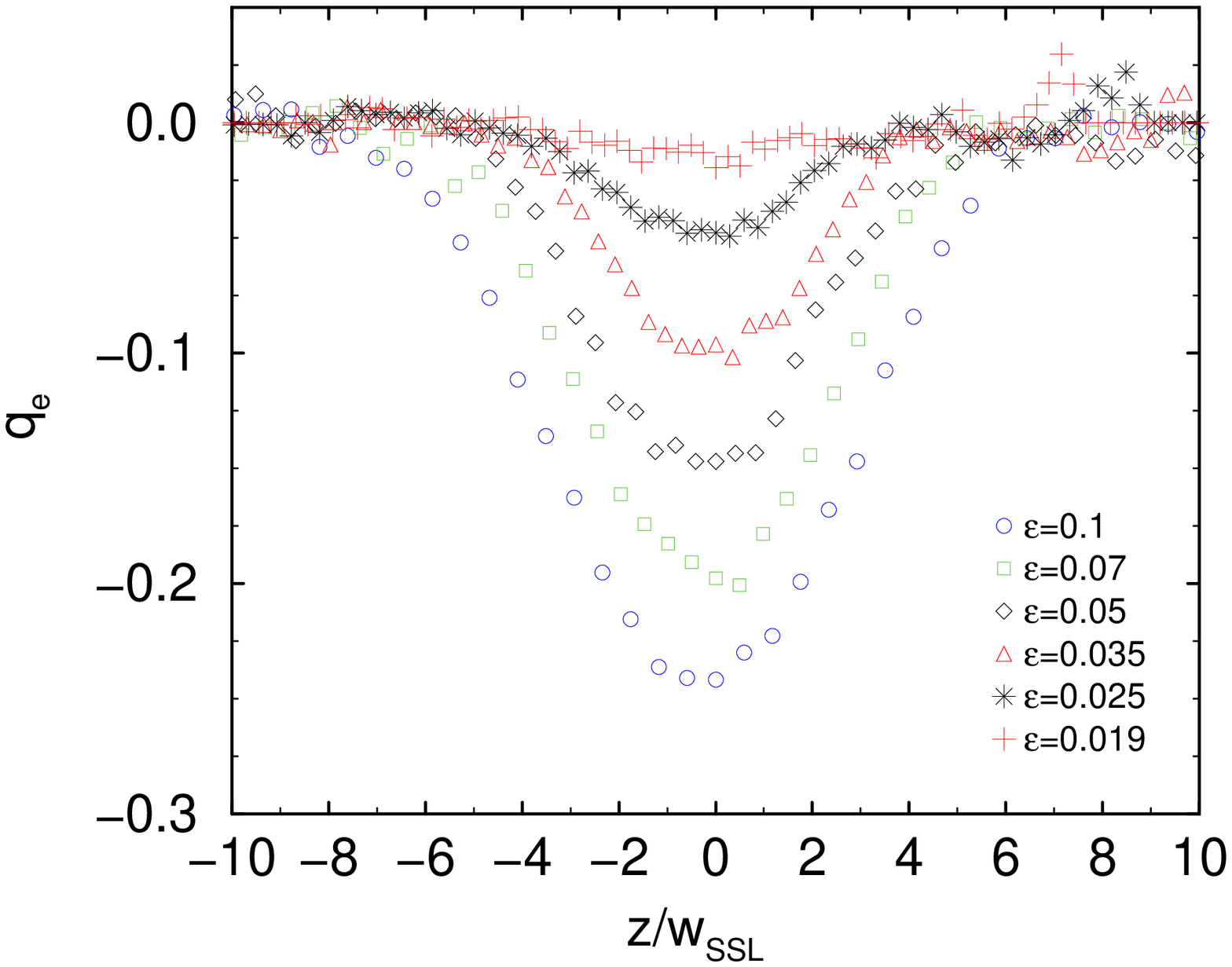}
	        }							         
     \end{minipage}%
     \hfill%
     \begin{minipage}[b]{110mm}%
\caption{({\bf a}): Orientation of bond vectors as a function of the distance 
	from the interface. Apparent profiles for chain length $N=32$ are shown 
	for various inverse temperatures $\epsilon$. The temperature ranges
	from the weak segregation limit to the strong segregation limit. The 
	inset presents self-consistent field calculations in the framework of the
	worm-like chain model for different values of the bending rigidity $\eta=0.1,0.5$  and $1$.
	and $\epsilon=0.1$. 
	({\bf b}): Apparent profile for the orientation of the chain's 
	end-to-end vector as a function of the distance from the interface. 
	Polymers align parallel to the interface. Note that the effect 
	is much more pronounced than for the bond vectors.
	}
	   \label{fig:9}
     \end{minipage}%
\end{figure}

These thermodynamic properties do not shed light on the microscopic structure of 
the interface\cite{COP2}. However, factors like the width of the interfacial 
region, the orientations of polymers on different length scales and the 
enrichment of vacancies or solvent influence the materials properties. 
Experiments indicate\cite{KRAMERC}, that entanglements in the interfacial
zone are of major importance for the mechanical properties of the composite. 
The distribution of chain ends are important for the interdiffusion and healing 
properties at interfaces between long polymers. They also play an important role 
for reactions at interfaces. Chain ends are enriched in the center of interfaces,
and this effects goes along with a depletion at a distance $R_g$ away from the 
midpoint of the profile. The shape of a polymer is a prolate ellipsoid, which
is oriented in the presence of an interface. 
This is quantified by the orientational parameter
\begin{equation}
q_X(z) \equiv \frac{3 \langle X_z^2  
\rangle_z - \langle \vec{X}^2 \rangle_z}{ 2  \langle \vec{X}^2 \rangle_z},
\end{equation}
where the outer index $z$ at the brackets denotes the $z$ coordinate of the 
vector's midpoint, and the inner indices the Cartesian components of the vector 
$\vec{X}$. The apparent orientation profiles for the end-to-end vector 
$\vec{R}_e$ and the bond vectors $\vec{b}$ are presented in Fig. \ref{fig:9}.
Both the bond vectors and the end-to-end vector align parallel to the interface.
Most notably, the orientation effects are the stronger the larger the length 
scale\cite{MBO}. The Gaussian chain model does not predict any orientation of
bond vectors, and the observed effect is rather small. The inset shows the 
predictions\cite{SM} of a worm-like chain model for the bond orientation. A 
rather small bending stiffness is sufficient to describe the simulation data
(for chains with bending potential $f=0$). On the end-to-end vector, however,
the interface has a similar effect than a free surface or a hard wall, 
and induces pronounced alignment.

\begin{figure}[htbp]
    \begin{minipage}[t]{110mm}%
       \setlength{\epsfxsize}{9cm}
       \mbox{\epsffile{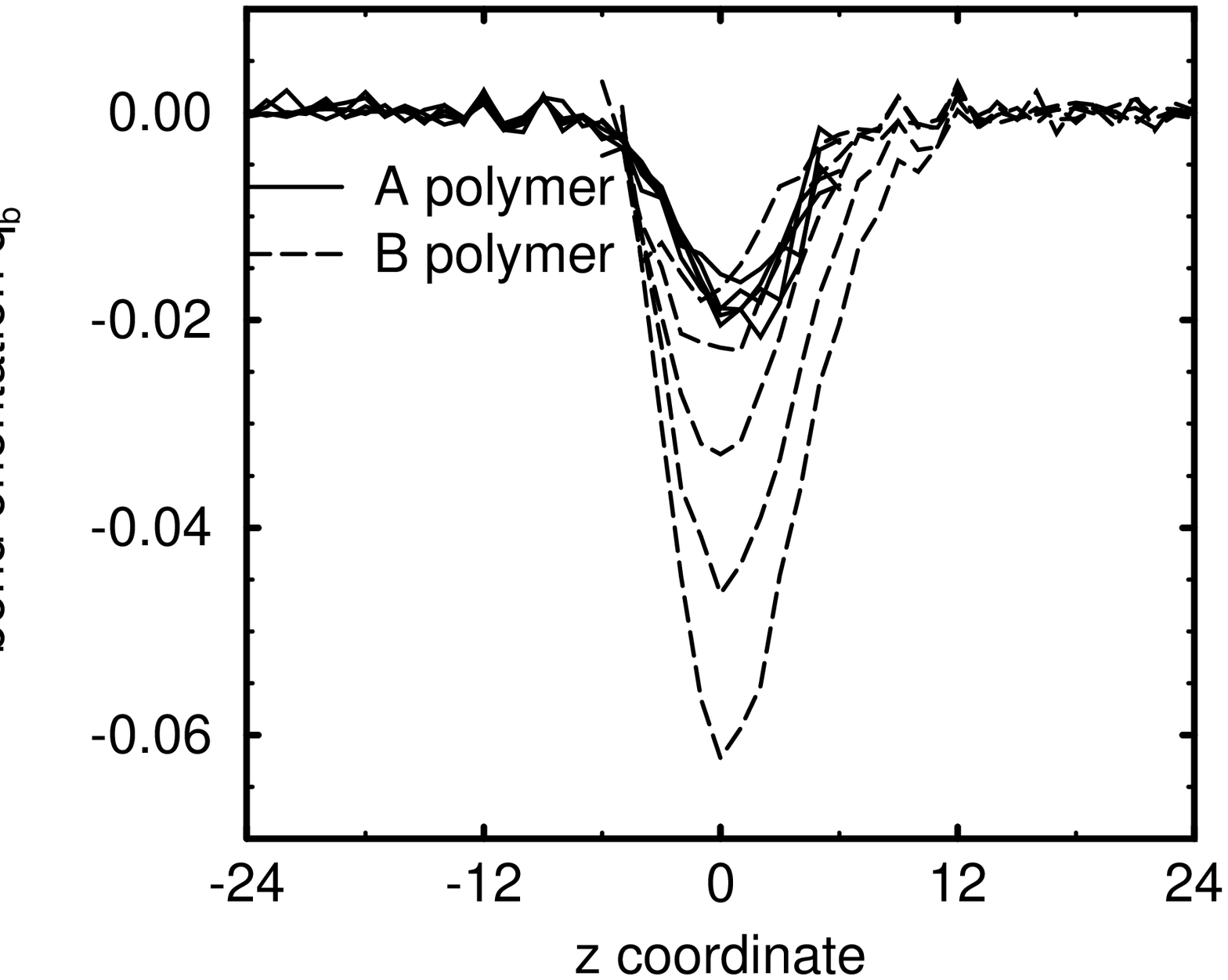}}
       \setlength{\epsfxsize}{9cm}
       \mbox{\epsffile{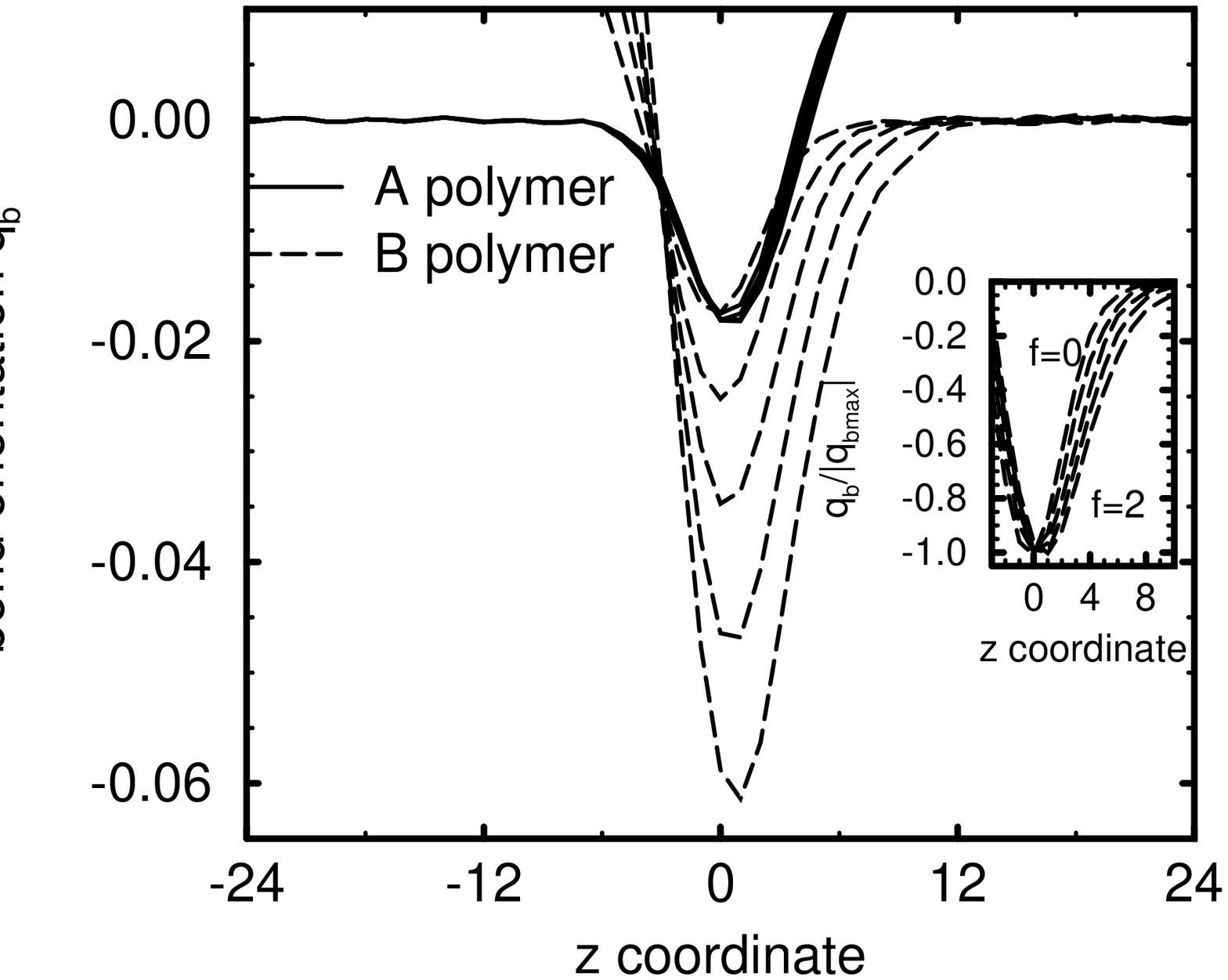}}
    \end{minipage}%
    \hfill%
    \begin{minipage}[b]{110mm}%
\caption{Orientation of bond vectors $q_b$, for flexible $A$-polymers (solid lines) and 
	semi-flexible $B$-polymers (dashed lines) with bending energies $f=0, 0.5, 1, 1.5$ and 2.
	The orientation of the $B$-bonds and the length scale of ordering
       increases upon increasing $f$.
       ({\bf a}) Monte Carlo results, ({\bf b}) self-consistent field calculations
       The inset shows the self-consistent field results normalised to the 
	maximum of $q_b$. Note, that the width of the orientation profile grows 
	upon increasing $f$.
        }
	From Ref. \cite{STIFF2}.
       \label{fig:Sqb}
    \end{minipage}%
\end{figure}

In blends with stiffness disparity\cite{STIFF2} the persistence length of the 
semi-flexible component introduces a second microscopic length scale 
into the interfacial profile, which is independent from the interfacial width. 
In the weak segregation limit, the interfacial width is larger than the 
persistence length of the semi-flexible component and the interfacial behaviour 
can be described appropriately within the Gaussian chain model, when the stiffness
dependence of the statistical segment length $b(f)$ is accounted for. Upon increasing 
the stiffness disparity at fixed incompatibility, the interfacial width increases too. 
In the strong segregation limit, the interfacial widths becomes narrower than the persistence 
length and the local chain architecture becomes important. In this highly 
incompatible regime, the interfacial width decreases upon increasing the 
stiffness disparity at fixed $\chi N$, in qualitative contrast to the predictions 
of the Gaussian chain model.
Self-consistent field calculations which take due account of the chain 
architecture on all length scales describe the simulation data almost 
quantitatively\cite{STIFF2}. Figure \ref{fig:Sqb} shows bond orientational 
profiles in a temperature region, where the stiffness disparity hardly affects
the width of the composition profile. Upon increasing the bending energy $f$ of 
the semiflexible component, the orientation of the semiflexible $B$ polymers 
increases, whereas the flexible component remains unaffected. The
width of the orientational profiles increases upon increasing $f$. The numerical 
self-consistent field calculations and the reduced interfacial profiles agree 
almost quantitatively.

Moreover, we find enrichment of vacancies  at the interface of our compressible 
blend. This segregation of vacancies can be qualitatively estimated as follows: 
By reducing the monomer density $\delta \varrho$, the system lowers the number of unfavourable 
contacts between unlike monomer species. This reduces the excess energy density 
by $\epsilon w z_c \delta \varrho/2$. The decrease of contact energy contrasts 
the entropy loss due to the creation of a density fluctuation. The latter can be 
estimated as $w\delta \varrho^2/2 \varrho \kappa_T$, where $\kappa_T$ denotes the 
isothermal compressibility. Balancing both contributions\cite{MBO}, we estimate  
the reduction of the total monomer density $\delta \varrho$ to be of the order 
$\chi \kappa_T \varrho/2$. This simple argument is compatible with the simulation 
data for the total monomer concentration profile, and with self-consistent field 
calculations for compressible polymer blends.

\begin{figure}[htbp]
    \begin{minipage}[t]{110mm}%
           \mbox{
	              \setlength{\epsfxsize}{9cm}
	              \epsffile{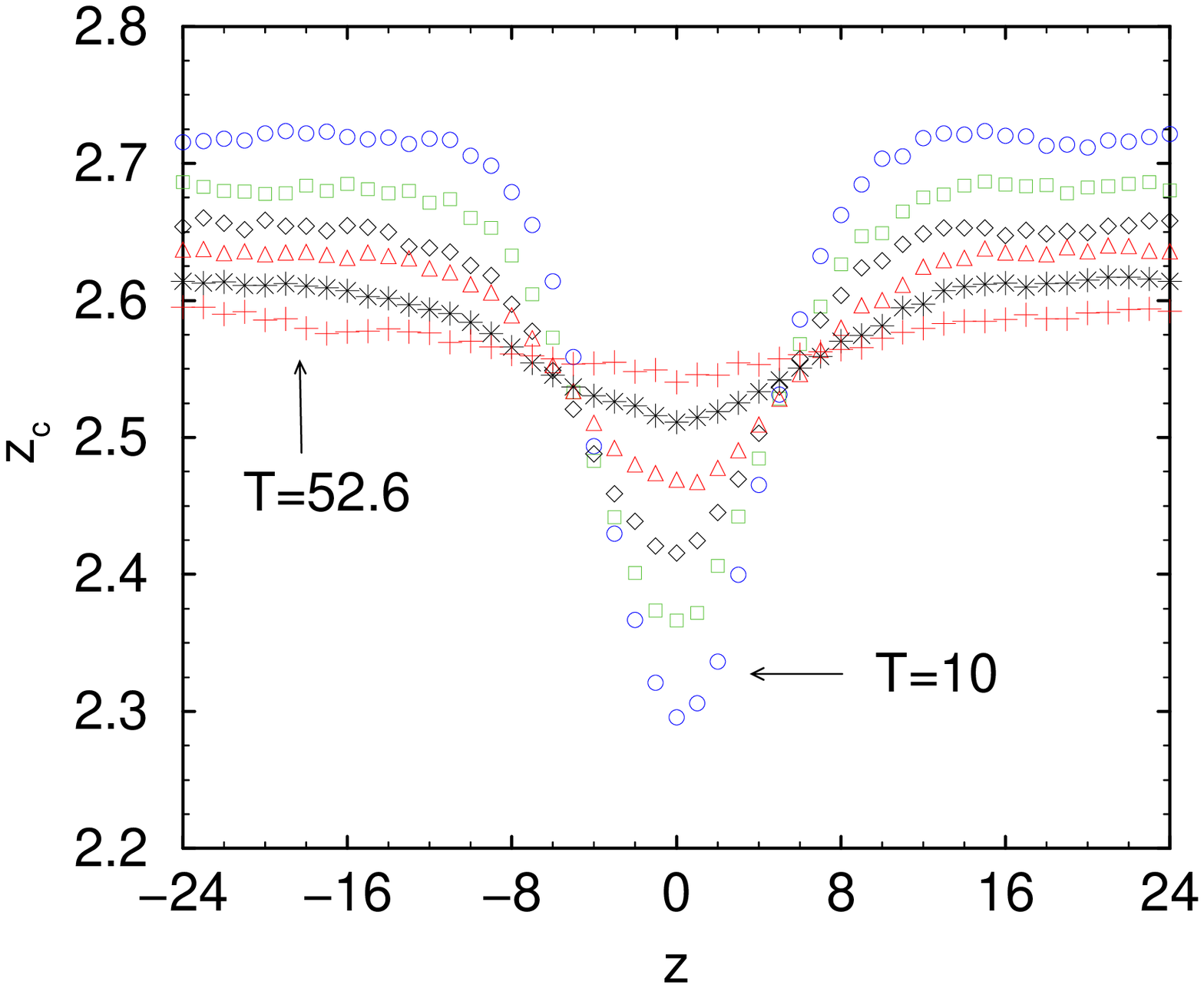}
	        }							         
     \end{minipage}%
     \hfill%
     \begin{minipage}[b]{110mm}%
\caption{Apparent profiles of the intermolecular contacts for inverse temperatures
	$\epsilon=0.1,0.07,0.05,0.035,0.025,0.019$ and chain length $N=32$. Note 
	the pronounced reduction of intermolecular contacts at the center of the 
	interface in the strong segregation limit.
	}
	From Ref. \cite{MBO}.
	   \label{fig:10}
     \end{minipage}%
\end{figure}

Finally in this section we discuss the profile of the number of intermolecular
contacts across the interface (Fig. \ref{fig:10} \cite{MBO}).
In the weak segregation limit, the number of contacts $z_c$ is independent of the 
distance from the interface, whereas it is strongly reduced in the center of the 
interface at high segregation. This effect does not simply reflect the
density reduction at the center of the interface, but is even more pronounced.
It results from orientation and  conformational changes of the macromolecules at 
the interface\cite{MBO}: Polymers rearrange as to exchange energetically 
unfavourable intermolecular contacts with intramolecular contacts. The latter 
number is increased at the interface.

\subsection{Copolymers at interfaces}

We will now discuss the effect of adding copolymers to the interface.
Such amphiphilic molecules are, of course, most likely to be found at the 
interface between $A$ and $B$ homopolymer regions, where the different blocks can 
extend into the appropriate volume. This reduces not only their own enthalpy, but 
also that of the homopolymers which it displaces from the interface. 
The reduction of the interfacial tension has a strong influence on the morphology 
of the blend, whereas the entanglement of the individual blocks with the 
corresponding homopolymers increase the mechanical stability of the composite.
The tendency to adsorb at the interface is balanced by the loss of entropy of 
mixing and of conformational entropy as the copolymers stretch to accommodate a 
greater areal density. Thus the adsorption at the interface competes with the 
creation of micelles in the bulk phases. Mean field theories predict that the 
addition of copolymers drives the system to compatibility in the weak segregation 
limit, while one encounters a first oder transition to a lamellar phase at strong 
segregation. In the intermediate regime, a complex phase behaviour is 
predicted by the theory.

\begin{figure}[htbp]
    \begin{minipage}[t]{110mm}%
           \mbox{
	              \setlength{\epsfxsize}{9cm}
	              \epsffile{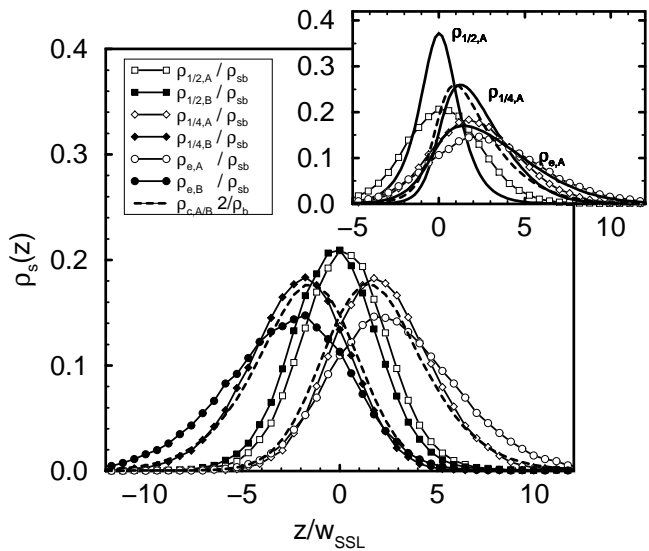}
	        }							         
     \end{minipage}%
     \hfill%
     \begin{minipage}[b]{110mm}
\caption{Copolymer segment density profiles for a symmetric diblock copolymer of 
	length 32 at a highly segregated homopolymer interface $N=32$ and 
	$\epsilon=0.1$. Apparent profiles are shown for the density of $A$ and 
	$B$ monomers in the middle of the chain ($\rho_{1/2}$, squares), at the 
	end of the chain ($\rho_e$, circles), and at one and three-fourths of the 
	chain ($\rho_{1/4}$, diamonds), and for the density of all copolymer 
	monomers ($\rho_c$, broken line). 
	The inset shows the predictions of the self-consistent field theory for 
	$A$ monomers. Full lines show the predictions for segment density profiles 
	$\rho_{1/2}, \rho_e$, and $\rho_{1/4}$, while the broken line presents 
	the total density profile $\rho_c$, and symbols compare with the MC 
	results (notation like above). The distance from the interface is 
	measured in units of the predicted width $w_{\rm SSL}$ in the strong segregation limit and densities are normalised 
	appropriately.
	}
	From Ref. \cite{COP2}.
	   \label{fig:copdens}
     \end{minipage}%
\end{figure}

First, we discuss the adsorption of a small amount of copolymers at a strongly 
segregated homopolymer interface. The areal density $\nu$ of copolymers is so small, 
that they do not overlap $\pi \nu b^2 N/12 \approx 0.3$. The apparent profiles of the individual segments are 
presented in Fig.\ref{fig:copdens}. Chain ends of the homopolymers $\rho_e$
are enriched at the interface, whereas the density of middle segments is reduced 
compared to the total homopolymer density. The copolymers, however, exhibit the 
opposite trend. The middle segment, which joins the distinct blocks, is enriched 
at the center of the interface, whereas the chain ends stretch out into their 
corresponding bulk phases. The inset presents the results of the self-consistent 
field calculations, which qualitatively agree with the simulation data, except
for the fact that they do not capture the capillary-wave broadening. The 
qualitative behaviour is also born out in  experimental findings of 
Russell\cite{COP_EXP} and co-workers. 

\begin{figure}[htbp]
    \begin{minipage}[t]{110mm}%
           \mbox{
	              \setlength{\epsfxsize}{9cm}
	              \epsffile{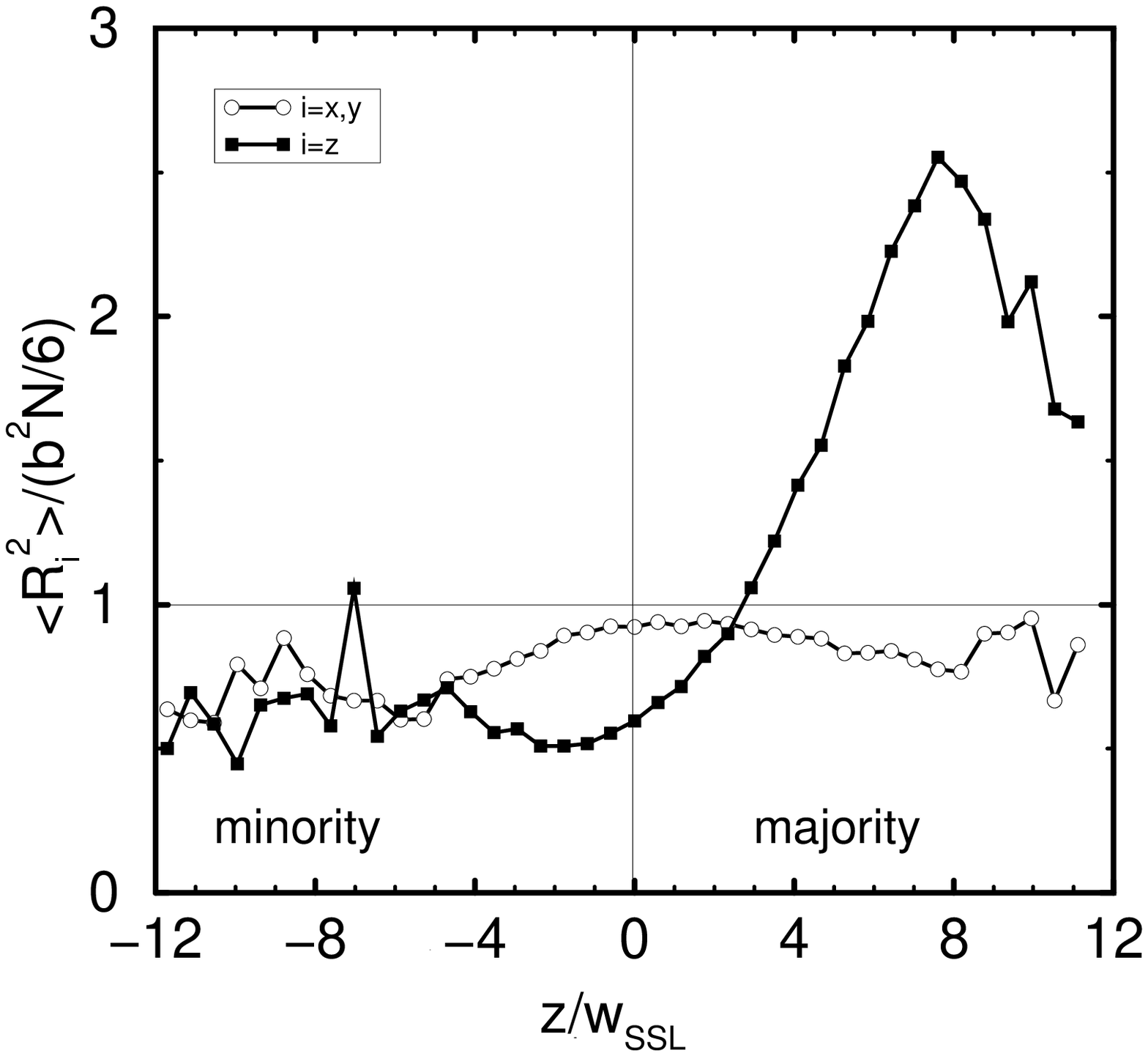}
	        }							         
     \end{minipage}%
     \hfill%
     \begin{minipage}[b]{110mm}%
\caption{Mean square end-to-end vector components of the copolymer {\em blocks} 
	in their minority phase ($A$ block in $B$-rich phase or vice versa) and 
	majority phase, in units of the average bulk value $b^2N/6$, plotted as 
	a function of the distance of the midpoint of the end-to-end vector from 
	the interface.
	}
	From Ref. \cite{COP2}.
	   \label{fig:11a}
     \end{minipage}%
\end{figure}

It is instructive to investigate the orientations of the copolymers at the 
interface. The normalised extensions of the end-to-end vector components parallel
and perpendicular to the interface of the individual blocks are shown in 
Fig.\ref{fig:11a}. Right at the interface ($z=0$) the individual blocks
align parallel to the interface, similar to the behaviour of the homopolymers. 
In the minority phase, the block shrinks, in order to reduce the number of
unfavourable interchain contacts. Deep inside the majority phase, the block 
aligns perpendicular, because the other block pulls it towards the interface. 
On average, the orientation of individual blocks is however parallel to
the interface.

\begin{figure}[htbp]
    \begin{minipage}[t]{110mm}%
           \mbox{
	              \setlength{\epsfxsize}{9cm}
	              \epsffile{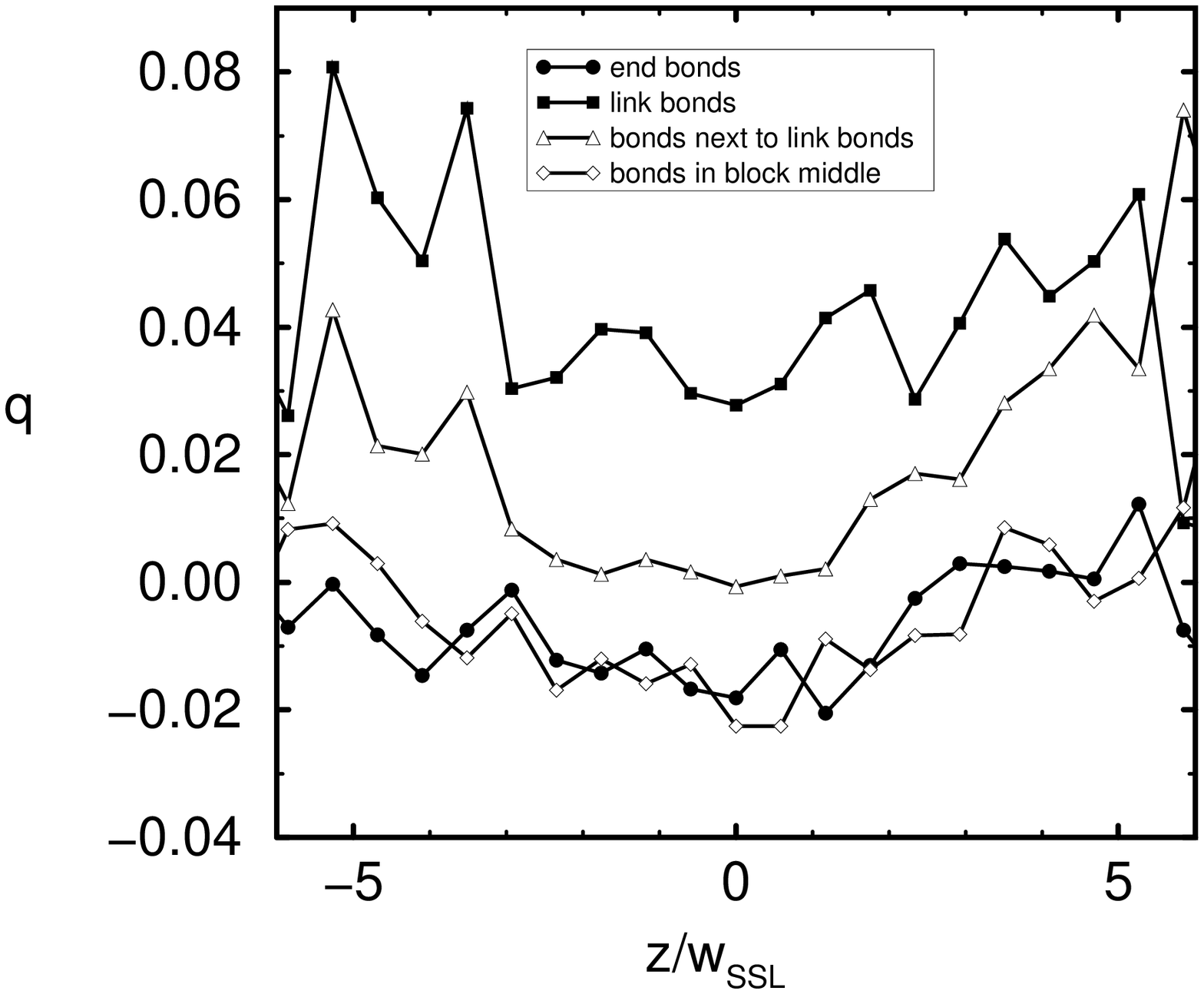}
                }\\
     \end{minipage}%
     \hfill%
     \begin{minipage}[b]{110mm}%
\caption{Orientation of bond vectors in a symmetric copolymer of length $N=32$ 
	at a highly segregated homopolymer interface ($\epsilon=0.1$). The 
	orientational order parameter $q$ for end bonds, link bonds, which join 
	the two blocks, and neighbours of the link bonds are presented.
        }
	From Ref. \cite{COP2}.
	   \label{fig:11b}
     \end{minipage}%
\end{figure}

The orientation of the copolymer bonds is shown in Fig.\ref{fig:11b}. The link 
bonds between the different blocks are strongly oriented perpendicular 
to the interface. Their orientations increase as the distance from the center of 
the interface grows. The orientation of the bond vectors, however,
rapidly decreases as we approach the ends of the blocks. The bonds in the middle 
of the blocks behave similar to the bond vectors in homopolymers. 

In sum, the shape of a copolymer resembles a dumbbell, the individual blocks are 
only mildly perturbed and resemble the homopolymer conformations at the interface. 
The vector, which connects the centers of the distinct blocks, however, has
a strong orientation perpendicular to the interface.

\begin{figure}[htbp]
    \begin{minipage}[t]{110mm}%
           \mbox{
	              \setlength{\epsfxsize}{9cm}
	              \epsffile{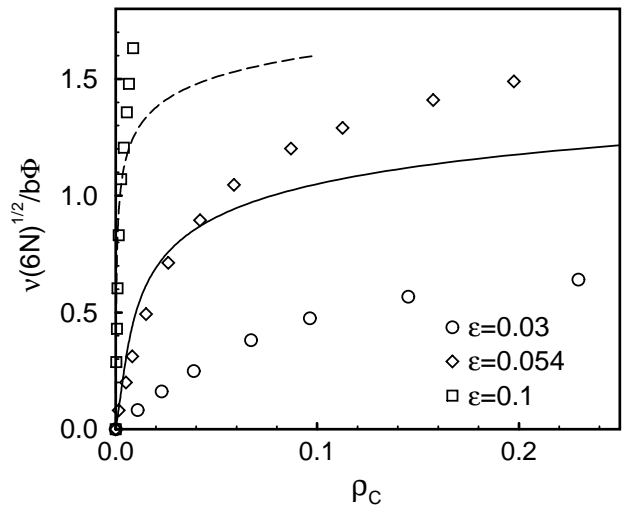}
                }\\
     \end{minipage}%
     \hfill%
     \begin{minipage}[b]{110mm}%
\caption{Segregation of symmetric copolymers of length $N=32$ at a homopolymer 
	interface. The dependence of the excess areal density $\nu$ of the 
	copolymers on the bulk concentration is presented for various 
	temperatures. The simulation data are compared without any adjustable 
	parameter to Semenov's theory, which interpolates between the mushroom 
	regime at low $\nu$ and the brush-like regime for large excess densities.
	}
	From Ref. \cite{COP}.
	   \label{fig:12}
     \end{minipage}%
\end{figure}

The interplay between the interfacial thermodynamics and the phase behaviour of 
the ternary homopolymer-copolymer mixture is best investigated in the 
semi-grandcanonical ensemble, which allows the simultaneous measurement of the 
interfacial free energy and excess quantities.
Following Leibler\cite{LUDWIG}, we estimate the chemical potential of the 
copolymers in the bulk by:
\begin{equation}
N\delta \mu = \ln \rho_c + \frac{1}{2} \chi N,
\end{equation}
where the first term represents the translational entropy of the copolymer, and 
the second one the enthalpic repulsion between the homopolymer and the copolymer.
The potential of a copolymer adsorbed at the interface, with the $A$ and $B$ block
extended into the appropriate homopolymer regions, is determined by the 
two-dimensional translational entropy of the joints at the interface,
\begin{equation}
N \delta \mu \approx \ln \frac{\nu N}{\varrho w_C},
\end{equation}
where $w_C$ is the width over which the copolymer joints are confined to the 
interface, and $\nu$ is the excess copolymer density at the interface.
As shown in the segmental profiles (cf.\ Fig.\ref{fig:copdens}), the width is of the same order than the 
interfacial width. Semenov\cite{SEM_COP} has given the estimate 
$w_C = \pi w_{\rm SSL}/2$ for strong segregation. In equilibrium, the chemical 
potential of the copolymers at the interface and in the bulk are equal, hence
\begin{equation}
\nu \approx \frac{\varrho w_C}{N} \exp \left( \frac{1}{2} \chi N \right).
\end{equation}
The above considerations neglect the stretching penalty upon crowding copolymers 
at the interface. This becomes important when the copolymer concentration at the 
center of the interface is large, {\em i.e.},
\ $\sqrt{6N}\nu/\varrho b \sim {\cal O}(1)$.
For higher interfacial excess, the number of copolymers does not increase linearly
with the bulk concentration. The dependence of the interfacial enrichment on the 
bulk concentration for different temperatures is presented in Fig.\ref{fig:12} 
and compared with an analytical expression due to Semenov\cite{SEM_COP}. 
The interfacial excess increases upon adding copolymers in the bulk. At low bulk 
concentrations, we find rather good agreement with the theoretical description,
without any adjustable parameter. The saturation at higher copolymer 
concentrations is described only qualitatively. In this regime of intermediate 
segregation, the copolymers do not form a well defined stretched brush\cite{COP}, 
but rather an unstructured thick layer. Hence the theory overestimates the amount 
of stretching and the associated entropy loss. This goes along with an 
underestimation of the copolymer excess.

\begin{figure}[htbp]
    \begin{minipage}[t]{110mm}%
           \mbox{
	              \setlength{\epsfxsize}{9cm}
	              \epsffile{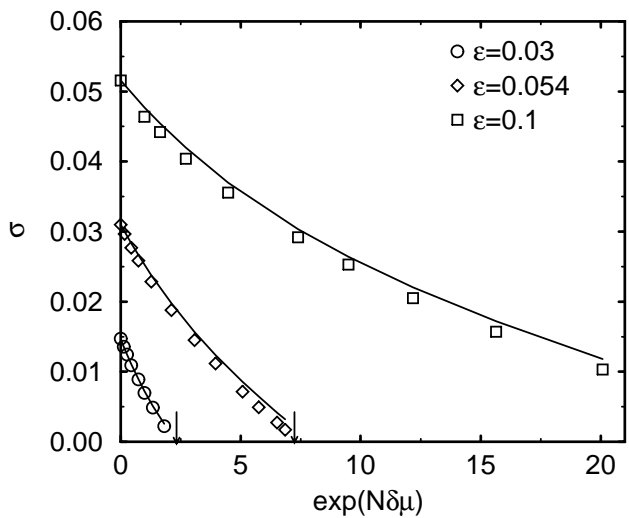}
                }\\
     \end{minipage}%
     \hfill%
     \begin{minipage}[b]{110mm}%
\caption{Reduction of the interfacial tension upon increasing the 
	copolymer's chemical potential $\delta \mu$. Symbols represent the value 
	of the interfacial tension measured via the reweighting scheme, and
	lines give the result of the Gibbs isotherm, using the data from the 
	previous figure. The stability limits of the coexisting homopolymer-rich 
	phases are indicated by arrows. Note that the addition of copolymers can 
	reduce the interfacial tension by roughly two orders of magnitude, which
	is a typical value for amphiphilic systems forming microemulsions.
	           }
	From Ref. \cite{COP}.
	   \label{fig:12b}
     \end{minipage}%
\end{figure}

The adsorption of copolymers also leads to a decrease of the interfacial 
tension $\sigma$. This quantity is directly accessible via the probability
distribution of the composition or the Gibbs adsorption isotherm
\begin{equation}
\frac{d\sigma}{k_BT} \approx -\nu_C N \delta \mu,
\end{equation}
where we have neglected the compressibility of the blend. The results of the 
direct measurement and the Gibbs adsorption isotherm are presented in 
Fig.\ref{fig:12b}. Both estimates agree nicely, indicating that the 
semi-grandcanonical simulation scheme establishes equilibrium between the bulk 
and the interface\cite{COP}. From the slope at vanishing copolymer concentration,
we can determine the effective width of the copolymer joint profile $w_C$, which 
agrees rather well with Semenov's estimate $w_C = \pi w/2$. In the weak segregation 
limit, the adsorption of copolymers leads to a compatibilisation of the blend. 
This second-order transition is indicated by the arrows in the figure. At higher 
segregation, we encounter a first order transition at which two homopolymer rich 
phases coexist with a spatially structured copolymer-rich phase. Both regimes are 
separated by a tricritical point\cite{COP}. The simulations indicate that the
spatially structured phase is a swollen lamellar phase at high incompatibilities,
and a microemulsion at intermediate incompatibilities. The latter has been
established by inspection of the composition correlation functions, which
are found to oscillate in space.

\begin{figure}[htbp]
    \begin{minipage}[t]{110mm}%
           \mbox{
	              \setlength{\epsfxsize}{9cm}
	              \epsffile{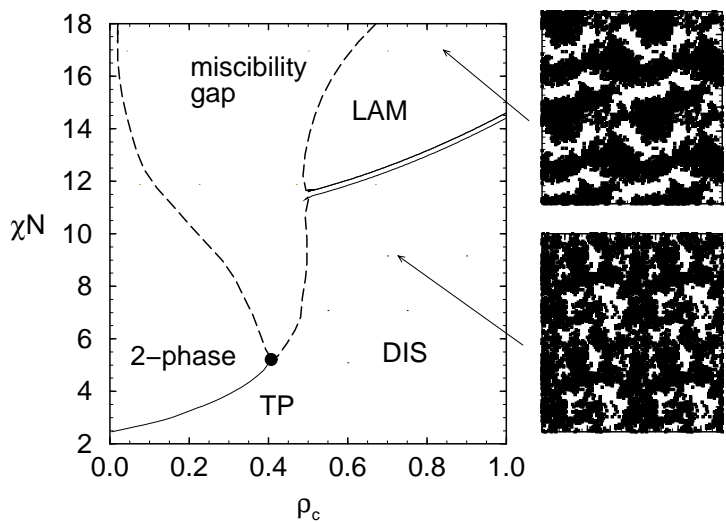}
	        }							         
     \end{minipage}%
     \hfill%
     \begin{minipage}[b]{110mm}%
\caption{Tentative phase diagram of a symmetric homopolymer-copolymer mixture of 
	chain length $N=32$ on the isopleth ($\Delta \mu=0$). The diagram includes the coexistence 
	between two unstructured homopolymer rich phases (2-phase) and a 
	copolymer-rich disordered phase (DIS) at low incompatibilities. The 
	disordered phase (DIS) is spatially structured (microemulsion) in the 
	vicinity of the triple point (TP). At high incompatibility, we find a 
	strong first order transition (with a large miscibility gap) between two 
	homopolymer-rich phases and a lamellar copolymer-rich phase (LAM). The 
	transition between the disordered phase (DIS,microemulsion) and the 
	lamellar phase could not be located in the simulations, and the double 
	solid lines are meant schematically. Typical snapshots of the $A$ monomer 
	density in a thin slice through the systems are also presented. For 
	clarity the simulation box ($L=80$) and three periodic images are shown. 
	The arrows indicate the locations in the phase diagram.
	}
	   \label{fig:5}
     \end{minipage}%
\end{figure}

A detailed study of the bulk thermodynamics in the symmetric ternary homopolymer 
copolymer blend yields a tentative phase diagram over the whole range of 
copolymer concentrations and incompatibilities (see Fig.\ref{fig:5}) It includes 
a two phase region at low copolymer concentration, where two homopolymer-rich 
phases coexist, and a disordered phase at low incompatibilities and high 
copolymer concentration. The disordered phase near the tricritical point (TP) is 
a microemulsion. A swollen lamellar phase is found at higher incompatibilities 
and copolymer content. The first order phase transition between the disordered 
phase and the lamellar phase is indicated by the double solid line.
Unfortunately the transition between the microemulsion (DIS) and the (highly 
defective) lamellar morphology (LAM) could not be located accurately due to the 
finite simulation box. Hence, the double solid line is just meant schematic. 
However, the typical snapshots of the morphology also shown in the figure 
illustrate the qualitative difference between the lamellar phase and the region 
where a polymeric microemulsion is stable.

In the weak and strong segregation limit, the simulations agree with 
self-consistent field calculations. At intermediate segregation, however, the
numerical self-consistent field calculations by Janert and Schick\cite{PHILIP} 
predict the coexistence of two asymmetric highly swollen lamellar phases in the 
region, where the simulations find a polymeric microemulsion\cite{COP}. 
Furthermore, the self-consistent field calculations predict a multi-critical 
Lifshitz point instead of a tricritical point. 

\begin{figure}[htbp]
    \begin{minipage}[t]{110mm}%
           \mbox{
	              \setlength{\epsfxsize}{9cm}
	              \epsffile{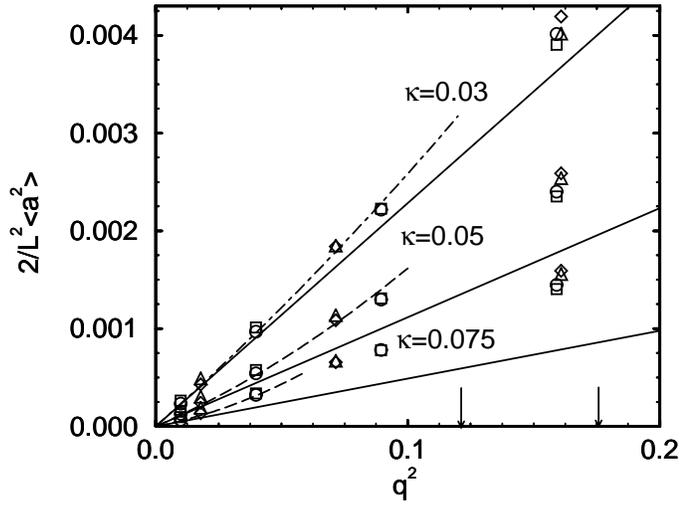}
	        }							         
     \end{minipage}%
     \hfill%
     \begin{minipage}[b]{110mm}%
\caption{Spectrum of interfacial fluctuations for a symmetric 
	homopolymer-copolymer mixture at $\epsilon=0.054$, {\em i.e.}, 
	intermediate segregation for three different chemical potentials of the 
	copolymer $\delta \mu = 0.25, 1.375, 1.75$ from top to bottom. 
	The solid lines shows results of the capillary-wave approximation 
	$\kappa=0$ using the independently determined interfacial tension 
	$\sigma$. The arrows mark Semenov's estimate for the cut-off 
	$q_{\rm max}=2/w$ for $\delta \mu=1.375$ and $1.75$. The dashed lines 
	correspond to fits according to the Helfrich Hamiltonian, using the same 
	interfacial tension as before. The fit values of the bending rigidity 
	are indicated.
	}
	From Ref. \cite{COP}.
	   \label{fig:6c}
     \end{minipage}%
\end{figure}

A discussion of the bending rigidity $\kappa$ of the interface provides a 
tentative explanation for the stability of the microemulsion observed in the 
simulations: At intermediate segregation, the interfacial tension $\sigma$ 
becomes very small in the vicinity of the triple line. Thus the bending 
rigidity controls the interfacial fluctuations, and the persistence length 
$\xi_p$ of the interface is of the order 
$\xi_p \sim b \exp(2\pi \kappa/k_BT)$ \cite{TAUPIN}. 
The bending rigidity of the interface is accessible in the simulations {\em via} 
the spectrum of interfacial fluctuations\cite{COP} (see Sec. 3). It is presented 
in Fig.\ref{fig:6c}. The fluctuation spectrum for the pure homopolymer interface 
is well describable by the capillary-wave Hamiltonian (\ref{eqn:capwave}) using
the independently measured interfacial tension, and this description accounts also 
rather well for interfaces with a small amount of copolymer adsorbed. Upon
increasing the copolymer excess, the interfacial tension decreases, and the 
deviations from the capillary-wave approximation (\ref{eqn:spectrum}) increase 
in turn. This indicates that the bending rigidity becomes important, and
a satisfactory description of the simulation data is only possible with the
full Helfrich Hamiltonian (\ref{eqn:helf1}). In order to determine the bending 
rigidity $\kappa$, we fit the simulation data to Eqn.\ (\ref{eqn:helf2}), fixing 
the interfacial tension to the value determined via the reweighting scheme.  
Unfortunately, the bending rigidity cannot be extracted with a high accuracy, 
yet for our purpose here we only need an order of magnitude estimate. The 
values of the bending rigidity are small ( $2\pi\kappa/k_BT < 0.5$) and increase 
with growing copolymer concentration. The order of magnitude and the way they
depend on the copolymer concentration agree qualitatively with analytical 
calculations on similar systems\cite{BENDINGPOLY}. 

The small value of the bending rigidity of a copolymer-laden interface suggests 
that interfacial fluctuations destroy the lamellar order predicted by the mean 
field theory, and lead to the formation of a microemulsion at intermediate 
segregation. This interpretation is in accord with Monte Carlo simulations of 
Gompper and Kraus\cite{GOMPPER}, who found that the transition between a
microemulsion and a lamellar phase takes place at about 
$2\pi\kappa/k_BT \approx 16$. Note that the chain length $N=32$ corresponds to a 
degree of polymerisation of about 150 on an atomistic level. Our results are 
thus typical for rather short polymers, and the simulation data exhibit
a behaviour between amphiphiles and high molecular weight polymers.

On increasing the chain length $N$, the interfacial tension even decreases like 
$1/\sqrt{N}$ at constant $\chi N$, whereas the bending rigidity $\kappa$
is expected to grow\cite{KAPPA}. Thus $\kappa$ dominates the fluctuation spectrum in the long 
chain length limit. De Gennes and Taupin\cite{TAUPIN} argued that the lamellar 
phase becomes more favourable for long chain lengths. Therefore, the phase diagram 
should become more mean field like for longer chains. However, in the ultimate 
vicinity of the tricritical point, both the interfacial tension and the bending 
rigidity vanish and hence interfacial fluctuations are expected to become 
important for {\em all} chain lengths. Indeed, general arguments indicate that 
isotropic Lifshitz points are unstable in three-dimensional systems \cite{LIF}. 
For very long chain lengths, M\"uller and Schick\cite{COP} have speculated that 
there might be a crossover from Lifshitz tricritical behaviour ($\xi_p$ large, 
swollen lamellae) to ordinary tricritical behaviour ($\xi_p$ small, microemulsion)
as one approaches the tricritical point.
Recent experiments by Bates et al.\ \cite{BATES} found experimental evidence for 
a bicontinuous polymeric microemulsion in an asymmetric homopolymer copolymer 
mixture in a region where the self-consistent field theory predicts an isotropic 
Lifshitz point. In fact, the experimental phase diagram of the copolymer 
homopolymer mixture closely resembles the phase diagram of small molecule mixtures 
of oil, water and amphiphiles\cite{MICROB}.

\subsection{Thin films}

The behaviour of confined complex fluid is of practical importance for various 
applications, {\em e.g.}, adhesives, coatings, and lubrication.
Confining surfaces may alter the phase behaviour profoundly, and the interplay
between interfaces separating coexisting bulk-like phases of a binary mixture
and the confining walls has attracted longstanding theoretical 
interest\cite{WET1,WET2,WET3}. 

In the following, we discuss the behaviour of a binary polymer blend, which is 
confined between two hard, impenetrable walls. Neutral walls, which do not 
preferentially interact with any of the components, reduce the critical 
temperature of a symmetric polymer blend and likewise the miscibility 
gap\cite{NAKANISHI}. This has been investigated through extensive Monte Carlo 
simulations by Kumar\cite{KUMAR} and collaborators, and by Rouault\cite{YR} et 
al.  The coexistence curve at the critical point is flattened compared to the 
bulk binodal, which indicates that the critical point belongs to the 2D Ising 
universality class. Upon approaching the critical point, the system crosses 
from mean field to Ising critical behaviour and passes from three-dimensional 
to two-dimensional critical behaviour. 

In general, one component may adsorb preferentially at the surface, such that the 
wall is coated with a layer of the component with the lower surface free energy.
The structural and thermodynamical properties of these wetting layers are of 
practical importance and of fundamental interest in the statistical mechanics of 
condensed matter.
At phase coexistence of the binary mixture, the surface free energy in the 
semi-infinite system undergoes a transition, at which the thickness of the 
adsorbed layer diverges. This wetting transition \cite{WET1,WET2,WET3} may be 
continuous (second-order wetting) or the thickness may jump from a finite value 
to infinity at the wetting transition temperature. While the unmixing temperature 
grows linearly with the chain length, the wetting transition in symmetric blends 
is independent of the chain length. Unlike the generic situation in mixtures
of small molecules, wetting in polymeric systems thus occurs far below the 
critical point\cite{SB}. The wetting behaviour in a binary polymer blend has 
been studied by Wang\cite{WANG} and co-workers {\em via} Monte Carlo simulations 
in the semi-grandcanonical ensemble.

\begin{figure}[htbp]
    \begin{minipage}[t]{110mm}%
           \mbox{
	              \setlength{\epsfxsize}{5cm}
	              \epsffile{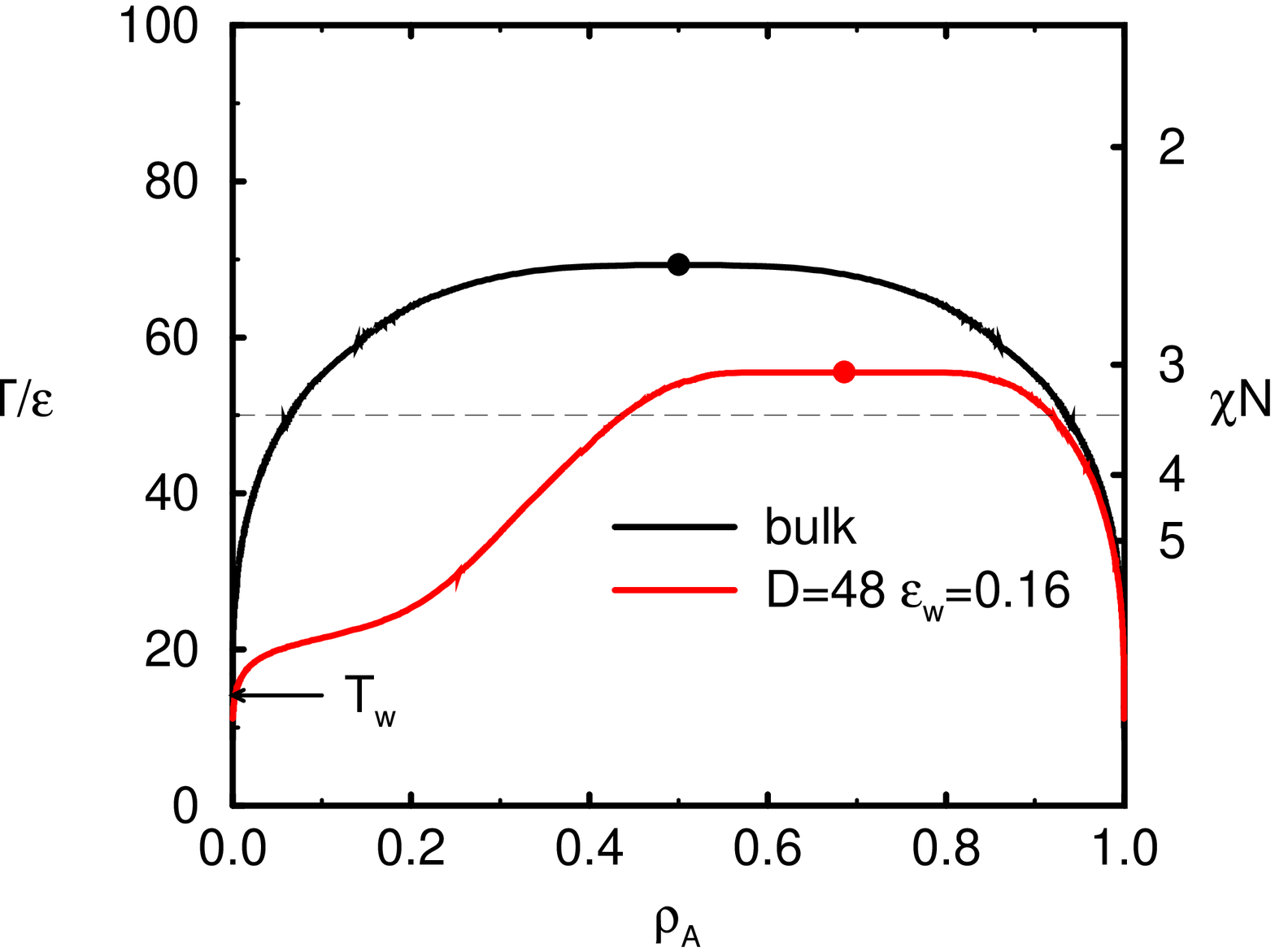} \hspace*{1cm}
	              \setlength{\epsfxsize}{5cm}
	              \epsffile{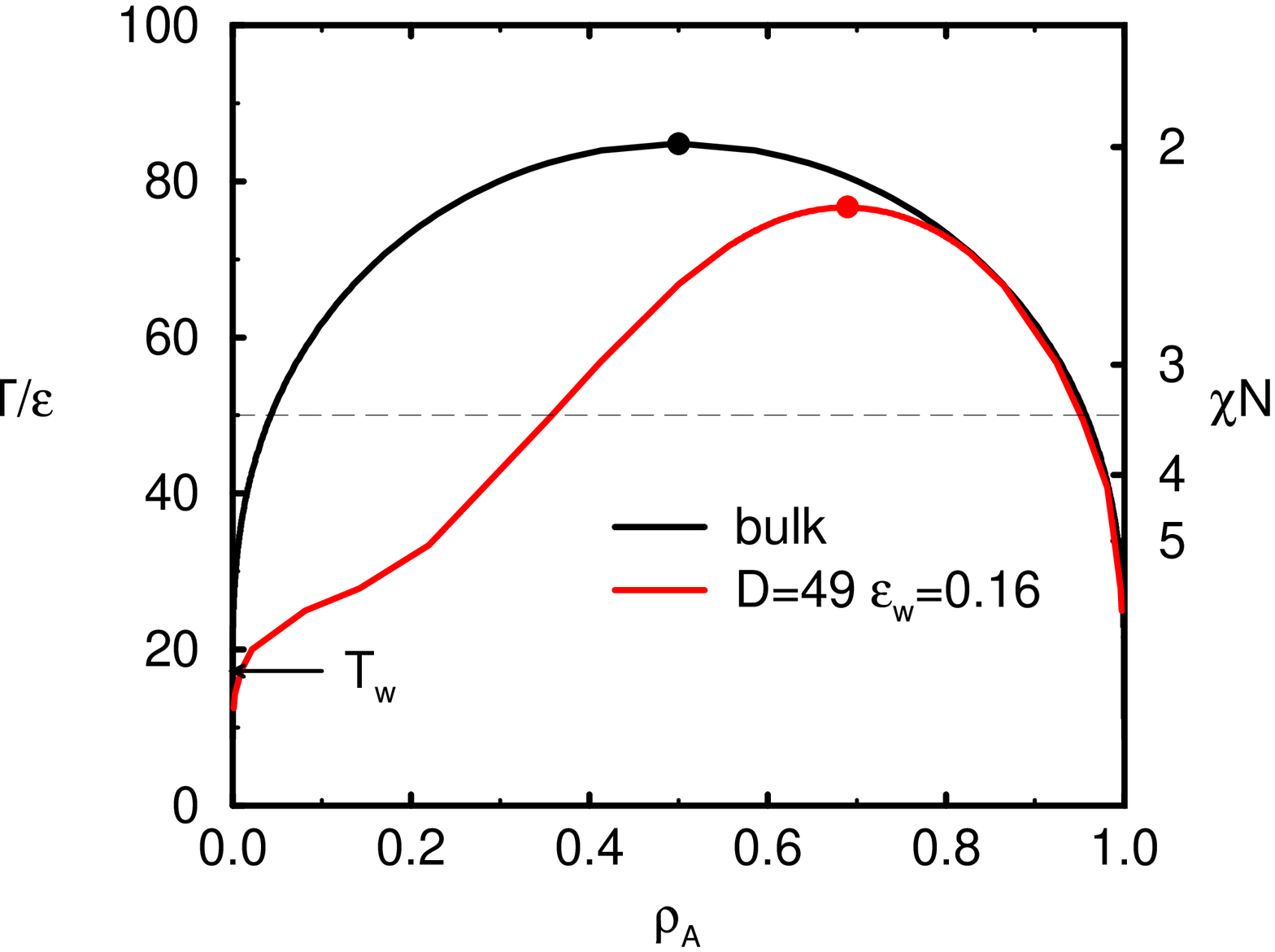}
	        }							         
     \end{minipage}%
     \hfill%
     \begin{minipage}[b]{110mm}%
\caption{Comparison between the Monte Carlo results (left figure) and the 
	self-consistent field prediction (right diagram) for the phase behaviour 
	of a symmetric, binary polymer blend of chain length $N=32$ confined 
	into a thin film of width $D=48$ (MC) and $D=49$ (SCF). Both surfaces 
	attract the $A$ component with equal strength $\epsilon_w=0.16$.
	The wetting temperature of the semi-infinite system is indicated by an 
	arrow. The figures also include the bulk phase diagrams for comparison.
	}
	From Ref. \cite{MBWET}.
	   \label{fig:16}
     \end{minipage}%
\end{figure}

In a thin film with non-neutral walls, two extreme cases are of special interest: 
In the asymmetric case, one surface attracts the $A$ component, whereas the other 
wall prefers the $B$ component. In the symmetric situation, both walls attract 
the same component with equal strength. We discuss their phase behaviour in turn:

In the symmetric situation, the short range potential, which attracts 
preferentially the $A$ component, shifts the coexistence chemical potential 
$\Delta \mu$ away from its bulk value\cite{NAKANISHI}. At coexistence, the 
confined system phase separates {\em laterally} into $A$ rich domains in
coexistence with regions, in which $A$-rich layers cover the surfaces, but
the $B$-component prevails in the center of the film. 
In the temperature range between the critical temperature of the film and the 
wetting temperature, the thickness of the wetting layer at coexistence is 
determined by the interplay of the repulsion between the wall and the 
$AB$-interface, which favours a thick wetting layer, and the shift of the 
coexistence chemical potential, which suppresses the total amount of the $A$ 
component in the film\cite{PARRYSYM}. The semi-grandcanonical ensemble
in junction with the reweighting methods permits an accurate location of the 
coexistence chemical potential, and a good characterisation of the phase 
behaviour over the whole temperature range. The phase diagram in the bulk, and 
in a thin film is presented and compared to self-consistent field calculations
in Fig. \ref{fig:16}\cite{MBWET}. The simulations and the self-consistent 
field calculations agree qualitatively: The critical temperature is reduced, and 
at the critical point, the component favoured by the walls is enriched. We find 
a strong first order wetting transition far below the critical temperature 
$T_w \approx 0.2 T_c$. Most notably, the $A$-poor binodal is convex in an 
intermediate temperature regime. This curvature is the signature of the wetting 
transition in the semi-infinite system.

In the asymmetric case, the effect of the wetting transition is even more 
fundamental\cite{PARRYASYM,PARRYFLUC}: There always exists at least one
interface which runs parallel to the walls and separates an $A$-rich phase
from a $B$-rich one at the appropriate walls. In the temperature regime between 
the bulk critical temperature and the interface localisation-delocalisation 
temperature $T_c(D)$ of a film of width $D$, it is localised in the middle of 
the film. The total composition of this ``soft-mode'' phase fluctuates around 1/2.
Hence, there is no symmetry breaking, and the system is in a one phase region. 
Below $T_c(D)$, two phases coexist in which the interface is localised close
to one of the walls. The transition temperature $T_c(D)$ is smaller than the 
wetting temperature $T_w$ of the semi-infinite system, and approaches $T_w$ 
exponentially fast as the film thickness $D$ grows. 

In the soft-mode phase, the deviations of the local interfacial position $u$ from 
the middle of the film can be described by a capillary-wave Hamiltonian 
(cf.\ Eqn.\ \ref{eqn:capwave}) augmented by the effective potential exerted by 
the walls onto the interface. We take the effective interaction between the 
interface and both walls, which binds the interface to the middle of the film, 
to be of the form $V(u) = a (T-T_c(D)) \exp(-\lambda D/2) u^2$. Here $1/\lambda$ 
denotes the effective range of the exponentially decaying potential between the 
interface and the wall, hence, $ \exp(-\lambda D/2)$ sets the strength of the 
interaction in the middle of the film. Therefore, the effective Hamiltonian 
reads
\begin{equation}
{{\cal H}} = \int dx\;dy\; \left\{ \frac{\sigma}{2} 
(\nabla u)^2 + a (T-T_c(D)) \exp(- \lambda D/2) u^2\right\}.
\label{eqn:asym}
\end{equation}
After Fourier decomposing the local interfacial position, we can calculate the 
spectrum of interfacial fluctuations
\begin{equation}
\frac{2}{L^2\langle a(q)^2\rangle}  
= \frac{\sigma}{k_BT} \left\{ q^2 + \left(\frac{1}{\xi_\|}\right)^2 \right\}.
\end{equation}
As in the case of the free interfaces (Eqn. (\ref{eqn:capwave2})), we integrate 
the spectrum of capillary waves from $q_{\rm min}=2\pi/L$ to $q_{\rm max}$,
and obtain
\begin{equation}
\label{ufs}
\langle u^2(x,y) \rangle = 
\frac{k_BT}{2\pi \sigma} \ln 
\sqrt{\frac{1+(q_{\rm max}\xi_\|)^2}{1+(q_{\rm min}\xi_\|)^2}}.
\end{equation}
Note that the parallel correlation length $\xi_\|$ plays a similar role than 
$q_{\rm min}$ in the previous analysis\cite{WET1}. It is given by
\begin{equation}
\xi_\| = \frac{1}{\sqrt{\sigma \frac{\partial^2V}{\partial u^2}}}  
= \frac{\exp(+\lambda D/4)}{\sqrt{2\sigma a (T-T_c(D))}}.
\end{equation}
Most notably, $\xi_\|$ grows exponentially with the film thickness $D$. This is 
the characteristic signature of the soft mode phase\cite{PARRYASYM}.
In a finite system\cite{andreas}, $\xi_\|$ can exceed the lateral system 
dimensions. Depending on whether $q_{\rm min}\xi_\|=2\pi \xi_\|/L$ is much larger 
or much smaller than one, the spectrum is then cut off by the finite lateral 
system size $L$, or by the parallel correlation length $\xi_\|$. 

The first limit, $q_{\rm min}\xi_\| \to \infty$, corresponds to interfacial 
fluctuations of a free interface. In the canonical ensemble, the constraint on 
the composition fixes the laterally averaged position $\bar{u}$ of the 
interface and Eqn. (\ref{ufs}) can be used to determine the interfacial width.
In the semi-grandcanonical ensemble, the average position of the interface is 
unconstrained, and one has an additional $q=0$-mode, which corresponds to a 
translation of the interface as a whole, and which dominates the broadening. 
With the effective Hamiltonian (\ref{eqn:asym}), the fluctuations of the average 
interfacial position are then given by 
$\langle \bar{u}^2 \rangle = k_BT \xi_\|^2/(L^2\sigma)$. 
In the limit $L \ll \xi_\|$, the fluctuations are of the order $D^2$ 
itself, and thus $w = D/2$ in the semi-grandcanonical ensemble.

In the second limit, $q_{\rm min} \xi_\| \to 0$, the dependence of the parallel 
correlation length on the film thickness leads to an anomalous size 
dependence\cite{BSIM,andreas} of the apparent interfacial width. Using the 
convolution approximation (\ref{eqn:apparent}) and the above equation for the 
lateral correlation length, we get
\begin{equation}
w^2 \approx w^2_0 + \frac{k_BT}{4\sigma} \ln q_{\rm max}\xi_\| 
= w^2_0 + \frac{k_BT\lambda D}{16\sigma} + {\rm const}.
\label{eqn:andreas}
\end{equation}
Hence, the width of the apparent profiles increases like $\sqrt{D}$ for large 
lateral extensions. The dramatic dependence on the film thickness demonstrates 
the importance of careful assessment of interfacial fluctuations, when analysing 
apparent interfacial profiles. This finding is also relevant to interfaces in
mixtures of of small molecules, and in particular to the interpretation of 
experimental data\cite{KERLE}. In the symmetric case (capillary condensation) 
or in the presence of long ranged (van der Waals) interactions, the dependence 
of the interfacial width on the film width $D$ is, however, only logarithmic.

\begin{figure}[htbp]
    \begin{minipage}[t]{110mm}%
           \mbox{
	              \setlength{\epsfxsize}{9cm}
	              \epsffile{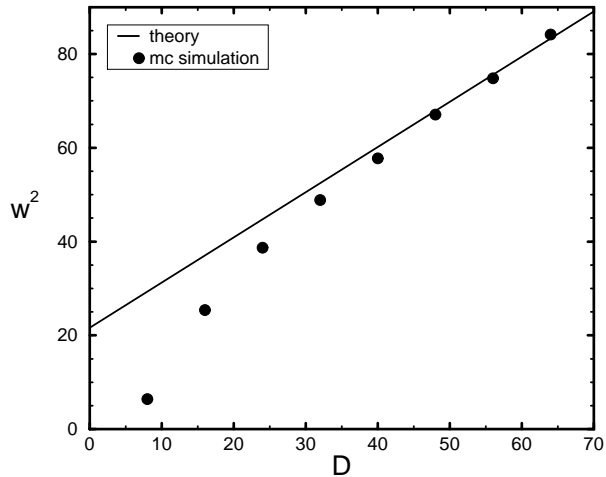}
	        }							         
     \end{minipage}%
     \hfill%
     \begin{minipage}[b]{110mm}%
\caption{Monte Carlo results for the apparent interfacial profile width of a symmetric, 
	binary polymer blend $N=32$ in a thin film of thickness $D$. The 
	confining surfaces attract different components of the blend with 
	an interaction potential $\epsilon_w=0.1$. The temperature 
	$\epsilon=0.03$ is well above the wetting temperature. Shown is the 
	squared interfacial width $w^2$ as a function of the film thickness $D$ 
	for large lateral system extension $L=256$. The straight line shows
	the expected a capillary-wave broadening, when the interfacial 
	fluctuation spectrum is cut off by the thickness dependent parallel 
	correlation length $\xi_{\|}$.
        }
	From Ref. \cite{andreas}.
	   \label{fig:14}
     \end{minipage}%
\end{figure}

\begin{figure}[htbp]
    \begin{minipage}[t]{110mm}%
           \mbox{
	              \setlength{\epsfxsize}{9cm}
	              \epsffile{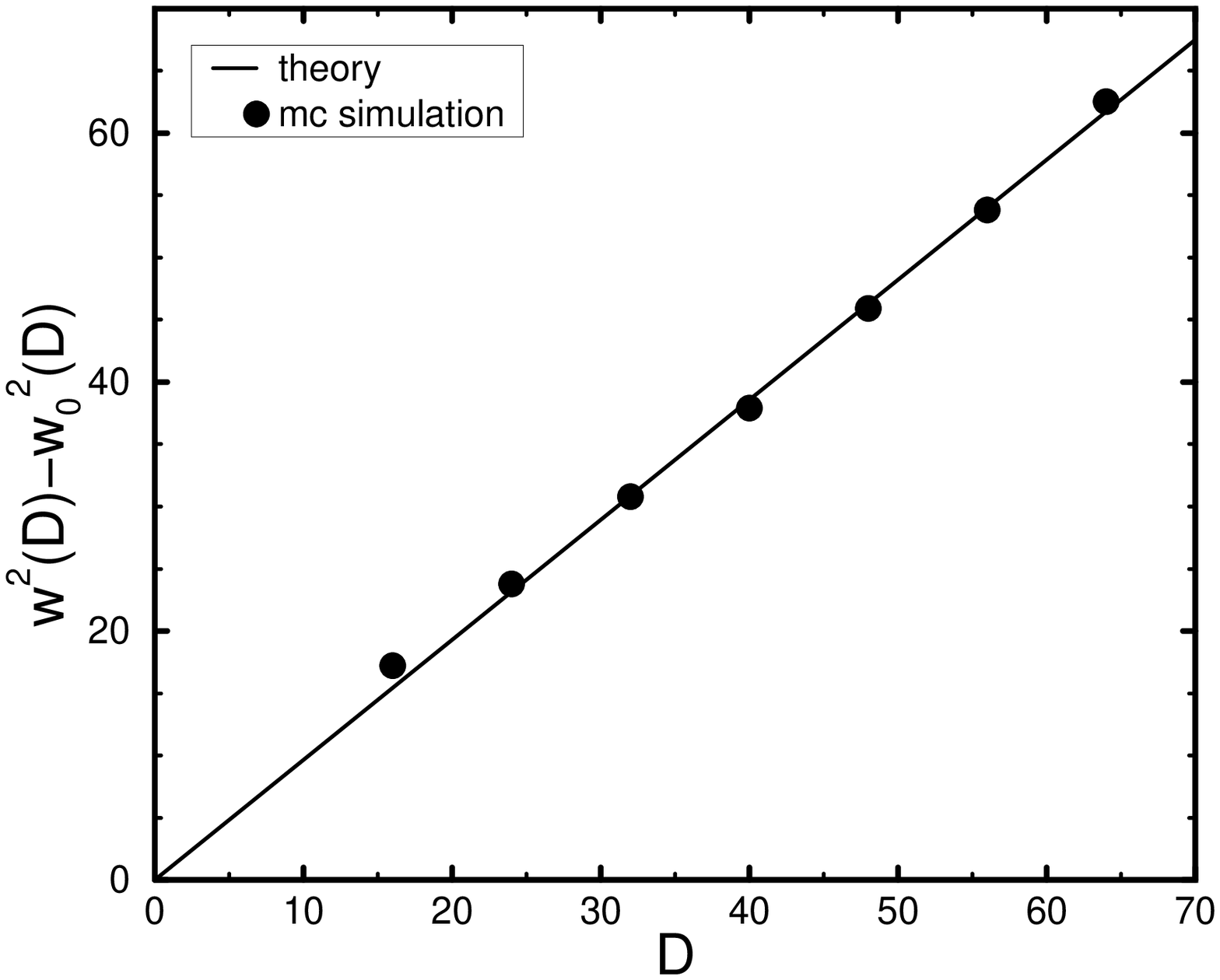}
	        }							         
     \end{minipage}%
     \hfill%
     \begin{minipage}[b]{110mm}%
\caption{Same data as in the previous figure, but the squeezing of the intrinsic 
	interfacial width due to the confining surfaces is accounted for by 
	subtracting the estimated intrinsic width $w_0^2(D)$. The straight line 
	corresponds to the expected capillary-wave broadening.
	           }
	From Ref. \cite{andreas}.
	   \label{fig:15}
     \end{minipage}%
\end{figure}

The structure and profiles across the film in the soft-mode phase have been 
studied with extensive Monte Carlo simulations by Werner et al.\cite{andreas}.
The system considered is a symmetric, binary blend of polymers with chain length 
$N_A=N_B=32$, confined to a thin film of width $D$. The left surface attracts $A$ 
monomers, which are less than 2 lattice spacings away, and repels $B$ monomers, 
with an interaction of strength $\epsilon_w=0.1$. Likewise, the right surface 
prefers $B$ monomers and repels $A$ segments. 
The temperature $1/\epsilon=33.3$ was chosen higher than the wetting temperature 
in the semi-infinite geometry, hence the system is without doubt in the 
soft-mode phase. In Fig. \ref{fig:14}, the apparent interfacial width $w$ is 
shown as a function of the film thickness $D$ for very large lateral system size 
$L=256$ in the semi-grandcanonical ensemble. The large lateral system 
extension ensures, that the fluctuation spectrum is cut off by the parallel 
correlation length $\xi_\|$. The Monte Carlo data are compared to the prediction 
(\ref{eqn:andreas}), where the range of the interfacial interaction $\lambda$ is 
taken to be $1/\lambda = \xi (1+k_BT/8\pi\xi^2\sigma)$\cite{PARRYFLUC,FLUC} with the bulk 
correlation length $\xi$, and the constant term is taken to be zero. Both, the
bulk correlation length and the interfacial tension have been measured independently.
The intrinsic width has been extracted from a self-consistent field calculation 
in an infinite system\cite{SM}. The Monte Carlo data and the predictions agree 
nicely for thick films; however, there are deviations for very thin films.
This can be understood from the following consideration:
If the film width $D$ is not very much larger than the radius of gyration 
$R_g(N=32) \approx 7$, the intrinsic profile $\rho_0$ is squeezed. Such an
effect has been observed in simulations\cite{andreas} and in self-consistent field 
calculations\cite{MBWET}. It can be analysed {\em via} a block analysis (see
Sec. 3) by inspection of the width of the reduced profiles as a function of the 
block size $B$. If $B$ is larger than the short wavelength cut-off of the 
interfacial fluctuation spectrum $1/q_{\rm max}$, the reduced profiles are 
broadened by capillary waves. One can attempt to identify an effective intrinsic 
width with the value $w_0(D)$ attained at the subsystem size $B = 8 \approx R_g$. 
As shown in Fig. \ref{fig:15}, Eqn.(\ref{eqn:andreas}) in combination with this 
effective width $w_0(D)$ quantitatively accounts for the deviations at small 
film thicknesses.

The relationship between the width of the apparent interfacial profiles and the 
film thickness follows from general considerations on the statistical mechanics of 
interfaces, and is not restricted to polymeric systems. Similar effects
have been observed previously in simulations of simple Ising models\cite{BSIM}. 
However, polymer blends are particularly suited for investigating 
phenomenological concepts of surface enrichment, wetting, interfacial 
localisation-delocalisation transitions and capillary condensation, because 
the chain length $N$ constitutes an additional control parameter\cite{SB}. 
On increasing the chain length $N$, one can reduce bulk composition 
fluctuations, which are ignored in most phenomenological approaches. 
There exist powerful analytical tools to describe the bulk and interfacial 
behaviour in the long chain length limit. In addition, the phenomena occur on 
larger length scales due to the large size of the polymer coils, which 
facilitates applications of several experimental techniques. In fact,
the anomalous size dependence of the apparent interfacial width has first been
observed experimentally in a polymer system by Kerle et al.\cite{KERLE}.

\section{Conclusions and outlook}

We have reviewed extensive computer simulations on the bulk thermodynamics and 
interfacial structure of polymer blends in the melt state. 
Simulations of polymer blends and interfaces are considerably more exacting in 
computational terms than those of small molecular fluids or magnetic systems. 
The difficulties stem from the difficulty of dealing with the widely spread time 
and length scales caused by the extended structure of the macromolecules. The 
accurate determination of the macroscopic behaviour while retaining the detailed 
atomistic chemical structure is not feasible even with state-of-the-art 
supercomputers. Yet, by a careful choice of the statistical ensemble, use of
recently developed simulation techniques and sophisticated data analysis methods, 
simulations of coarse-grained polymer models yield valuable insight into the 
structural and thermodynamic properties of polymeric composites. 

Simulations contribute to the identification of important effective parameters, 
which determine qualitatively the bulk and interfacial properties. They can 
examine the influence of the coarse-grained chain architecture on the phase 
and interfacial behaviour. As an example, we have discussed how stiffness 
disparity\cite{STIFF1} between the components gives rise to a positive 
contribution to the Flory-Huggins parameter $\chi$. This finding is in agreement 
with field theoretical considerations\cite{liu} and P-RISM 
calculations\cite{singh}. Furthermore, we have shown that the persistence length 
affects the interfacial behaviour in  highly incompatible blends. For weak 
incompatibilities, the effect of the stiffness can be modelled within
the Gaussian chain model. This model incorporates two parameters, the 
incompatibility $\chi N$, and the length scale $R$ set by the polymer
radius of gyration. In the strong segregation limit, the width of the interface 
scales as $R/\sqrt{\chi N}$. However, for large incompatibilities the interfacial
width becomes of the order of the persistence length. On this length scale, the 
conformations are not Gaussian, but the bond vectors are strongly correlated. 
In this regime simulations and self-consistent field calculations show that the 
interfacial width decreases upon increasing the stiffness disparity, in 
qualitative contrast\cite{STIFF2} to the predictions of the Gaussian chain model.

Many static and dynamical properties on various length scales are simultaneously 
accessible in simulations. Therefore the simulations provide a detailed
picture of the interfacial properties in binary and ternary polymer mixtures. 
Both structural informations ({\em e.g.}, the orientation of bond vectors, or 
of whole chains\cite{MBO,SM,COP2}, the adsorbtion of copolymers\cite{COP}, the
enrichment of vacancies) as well as thermodynamic properties ({\em e.g.}, 
the interfacial tension or the bending rigidity) are obtained simultaneously 
and hence permit a quantitative comparison with other theoretical approaches 
like scaling descriptions\cite{degennes}, self-consistent field
calculations\cite{SCF1,SCF2,SCF3,scheutjens}, density functional 
theories\cite{FREED1,tang} and P-RISM calculations\cite{SCHWEIZERR}.

Simulations of the interfacial behaviour of symmetric block copolymers\cite{COP2} 
yield a detailed picture of the structure. If the areal density of copolymers 
adsorbed at the homopolymer interface is small, the copolymers assume
dumbbell-like conformations. The individual blocks extend into the appropriate 
bulk phases and resemble only mildly perturbed homopolymers, in particular
each block is aligned parallel to the interface. This finding is in qualitative 
agreement with available experimental data by Russell et al.\cite{COP_EXP}. 

In general, simulations and mean field treatments agree qualitatively, because 
the extended structure of the polymers suppresses fluctuations, and microscopic
details can be qualitatively described with a few effective, coarse-grained 
parameters. The simulation can access the way in which the mean field behaviour 
is approached upon increasing the chain length\cite{M0}. However, one important 
source of discrepancy between simulations and mean field theories are interfacial 
fluctuations and we have highlightened two consequences:

In ternary homopolymer copolymer blends\cite{COP}, interfacial fluctuations 
destroy the order in highly swollen lamellar phases and lead to the 
fluctuation-induced formation of a microemulsion. Simulations find tentative 
evidence for a tricritical point, where the mean field treatment\cite{PHILIP} 
predicts a Lifshitz tricritical point in symmetric systems. 
Upon increasing the chain length, however, there may be a crossover from 
Lifshitz criticality to ordinary tricritical behaviour\cite{COP}.
This ``catastrophic failure of mean--field theory''\cite{BATES} is also observed 
in recent experimental studies by Bates and coworkers. These experiments 
on an asymmetric ternary blend find a bicontinuous microemulsion close to the 
region, where the mean--field treatment predicts an isotropic Lifshitz point. 
The mesoscopic structured yet macroscopic homogeneous composites possibly have 
unique and useful mechanical or electrical materials properties\cite{BATES}.

Interfacial fluctuations profoundly influence apparent profiles
measured in simulations\cite{andreas}. They lead to a pronounced broadening 
of the width of the measured profiles. In a thin film with asymmetric
surfaces attracting different components, there exists a temperature 
regime\cite{PARRYASYM}, in which one interface separating an $A$ rich domain 
and a $B$ rich one runs parallel to the surfaces in the middle of the film.
Effective interactions between the interface and both walls pin this
interface to the middle of the film, and generate a parallel correlation 
length $\xi_\|$ which grows exponentially with the film thickness $D$ in
in the case of short range interactions. This large length scale acts as a 
cut-off for the spectrum of capillary waves, and consequently limits the 
broadening. The apparent interfacial width increases like $\sqrt{D}$.
If we were to include  long range van der Waals interaction in our simulations, 
we would expect a logarithmic dependence of the apparent interfacial width on 
the film thickness. The effect is also pertinent to the interpretation of recent 
experimental studies\cite{KERLE,JONES_CAP,TOCOME}.

In view of these findings, it would be of interest in future simulation studies
to investigate the role of the short length scale cut-off of the interfacial 
fluctuations spectrum. A study of the way how the capillary-wave broadening
depends on the chain length, and on the incompatibility, might help to solve
the question, whether and how the length scales of the lateral fluctuations
contributing to the intrinsic profile or to the capillary-wave broadening
can be separated from each other. This might contribute in general
to the interpretation of intrinsic profiles calculated in analytical 
approaches. Another, more subtle point, concerns the way the {\em shape}
of the profiles is affected by the capillary-wave broadening. The 
intrinsic profiles obtained, {\em e.g.}, with a block analysis, should
be reasonably well described by a tanh profile, at least in the weak 
segregation limit. Preliminary results indicate that
the intrinsic profiles are indeed closer to a tanh shape than to an error
function shape as the critical point is approached\cite{TOCOME}.
The shape of the apparent profiles is given by the
convolution of the intrinsic profile and the gaussian curve associated
with the capillary broadening, which is closer to an error function in large 
enough systems. The latter error function shape has been established
for the case of capillary-wave broadening in Ising models only recently
by Moseley\cite{MOSELEY}. 

Moreover, the dynamical evolution of interfacial profiles, after two 
flat interfaces have been brought into contact, is relevant in some experimental 
situations. Simulations might investigate whether there is a separation of time 
scales on which an intrinsic profile develops which is then, on a different time 
scale, broadened by capillary waves.

A few other interesting questions shall be sketched briefly:
The interfacial properties of random or branched copolymers have attracted 
considerable interest because of the extensive commercial applications. Recent 
interest stems from experimental progress of the reactive polymer blending 
techniques\cite{REACEXP}. This has been subsequently studied 
theoretically\cite{REACTION1,REACTION2,REACTION,DOIREACTION} for rather 
idealised situations. Other important problems are related to the single chain 
dynamics in blends. In the miscible case, {\em e.g.}, simulations could test 
recent theoretical predictions by Schweizer\cite{SCHWEIZERDYN} and co-workers on 
the coupling between the thermodynamics and the single chain dynamics.

Furthermore, recent attempts to construct even more coarse-grained 
model\cite{MURAT}, which incorporate the instantaneous shape of a polymer only 
{\em via} a continuous distribution of its monomer density, might offer promising 
routes to investigate the large length scale properties of macromolecular 
composites. This might shed some light onto the mechanical properties of 
polymeric interfaces, which are of long standing practical and experimental 
interest.

\subsection*{ Acknowledgements }

It is a great pleasure to thank K. Binder, M. Schick, N.B. Wilding, A. Werner and 
P. Janert for enjoyable and fruitful collaborations and many stimulating 
discussions, and K. Binder and A. Werner in particular for a critical
reading of this manuscript.
Financial support by the grants DFG Bi 314/3-4, BMBF  03N8008C, NSF DMR9220733 
and the Alexander von Humboldt foundation are gratefully acknowledged, as well as 
generous grants of computing time at the computing centers CINECA 
(Bologna, Italy), EPFL (Lausanne, Suisse),  SDSC (San Diego, USA), the Johannes
Gutenberg university (Mainz, Germany), HLRZ (Juelich, Germany), and  RUS 
(Stuttgart, Germany).

\vspace{1cm}



\begin{thebibliography}{99}
\bibitem{degennes}
  P. G. de Gennes, Scaling Concepts in Polymer Physics
   (Cornell University Press, Ithaca, New York, 1979).
\bibitem{taylor1}
  G. I. Taylor, Proc. R. Soc. London, Ser. A 138, 41(1932).
\bibitem{MORPHOLOGY2} S. Wu, Polym.Eng.Sci. 27, 335(1987).
\bibitem{BATEST} 
  F. S. Bates, G. H. Fredrickson, Ann. Rev. Phys. Chem. 41, 512(1990).
\bibitem{COPINTER}   
  S. H. Anastasiadis, I. Gancarz, J. T. Koberstein, 
  Macromolecules 22, 1449(1989).
\bibitem{beck}
  N. C. Beck Tan, S.-K. Tai, R.M. Briber, Polymer 37, 3509(1997).
\bibitem{MILNER}
  S. T. Milner, H. Xi, J. Rheology 40, 663 (1996);
  S. T. Milner, MRS Bulletin 22, 38(1997).
\bibitem{BROWN}
   H. R. Brown, V. R. Deline, P. F. Green, Nature 341, 221(1989);
   Macromolecules 26, 4155(1993); H. R. Brown, Macromolecules 22, 2859(1989).
\bibitem{phdexp}
  D. J. Kinning, K. I. Winey, E. L.  Thomas, 
    Macromolecules 21, 3502(1988);
  D. J. Kinning, J. Chem. Phys. 90, 5806 (1989);
  K. I. Winey, E. L. Thomas, L. J. Fetters, J. Chem. Phys. 90, 9367(1991);
\bibitem{BATES}      
  F. S. Bates, W. W. Maurer, P. M. Lipic, P. M. A. Hillmyer, K. Almdal, 
  K. Mortensen, G. H. Fredrickson, T. P. Lodge, Phys. Rev. Lett. 79, 849(1997).
\bibitem{phdth}
  L. Leibler, P. A. Pincus, Macromolecules 17, 2922(1984);
  A. N. Semenov, Macromolecules 26, 2273(1993);
  M. W. Matsen, F. S. Bates, Macromolecules 29, 1091 (1996).
\bibitem{PHILIP}
  P. K. Janert, M. Schick, Macromolecules 30, 3916(1997);
  P. K. Janert, M. Schick, Macromolecules 30, 137(1997).
\bibitem{SCIENCE1} 
  M. Templin, A. Franck, U. Chesne, H. Leist, Y. Zhang, R. Ulrich,
  V. Sch\"adler, U. Wisner, Science 278, 1795 (1997).
\bibitem{SCIENCE2}
  F.S. Bates, Science 251, 898 (1991).
\bibitem{WET1}   
  M. Schick, Les Houches lectures on ``Liquids at interfaces'', 
  Elsevier Science Publishers B.V. (1990);
  S. Dietrich, in {\em Phase Transitions and Critical Phenomena}, Vol. 12, 
  C. Domb and J.L. Lebowitz (eds) NY Academic Press (1988).
\bibitem{WET2}   P. G. de Gennes, Rev. Mod. Phys. 57, 827(1985);
\bibitem{WET3}   A. O. Parry, J. Phys.: Cond. Matt. 8, 10761(1996);
                 R. Evans, J.Phys.:Cond.Matt. 2, 8989 (1990);
		 R. Evans, U. Marini, B. Marconi, P. Tarazona, J.Chem.Phys. 84, 2376 (1986).  
\bibitem{GINZBURG}  
  V. L. Ginzburg, Sov. Phys. Solid state 1, 1824(1960);
  P.-G. de Gennes, J. Phys. Lett. (Paris) 38, L-441(1977);
  J. F. Joanny, J. Phys. A 11, L-117(1978;.
  K. Binder, Phys. Rev. A 29, 341(1984).
\bibitem{holyst} For a recent review, see
  R. Holyst, T. A. Vilgis, Macromol. Theory and Simulations 5, 573(1996).
\bibitem{sharon} 
  S. G. Glotzer, in {\em Annual Reviews of Computational Physics II},
  p. 1, World Scientific, Singapore (1995).
\bibitem{SCF1}       
  E. Helfand, Y. Tagami, J. Polym. Sci. B 9, 741(1971);
  J. Chem. Phys. 56, 3592(1971);  57, 1812(1972);
  E. Helfand, A. M. Sapse, J. Chem. Phys. 62, 1327(1975);
  E. Helfand, J. Chem. Phys. {62}, 999 (1975).
\bibitem{SCF2}       
  J. Noolandi, K. M. Hong, Macromolecules 14, 727(1981); 15, 483(1982);
\bibitem{SCF3}
  K. R. Shull, Macromolecules 26, 2346(1993); 
  K. R. Shull, E. J. Kramer, Macromolecules 23, 4769(1990).
\bibitem{scheutjens} 
  J. M. H. M. Scheutjens, G. J. Fleer, J. Chem. Phys. 83, 1619(1979);
  D. N. Theodourou, Macromolecules 22, 4578(1989).
\bibitem{FREED1}
  H. Tang, K. F. Freed, J. Chem. Phys.  94, 1572(1991);
  K. F. Freed, J. Chem. Phys. 103, 3230(1995).
\bibitem{FH}  
  P.J. Flory, J. Chem. Phys. 9, 660(1941);
  H. L. Huggins, J. Chem. Phys. 9, 440(1941).
\bibitem{flory}
  P. Flory, {\em Principles of Polymer Chemistry} 
  (Cornell University Press, Ithaca, New York, 1971).
\bibitem{FREED2}
  A. I. Pesci, K. F. Freed, J. Chem. Phys. 90, 2003 and 2017(1989);
  J. Dudowicz, K. F. Freed, Macromolecules 24, 5076, 5096 and 5112(1991);
  K. W. Foreman, K. F. Freed, I. M. Ngola, J. Chem. Phys. 107, 4688(1997).
\bibitem{FREED3}     
  J. Dudowicz and K. F. Freed, Macromolecules 23, 1519(1990);
  M. Lifschitz and K. F. Freed, J. Chem. Phys. 98, 8994(1993).
  K. W. Foreman and K. F. Freed, J. Chem. Phys. 102, 4663(1995).
\bibitem{KIKUCHI}
  F. Aguilera-Granja, R. Kikuchi, 
    Physica A 176, 514(1990); 182, 331(1991); 189, 81 (1992); 189, 108(1992).
\bibitem{SCHWEIZERR} For a recent review, see
  K. S. Schweizer, J. G. Curro, in {\em Advances in Chemical Physics},
  Vol XCVIII, I. Prigogine and S. A. Rice (eds.), Wiley, New York(1997).
\bibitem{SCHWEIZER}
  C. J. Grayce, K. S. Schweizer, J. Chem. Phys. 100, 6846(1994);
  C. J. Grayce, A. Yethiraj, K.S. Schweizer, J. Chem. Phys. 100, 6857(1994).
\bibitem{binder1}
  {\em Monte Carlo and Molecular Dynamics Simulations in Polymer Science},
  K. Binder edt., Oxford University Press, New York (1995).
\bibitem{tanny}
T. B. Liverpool, in {\em Annual Reviews of Computational Physics IV},
  p. 317, World Scientific, Singapore (1996).
\bibitem{freedr} K.F. Freed, {\em Renormalization Group Theory of Macromolecules}, Wiley-Interscience, NY(1987).
\bibitem{hilde}
 J. H. Hildebrand, R. L. Scott, {\em The solubility of non-electrolytes},
 Dover, New York, (1964).
\bibitem{singh}
 C. Singh, K. S. Schweizer, J. Chem. Phys. 103, 5814(1995);
 C. Singh, K. S. Schweizer, Macromolecules 30, 1490(1997);
\bibitem{liu}
  G. H. Fredrickson, A. J. Liu, F.S. Bates, Macromolecules 27, 2503(1994).
\bibitem{higgins}
  J. Higgins, H. Benoit, {\em Polymers an Neutron Scattering},
  (Oxford University Press, New York, 1994).
\bibitem{bates1}
  F. S. Bates, M. Muthukumar, G. D. Wignall, L. J. Fetters,
  J. Chem. Phys. 89, 535(1988);
  J. D. Londono, A. H. Narten, G. D. Wignall, K. G. Honnell, E. T. Hsieh,
  T. W. Johnson, F. S. Bates, Macromolecules 27, 2864(1994).
\bibitem{krishna}
  R. Krishnamoorti, W. W. Graessley, N. P. Balsara, D. J. Lohse,
  J. Chem. Phys. 100, 3894(1994).
\bibitem{taylor2}
  J. K. Taylor, P. G. Debenedetti, W. W. Graessley, S. K. Kumar,
  Macromolecules 29, 764(1996).
\bibitem{kumar1}
  S. K. Kumar, B. A. Veytsman, J. K. Maranas, B. Crist,
  Phys. Rev. Lett. 79, 2265(1997).
\bibitem{VILGIS2}    R. Holyst and T.A. Vilgis, Phys.Rev. { E 50}, 2087 (1994).
\bibitem{WEINHOLD}   
 J. D. Weinhold, S. K. Kumar, C. Singh, K. S. Schweizer, 
   J. Chem. Phys. 103, 9460(1995).
\bibitem{STIFF1} M. M\"uller, Macromolecules 28, 6556(1995).
\bibitem{GARY1} 
  G. S. Grest, M. D. Lacasse, K. Kremer, A.M. Gupta, 
    J. Chem. Phys. 105, 10583(1996);
  G. S. Grest, M. D. Lacasse, M. Murat, MRS Bulletin 22, 27(1997).
\bibitem{Deutsch}    
	H.-P. Deutsch, K. Binder, Europhys. Lett. { 17}, 697(1992);
	H.-P. Deutsch, K. Binder, Macromolecules { 25}, 6214(1992).
\bibitem{M0} M. M\"uller, K. Binder, Macromolecules 28, 1825(1995).
\bibitem{Schweizer93} 
K. S. Schweizer, Macromolecules 26, 6033(1993) {\em ibid} 6050(1993);
K. S. Schweizer, J. Chem. Phys. 98, 9053(1993) {\em ibid} 9080(1993).
\bibitem{MBO}        
M. M\"uller, K. Binder, W. Oed, Faraday Trans. { 91}, 3269(1995).
\bibitem{SM}         
F. Schmid and M. M\"uller, Macromolecules { 28}, 8639(1995).
\bibitem{STIFF2} M. M\"uller, A. Werner, J. Chem. Phys. 107, 10764(1997).
\bibitem{FREDDI} F. Schmid, J. Chem. Phys. 104, 9191(1996).
\bibitem{hansen}
J.-P. Hansen, I.R. McDonald, {\em Theory of simple liquids}, Academic
 Press, London (1990).  
\bibitem{YETH} 
A. Yethiraj, K. S. Schweizer, J. Chem. Phys. { 98}, 9053(1993).
\bibitem{B1}    K. Binder, Adv. Poly. Sci. { 112} 181(1994).
\bibitem{B2}    K. Binder, Adv. Poly. Sci. in press (1998).
\bibitem{B3}    K. Binder, Acta Polymer 46, 204(1995).
\bibitem{Gehlsen}
H. Gehlsen, J. Rosedale, F. S. Bates, G. Wignall, L. Hansen, K. Almdal, 
  Phys. Rev. Lett. 68, 2452(1992).
\bibitem{Deutsch2}   H.-P. Deutsch, J. Stat. Phys. 67, 1039(1992).
\bibitem{PACK}       A. Yethiraj, R. Dickman, J. Chem. Phys. { 97}, 4468(1992).
\bibitem{BFM}        I. Carmesin, K. Kremer, Macromolecules { 21}, 2819(1988);
		     H.-P. Deutsch, K. Binder, J. Chem. Phys. { 94}, 2294(1991).
\bibitem{Paul}       
W. Paul, K. Binder, D. W. Heermann, K. Kremer, J. Chem. Phys. 95, 7726(1991).
\bibitem{MAP1}       
V. Tries, W. Paul, J. Baschnagel, K. Binder, J. Chem. Phys. { 106}, 738(1997);
W. Paul, N. Pistoor, Macromolecules { 27}, 1249(1994);
W. Paul, K. Binder, J. Batoulis, B. Pittel, K. H. Sommer, 
 Makromol. Chem., Macromol.Symp. { 64}, 1(1993).
\bibitem{MAP2}       
P. Doruker and W. L. Mattice, Macromolecules { 30}, 5520(1997).
\bibitem{SOLVENT}    
N. B. Wilding, M. M\"uller, K. Binder, J. Chem. Phys. 105, 802(1996).
\bibitem{RING}       
M. M\"uller, J. P. Wittmer, M. E. Cates, Phys. Rev. { E 53}, 5069(1996).
\bibitem{JOERG}      
J. Baschnagel, K. Binder, Macromolecules { 28}, 6808(1995).
\bibitem{CIFRA1}     
G. Y. A. Ypma, P. Cifra, E. Nies, A. R. D. vanBergen, 
Macromolecules 29, 1252(1996);
P. Cifra, E. Nies, F. E. Karasz, Macromolecules { 27}, 1166(1994).
\bibitem{VIRIAL}     
R. B. Pandey, A. Milchev, K. Binder, Macromolecules { 29}, 1194(1997).
\bibitem{SPINODAL}   
P. Wiltzius, F. S. Bates, J. Chem. Phys. 91, 3258(1989);
H. Tanaka, Phys. Rev. Lett. 71, 3158 (1993);
J. S. Langer, M. Bar-on, H. D. Miller, Phys. Rev. A 11 , 1417(1975).
\bibitem{ASYM}        M. M\"uller, K. Binder, Comp.Phys.Comm. 84, 173 (1994).
\bibitem{SARIBAN}    A. Sariban, K. Binder, Macromolecules { 21}, 711 (1988);
		     A. Sariban, K. Binder, J.Chem.Phys. { 86}, 5859 (1987).
\bibitem{COP}        
  M. M\"uller, M. Schick, J. Chem. Phys. 105, 8885(1996).
\bibitem{HPDASYM}    H.-P. Deutsch, J.Chem.Phys. 99, 4825 (1993).
\bibitem{ROVERE}     
M. Rovere, D.W. Heermann, K. Binder, Europhys.Lett. 6, 585 (1988);
                     J.Phys.: Condensed Matter 2, 7009 (1990).
\bibitem{MMFSS}      M. M\"uller, N. B. Wilding, Phys. Rev. E 51, 2079(1995).
\bibitem{NIGEL}      N. B. Wilding, Phys. Rev. E 52, 602(1995);
		     N. B. Wilding, J. Cond. Matter 9, 585(1997).
\bibitem{HISTO}      
A. M. Ferrenberg, R. H. Swendson, Phys. Rev. Lett. { 61}, 2635(1989);
		     {\em ibid.} { 63}, 1195(1989); 
		     J. Bennett, J. Comput. Phys. { 22}, 245(1979).
\bibitem{REWEIGHT}   B. A. Berg, T. Neuhaus, Phys. Rev. Lett. { 68}, 9(1992).
\bibitem{SIGMA}      K. Binder, Phys.Rev { A 25}, 1699(1982).
\bibitem{MBWET}      M. M\"uller, K. Binder, preprint December 1997.
\bibitem{MCOMP}      M. M\"uller, EPFL Supercomputer Review 7, 21(1995).
\bibitem{PORE}       
  M. M\"uller, M. Schick, J. Chem. Phys. 105, 8282(1996).
\bibitem{andreas}    A. Werner, F. Schmid, M. M\"uller, K. Binder, 
J. Chem. Phys. { 107}, 8175(1997).
\bibitem{BUFF} 
   F. P. Buff, R. A. Lovett, F. H. Stillinger, Phys. Rev. Lett. 15, 621(1965).
\bibitem{HELFRICH}   W. Helfrich, Z. Naturforsch.  28c, 693(1973);
		     P. B. Canham, J. Teor. Biol.  26, 61(1970).
		     E. Evans, Biophys. J. 14, 923(1974).
\bibitem{SEM_CAP}    
  A. N. Semenov, Macromolecules 26, 6617(1993); 27, 2732(1994).
\bibitem{JASNOV}     D. Jasnov, Rep. Prog. Phys. 47, 1059 (1984).
\bibitem{KLEIN}  
F. Scheffold, E. Eiser, A. Budkowski, U. Steier, J. Klein, L. J. Fetters, 
J. Chem. Phys. 104, 8786 and 8796(1996).
\bibitem{EXP_GINZ} 
F. S. Bates, I. H. Rosedale, P. Stepanek, T. P. Lodge, P. Wiltzius, 
G. H. Fredrickson, R. P. Hjelm, Phys. Rev. Lett. 65, 1839(1990);
D. Schwahn, S, Janssen, T. Springer, J. Chem. Phys. 97, 8775(1992);
D. W. Hair, E. K. Hobbie, J. Douglas, C. C. Han, Phys. Rev. Lett. 68, 2476(1992).
\bibitem{Honeycutt} J.D. Honeycutt, Macromolecules 27, 5377(1994).
\bibitem{Schwahn} D. Schwahn, K. Mortensen, T. Springer, H. Yee-Madeira, 
R. Thomas, J. Chem. Phys. 87, 6078(1987).
\bibitem{ROD} R. Dickman Comput. Polym. Sci. 1, 206(1991).
\bibitem{KUMAR_PD}
  S. K. Kumar, Macromolecules 27, 260(1994); 30, 5085(1997).
\bibitem{kumar2}
  S. K. Kumar, J. D. Weinhold, Phys. Rev. Lett. 77, 1512(1996).
\bibitem{VILGIS}     
R. Holyst, T.A. Vilgis, J. Chem. Phys. { 99}, 4835(1993).
\bibitem{MORSE}
  D. C. Morse, G. H. Fredrickson, Phys. Rev. Lett. 73, 3235(1994).
\bibitem{WORM}       
   N. Saito, K. Takahashi, Y. Yunoki, J. Phys. Soc. Jpn. 22, 219(1967).
\bibitem{tang}
  H. Tang, K. F. Freed, J. Chem. 94, 6307(1991).
\bibitem{ER}
  A. V. Ermoshkin, A. N Semenov, Macromolecules 29, 6294(1996);
  A. N. Semenov, J. Phys. II 6, 1759(1997).
\bibitem{SANCHEZ}    
see, {\em e.g.}, {\em Physics of Polymer Surfaces and Interfaces}, 
ed. I.C. Sanchez, Butterworth-Heinemann, Boston(1992).
\bibitem{RUSSELL}    
T. P. Russell, Materials Science Reports { 5}, 171(1990).
\bibitem{NOSE}       
T. Nose, T. Tanabe, Macromolecules { 30}, 5457(1997).
\bibitem{EX1}    
M. Stamm, D. W. Schubert, Ann. Rev. Mat. Sci { 25}, 326(1995);
D. W. Schubert, V. Abetz, M. Stamm , T. Hack, W. Siol, 
Macromolecules { 28}, 2519(1995).
\bibitem{EX2}    
B. L\"owenhaupt, G. P. Hellmann, Colloid \& Polymer Sci. { 268}, 885(1990).
\bibitem{EX3}    
H. E. Hermes, J. S. Higgins, D. G. Bucknall, Polymer { 38}, 985(1997).
\bibitem{EX4}    
K. H. Dai, E. J. Kramer, Polymer { 35}, 157(1994);
{\em ibid} Macromolecules { 25}, 220(1992).
\bibitem{REITER}     
J. Reiter, G. Zifferer, O. F. Olaj, Macromolecules 23, 224(1990).
\bibitem{GARY2} 
M. D. Lacasse, G. S. Grest, A.J. Levine, Phys. Rev. Lett. 80, 309(1998).
\bibitem{cop}
 Y. Wang, W. L. Mattice, J. Chem. Phys. 98, 9881(1993); 
 Y. Wang, Y. Li, W. L. Mattice, J. Chem. Phys. 99, 4068 (1993);
 A. C. Balasz, D. Gersappe, R. Israels, M. Fasolera,
  Makrom. Theory Simul. 4, 585 (1995).
\bibitem{JENSUWE}
 J.-U. Sommer, G. Peng, A. Blumen, J. Physique II 6, 1061 (1996);
   J. Chem. Phys. 105, 8376 (1996). 
\bibitem{pan}
  T. Pan, K. Huang, A. C. Balasz, M. S. Kunz, A. M. Mayes, T. P. Russell,
    Macromolecules { 26}, 2860(1993).
\bibitem{1}          
E. Helfand, S. M. Bhattacharjee, G. H. Fredrickson, J. Chem. Phys. 91, 7200(1989).
\bibitem{2}          
D. Broseta, G. H. Fredrickson, E. Helfand, L. Leibler, 
Macromolecules 23, 132(1990).
\bibitem{COP2}   
A. Werner, F. Schmid, K. Binder, M. M\"uller, Macromolecules { 29}, 9241(1996).
\bibitem{KRAMERC} 
C. Creton, E.J. Kramer, G. Hadziioannou, Macromolecules { 24}, 1846(1991);
J. Washiyama, C. Creton, E. J. Kramer, Macromolecules { 25}, 4751(1992);
E. J. Kramer, Adv. Polym. Sci. 52/53, 1(1983).
\bibitem{COP_EXP} 
P. F. Green, T. P. Russell, Macromolecules { 24} 2931(1991);
K. R. Shull, A. M. Mayes, T. P. Russell, Macromolecules { 26} 3929(1993).
\bibitem{LUDWIG}     
L. Leiber, Makromol. Chem. Macromol. Symp. 16, 1(1988); 
Physica A 172, 258(1991).
\bibitem{SEM_COP}    
A. N. Semenov, Macromolecules 25, 4967(1992).
\bibitem{TAUPIN}     
C. Taupin, P.G. deGennes, J. Phys. Chem. 86, 2294(1982).
\bibitem{BENDINGPOLY} 
M. W. Matsen, M. Schick, Macromolecules 26, 3878(1993); {\em ibid} 27, 2317(1994).
\bibitem{GOMPPER}    
G. Gompper, M. Kraus, Phys. Rev. E 47, 4289 and 4301(1993).
\bibitem{KAPPA}      
R. Cantor, Macromolecules 14, 1186(1981).
Z. G. Wang, S. A. Safran, J. Phys. France 51, 185(1990).
\bibitem{LIF} 
N. S Tonchev, D. I. Uzunov, Physica { A 134}, 265(1985);
O. V. Vasilev, K. A. Dawson, Phys. Rev. { E 50}, 2115(1994);
R. M. Hornreich et al., Z.Phys. { B 35}, 91(1979);
A. Erzan, G. Stell, Phys. Rev. { B 16}, 4146 (1977).
\bibitem{MICROB} 
M. C. Barbosa, M. Frichembruder, Phys. Rev. { 51}, 4690(1995).
\bibitem{NAKANISHI} 
M. E. Fisher, H. Nakanishi, J. Chem. Phys. 75, 5857(1981);
H. Nakanishi, M. E. Fisher, J. Chem. Phys. 78, 3279 (1983).
\bibitem{KUMAR}  
S. K. Kumar, H. Tang, I. Szleifer, Mol.Phys. { 81}, 867(1994).
\bibitem{YR}     
Y. Rouault, J. Baschnagel, K. Binder, J. Stat. Phys. { 80}, 1009(1995);
Y. Rouault, B. D\"unweg, J. Baschnagel, K. Binder, Polymer { 37}, 297(1996).
\bibitem{SB}     
I. Schmidt, K. Binder, J. Phys. 46, 1631(1985);
H. Nakanishi, P.Pincus, J. Chem. Phys. 79, 997(1983).
\bibitem{WANG}       
J. S. Wang, K. Binder, J. Chem. Phys. { 94}, 8537(1991);
G. G. Pereira, J. S. Wang, J. Chem. Phys.  { 105}, 3849(1996);
G. G. Pereira, J. S. Wang, J. Chem. Phys.  { 104}, 5294(1996);
G. G. Pereira, J. S. Wang, Phys. Rev.  { E 54}, 3040(1996);
G. G. Pereira, J. Chem. Phys.  { 106}, 4282(1997);
G. G. Pereira, J. Chem. Phys.  { 107}, 3740(1997).
\bibitem{PARRYSYM}  A. O. Parry, R. Evans, J.Phys.A  25, 275(1992).
\bibitem{PARRYASYM}  A. O. Parry, R. Evans, Physica A 181, 250(1992).
\bibitem{PARRYFLUC}  A. O. Parry, J. C. Boulter, Physica A 218, 77 and 109(1995).
\bibitem{FLUC}   
E. Brezin, B. I. Halperin, S. Leibler, Phys. Rev. Lett. 50, 1387(1983);
R. Lipowski, D. M. Kroll, R. K. Zia, Phys. Rev. B 27, 4499(1983);
D. S. Fisher, D. A. Huse, Phys. Rev. B 32, 247(1985). 
\bibitem{BSIM}   
K. Binder, D. P. Landau, A. M. Ferrenberg, Phys.Rev. E 51, 2823(1995);
K. Binder, R. Evans, D. P. Landau, A. M. Ferrenberg, Phys. Rev. E 53, 5023(1996).
\bibitem{KERLE}  
T. Kerle, J. Klein, K. Binder, Phys. Rev. Lett. 77, 1318(1996);
T. Kerle, J. Klein, K. Binder, preprint, January 1998.
\bibitem{JONES_CAP} 
M. Sferrazza, C. Xiao, R. A. L. Jones, D. G. Bucknall, J. Webster, J. Penfold, 
Phys. Rev. Lett. 78, 3693(1997).
\bibitem{TOCOME}
 A. Werner {\em et al}, work in preparation.
\bibitem{MOSELEY}
 L. L. Moseley, Int. J. Mod. Phys. C 8, 583 1997).
\bibitem{REACEXP}     
C. E. Scott, C. W. Macosko, Polymer { 35}, 5442 (1994); see also 
Macromol.Symposia.:Polym. Blends { 112}, 141-175(1996).
\bibitem{REACTION1}  
B. O'Shaughnessy, U. Sawhey, Macromolecules { 29}, 7230(1996); 
Phys. Rev. Lett. { 76}, 3444(1996).
\bibitem{REACTION2}  
S. T. Milner, G. H. Fredrickson, Macromolecules { 29}, 7386(1996); 
G. H. Fredrickson, Phys. Rev. Lett. { 76}, 3440(1996).
\bibitem{REACTION}   M. M\"uller, Macromolecules { 30}, 6353(1997).
\bibitem{DOIREACTION} R. Haswgawa, M. Doi, Macromolecules { 30}, 5490(1997).
\bibitem{SCHWEIZERDYN} 
K. S. Schweizer, M. Fuchs, G. Szamel, M. Guenza, H. Tang, 
Macromol. Theory Simul. { 6}, 1037(1997).
\bibitem{MURAT}  M. Murat, K. Kremer, preprint, October 1997.







%
\end{thebibliography}
\end{document}